\documentclass[english]{article}
\pdfoutput=1
\usepackage[T1]{fontenc}
\usepackage{geometry}
\usepackage[latin9]{inputenc}
\usepackage{xspace}
\usepackage[english]{babel}
\usepackage[T1]{fontenc}
\usepackage{lmodern}

\usepackage{color}
\usepackage{xcolor}
\usepackage[hang]{caption}
\usepackage{float}
\usepackage{morefloats}
\usepackage{datetime}
\usepackage{epsf}
\usepackage{epsfig}
\usepackage{psfrag}
\usepackage{verbatim}
\usepackage{graphicx}
\usepackage{graphics}
\usepackage{subfig}
\usepackage{tikz}
\usepackage{amsmath}
\usepackage{amsfonts,bm,bbm}
\usepackage{amssymb}
\usepackage{wasysym}
\usepackage{amsopn}
\usepackage{mathtools}
\usepackage{pifont}
\usepackage{array}
\usepackage{tabularx}
\usepackage{multirow}
\usepackage{booktabs}
\usepackage{ctable}
\usepackage{indentfirst}
\usepackage{enumerate}
\usepackage{enumitem}

\title{Pre-asymptotic dynamics of the infinite size Neumann\\
 ($p=2$ spherical) model}

\author{
{\large Damien Barbier$^1$, Leticia F. Cugliandolo$^1$, Gustavo S. Lozano$^2$, } 
\\
{\large Nicol\'as Nessi$^2$, Marco Picco$^1$ and Alessandro Tartaglia$^1$}
\\
$^1$Sorbonne Universit\'e \& CNRS,
Laboratoire  de  Physique  Th\'eorique  et  Hautes  Energies, UMR 7589, 
\\
4, Place Jussieu, 75252 Paris Cedex 05, France
\\
$^2$Departamento de F\'{\i}sica,
FCEYN Universidad de Buenos Aires \& IFIBA CONICET, 
\\
Pabell\'on 1, Ciudad Universitaria, 1428 Buenos Aires, Argentina
}

\begin{document}

\maketitle

\abstract{In this contribution we further study the classical disordered
$p=2$ spherical model with Hamiltonian dynamics, or in integrable systems terms, the Neumann
model, in the infinite size limit. We
summarise the asymptotic results that some of us presented in a recent publication,
and we deepen the analysis
of the pre-asymptotic dynamics. We also discuss the possible description of the
asymptotic steady state with a Generalised Gibbs Ensemble.}

\tableofcontents

\clearpage

\section{Introduction}

Interest in the dynamics of quantum systems in {\it perfect isolation} has been re-boosted by the
large activity in cold atom experiments~\cite{Bloch08} and, in parallel, the exact solution of one-dimensional
models~\cite{Polkovnikov10,Pasquale-ed,Gogolin}. A huge amount of work has been performed in the quantum context and a
host of results are now available for a variety of low-dimensional, mostly integrable, models. In
particular, {\it quantum quench} protocols have been intensively used. Typically, these
consist in following  the dynamics of a state prepared with some prescription and evolved with a
Hamiltonian of which it is not an eigenstate.

Questions on the equilibration, or not, of these systems are among the most prominent ones addressed in theoretical and experimental studies.
Whether an equilibrium, or an equilibrium-like, description of the asymptotic dynamics of
isolated systems exists, is a question that strictly makes sense in the limit of an infinite number of
degrees of freedom only.

Very similar issues can be addressed in the classical, and not necessarily low-dimensional,
context. Therefore, we very recently initiated a line of research aiming at clarifying the evolution of classical
interacting isolated systems in the thermodynamic limit~\cite{CuLoNe17,CuLoNePiTa18}. The latter condition 
is fundamental for our purposes since we wish to analyse the possible thermal properties of the long-term behaviour of the systems after sudden changes in their parameters, the classical equivalent of a quantum quench

Focusing on such thermodynamic limit, two classes of systems have been distinguished:
integrable and non-integrable. While for the former, no canonical equilibrium is expected because of the infinitely many integrals of motion, for the latter, interactions are
expected to be sufficiently efficient to lead them to standard equilibration.
These issues have been and still are actively studied in the context of quantum
isolated systems, see {\it e.g.}~\cite{Polkovnikov10,Pasquale-ed,Gogolin} for reviews,
but, surprisingly enough, they have not been equally addressed in the
realm of classical mechanics.

In the context of classical disordered models, we have checked that there is equilibration of interacting disordered systems, by studying the quenched dynamics of the isolated $p\geq 3$-spin disordered model~\cite{CuLoNe17} (excluding glassy features for quenches to the so-called threshold~\cite{CuKu93}). In a subsequent publication, we started the study of the $p=2$ case~\cite{CuLoNePiTa18},
a model that, as we explain below, is integrable and therefore not expected to equilibrate to the
conventional Gibbs-Boltzmann measure.

Our intention in~\cite{CuLoNePiTa18} was  to analyse the dynamics of a  classical integrable system that is not simply mappable to an ensemble of independent harmonic oscillators. We were especially interested in the limit of an infinite
number of degrees of freedom, $N\to\infty$. We
wished to establish whether the long-term dynamics approach a steady state that could be
described with a Generalised Gibbs Ensemble, a classical version of the ensembles used to characterise the asymptotic states of integrable quantum systems~\cite{Polkovnikov10,Pasquale-ed,Gogolin,Rigol07,Rigol08}. With this aim, we needed to fully understand the dynamics of the classical model in the long-times limit.

The model that we choose to study is very well-known in the field of disordered systems and
it goes under the name of  $p=2$ spherical model~\cite{KoThJo76,ShSi81,CidePa88,CuDe95a,CuDe95b,BoPaPaRi96a,BoPaPaRi96a,ZiKuHo00,BeCuIg01,BeDeGu01,ChCuYo06,DeGuMa07,FyPeSc15}. In the Hamiltonian version that we will study here, it is a well-known integrable model, that was proposed by Neumann in the
19th century~\cite{Neumann} and studied by several authors~\cite{Uhlenbeck,AvTa90,BaTa92}
more recently.
In Ref.~\cite{CuLoNePiTa18} we studied its dynamics in great detail. In particular, we elucidated the dynamic phase diagram for quenches from initial configurations drawn from a
canonical equilibrium measure at an inverse temperature $T'$, after quenches that we specify below.
This choice of initial conditions corresponds to taking the system  in equilibrium with a bath at the chosen initial temperature, $T'$, at times $t<0$, and then suddenly
switching off the coupling at $t=0$. If no change in the Hamiltonian is done, the
dynamics remains in equilibrium at $T'$. If, instead, a change in the Hamiltonian is operated at
$t>0$, the system is subject to a sudden quench and the subsequent dynamics are out of equilibrium.

In~\cite{CuLoNePiTa18} we did not,
though, present a detailed study of the way in which the asymptotic states
are reached. We fill this gap here investigating the pre-asymptotic dynamics of the model for $N\to\infty$. We will elucidate the pre-asymptotic behaviour and distinguish it from the one of the
dissipative problem that is not able to equilibrate for certain choices of parameters either,
but for different reasons.

The article is organised as follows. In Sec.~\ref{sec:model} we recall the definition of the model
and we give some more details on the quench performed. Section~\ref{sec:dynamics} is
devoted to the study of the pre-asymptotic dynamics. Finally, in Sec.~\ref{sec:conclusions}
we present our conclusions and some discussion of lines for further research.

\section{The model}
\label{sec:model}

In this Section we briefly present the definition of the model, the initial conditions that we choose, and the kind of quench that we perform. A more detailed summary of the behaviour of the model in canonical equilibrium and its relaxational dynamics driven by the coupling to a thermal bath can be found in~\cite{CuLoNePiTa18}.

\subsection{The Hamiltonian}

The potential energy of the $p=2$ spin model is the one of
a system with two-spin interactions mediated by
quenched random couplings $J_{ij}$:
\begin{eqnarray}
H_{\rm pot}[\{s_i\}] &=& - \frac{1}{2} \sum^N_{i\neq j} J_{ij} s_{i} s_{j}
\; .
\label{eq:pspin-pot}
\end{eqnarray}
The exchanges $J_{ij}$ are independent identically distributed random variables drawn from a Gaussian probability distribution  function with average and variance
\begin{equation}
[ J_{ij } ] =0
\; , \qquad\qquad
[ J^2_{ij } ] = \frac{J^2}{N}
\; ,
\label{eq:disorder-statistics}
\end{equation}
respectively.
All couplings are symmetric under the exchange of their indices, $J_{ij}=J_{ji}$, and the parameter $J$ characterises the width of the Gaussian distribution.
We use continuous ``spin''
variables, $-\infty \leq s_i \leq \infty$ with $i=1, \dots, N$, globally forced to satisfy (on average) a spherical
constraint, $\sum_{i=1}^N s_i^2 = N$, with $N$ the total number of spins~\cite{KoThJo76}, that is imposed by a term
\begin{equation}
H_{\rm constr}[\{s_i\}, z] = \frac{z}{2} \ \left(\sum_{i=1}^N s_i^2 - N \right)
\end{equation}
added
to the Hamiltonian. $z$ is a Lagrange multiplier that depends on the system's configuration. The spins thus defined do not
have intrinsic  dynamics. In statistical physics applications their temporal evolution
is given by the coupling to a thermal bath, {\it via} a
Langevin equation~\cite{KoThJo76,ShSi81,CidePa88,CuDe95a,CuDe95b,ZiKuHo00,BeCuIg01,BeDeGu01,ChCuYo06,DeGuMa07,FyPeSc15} or a 
Monte Carlo rule~\cite{BoPaPaRi96a,BoPaPaRi96b}.

The model is endowed with conservative dynamics by changing the
``spin'' interpretation into a ``particle'' one~\cite{CuLoNe17,CuLoNePiTa18}. In this way, one defines a kinetic energy~\cite{CuLo98,CuLo99,CuGrLoLoSa02}
\begin{eqnarray}
H_{\rm kin}[\{\dot s_i\}] = \frac{m}{2} \sum_{i=1}^N (\dot s_i)^2
\; ,
\label{eq:pspin-kin}
\end{eqnarray}
that is added to the potential and constraint terms to yield
the total energy of the {\it Hamiltonian spherical $p=2$-spin model}
\begin{equation}
H_{\rm syst} = H_{\rm kin} + H_{\rm pot} + H_{\rm constr}
\; .
\label{eq:pspin-total-energy}
\end{equation}
The model represents now a particle constrained to move on an $N$-dimensional hyper-sphere
with radius $\sqrt{N}$. Its position  is given by the $N$-dimensional vector $\vec s=(s_1, \dots, s_N)$ and its velocity
by another $N$-dimensional vector
$\dot{\vec s}=(\dot s_1, \dots, \dot s_N)$ .
The $N$ coordinates $s_i$  are globally constrained
to lie, as a vector, on a hypersphere with radius $\sqrt{N}$. The velocity vector $\dot {\vec s}$ is, on average, perpendicular to
$\vec s$, due to the spherical constraint. The parameter $m$ is the mass of the particle.

\subsection{The dynamic equations}

The $N$ equations of motion for the isolated system are
\begin{equation}
m \ddot s_i(t) + z(t) s_i(t) =  \sum_{j (\neq i)} J_{ij} s_{j}(t)
\; ,
\label{eq:dynamic-p2}
\end{equation}
where the Lagrange multiplier needs to be time-dependent to enforce the spherical constraint all along the evolution.

Equation~(\ref{eq:dynamic-p2}) yields an identity between the  energy density and the Lagrange multiplier. Its product with $s_i(t_2)$ and the limit $t_2 \to t_1^-$ imply
\begin{equation}
\lim_{t'\to t^-} m \partial_{t^2} C(t,t') + z(t) = - 2 e_{\rm pot}(t)
\; .
\end{equation}
The first term can be rewritten as $m\lim_{t'\to t^-} \partial_{t^2} C(t,t') = - m \lim_{t'\to t^-}  \sum_{i=1}^N \dot s_i(t) \dot s_i(t') = - m  \sum_{i=1}^N (\dot s_i(t) )^2$, and one has
\begin{equation}
 z(t) = - 2 e_{\rm pot}(t) + 2 e_{\rm kin}(t)
 \; .
 \label{eq:relation-z-energies}
\end{equation}
The Lagrange multiplier takes the form of an action density, as a difference between kinetic and potential
energy densities.
Using now the conservation of the total energy, $e_f=e_{\rm pot}(t) + e_{\rm kin}(t)$, one obtains
\begin{equation}
 z(t) =2 e_f - 4 e_{\rm pot}(t) = - 2 e_f +  4 e_{\rm kin}(t)
 \; .
 \label{eq:relation-z-energies2}
\end{equation}

\subsection{The initial conditions}

The initial condition will be taken to be $\{s_i^0, {\dot s}_i^0\} \equiv \{ s_i(0), {\dot s}_i(0)\}$ and chosen in ways that we specify below.
We will be interested in using equilibrium initial states drawn from a Gibbs-Boltzmann measure at different temperatures
$T'$.

\subsection{The quench}

The quench is performed in such a way that $J_{ij}^0 \mapsto J_{ij} = J/J_0 \, J_{ij}^0$ instantaneously. This means that each interaction strength is rescaled by a parameter $J/J_0$
so quickly that neither the position not the velocity of the particle are changed. It is not hard to
prove that under these conditions energy is either injected ($J/J_0<1$)
or extracted ($J/J_0>1$)  from the system.

\section{Dynamics of the infinite size system}
\label{sec:dynamics}

In this paper we will only discuss the dynamics in the $N\to\infty$ limit.
Several methods can be applied to derive, in this limit, exact integro-differential
equations for only three observables, the Lagrange multiplier, the
self-correlation and the linear  response, that fully characterise the collective
dynamics of the system. The long times limit will only be taken next. The
relevant order of limits is, therefore,
\begin{equation}
\lim_{t\to\infty} \lim_{N\to\infty}
\; .
\end{equation}

\subsection{Correlations, linear response and Lagrange multiplier}

The self correlation and linear response are defined as
\begin{eqnarray}
&&
C_{ab}(t,t') = \frac{1}{N} \sum_{i=1}^N [\langle s^a_i(t) s^b_i(t')\rangle_{i.c.}]
\; ,
\\
&&
C_{ab}(t,0) = \frac{1}{N} \sum_{i=1}^N [\langle s^a_i(t) s^b_i(0)\rangle_{i.c.}]
\; ,
\\
&& R_{ab}(t,t') =
\left.
\frac{1}{N} \sum_{i=1}^N \frac{\delta [\langle s^a_i\rangle^{(h)}_{i.c.}]}{\delta h^b(t')} \right|_{h=0}
\; ,
\end{eqnarray}
where the averages $\langle \dots \rangle_{i.c.}$ are taken over the probability distribution of the initial
conditions. The $a$ and $b$ indices run over replica indices, since we will
use this method to enforce the initial conditions. The infinitesimal perturbation
$h$ is linearly coupled to the spin, $H \mapsto H-h \sum_i s_i$. The
upper-script ${(h)}$ indicates that the configuration is measured after
having applied the field $h$. Because of
causality, the linear response is non-zero only for $t>t'$. The dynamics is causal and, in
consequence, the linear response is proportional to a Heaviside theta function $\theta(t-t')$.

In the $N\to\infty$ limit, the choice of the interaction matrix $J_{ij}$ is irrelevant
in the sense that any one will be a typical one and the  dynamics will be statistically
the same. Still, in the calculation that we use to
derive the Schwinger-Dyson equations an average over
quenched randomness is taken.  The square brackets denote here and
everywhere in the paper this average.

The Lagrange multiplier $z(t)$ is fixed by  the condition
$C_{aa}(t,t)=1$ and we have already explained that it equals twice the difference
between kinetic and potential energy densities.

\subsection{Dynamic equations}

In the $N\to\infty$ limit exact causal Schwinger-Dyson equations
for the conservative dynamics for initial conditions drawn from the
Gibbs-Boltzmann probability measure with Hamiltonian $H_0$ and
inverse temperatrue $\beta'$ are derived. They determine the
evolution of the time-delayed self-correlation and linear response
\begin{eqnarray}
&&
\!\!\!\!\!
[m\partial^2_t-z(t)] R(t,t_w) = \int dt'  \; \Sigma(t,t') R(t',t_w) +\delta(t-t_w)
\; ,
\\
&&
\!\!\!\!\!
[m\partial^2_t-z(t)] C(t,t_w) =
\int dt'  \left[ \Sigma(t,t') C(t',t_w) + D(t,t') R(t_w,t') \right]
\\
&&
\qquad\qquad\qquad\qquad\qquad
+ \frac{\beta'J_0}{J}  \sum_{a=1}^n D_a(t,0) C_a(t_w,0)
\; ,
\\
&&
\!\!\!\!\!
[m\partial^2_t-z(t)] C_a(t,0) =
\int
\! dt'  \,
\Sigma(t,t') C_a(t',0)
+ \frac{\beta' J_0}{J}  \sum_{a=1}^n D_b(t,0) Q_{ab}
\; ,
\end{eqnarray}
where the replica indices are $a=1,\dots, n\to 0$
since we have used the replica method to deal with $e^{-\beta' H_0}$
and fix the overlap replica matrix $Q_{ab}$~\cite{HoJaYo83,FrPa95,BaBuMe96}.
The two kernels are the self-energy $\Sigma$ and the vertex $D$ and for the
$p=2$ potential they take the forms
\begin{eqnarray}
&&
D(t,t_w) \; = \; J^2 \ C(t,t_w)
\; ,
\\
&&
D_a(t,0) \; = \; J^2  \ C_a(t,0)
\; ,
\\
&&
\Sigma(t,t_w) \; = \; J^2 \  R(t,t_w)
\; .
\end{eqnarray}
The causality of the linear response, and hence of the self-energy that is proportional to
the linear response, imposed the upper limit of  the integrals to be either $t'$ or $t$. The
lower limit is always the initial time that we call $t=0$.
The equation for the Lagrange multiplier can be written as
\begin{eqnarray}
z(t) =
-m\partial_{t}^{2}C(t,t')\vert_{t'\rightarrow t^{-}}
 +
2J^2\int dt'' \ R(t,t'')C(t,t'')
+ \frac{J J_0}{T'} \sum_a (C_a(t,0))^2
\label{eq:dyn-eqs-z-0}
\; .
\end{eqnarray}
	
\subsection{Initial conditions}

We have chosen to work with initial conditions drawn from the canonical
equilibrium probability distribution built with the Hamiltonian $H_0$, with the
$p=2$ form, and variance of the coupling strengths proportional to $J^2_0$.
The canonical equilibrium properties of the model~\cite{KoThJo76}
are such that there is a phase transition at $T_c=J_0$ between a disordered
paramagnetic phase at high temperature and an ordered phase at low temperature.
In the replica formalism the two phases are distinguished by the structure of the
replica matrix $Q_{ab}$~\cite{KoThJo76}:

Disordered initial conditions $\;\;\;T'>J_0$
\hspace{1cm} $Q_{ab} = \delta_{ab} \; .$

Ordered initial conditions $\;\;\;\;\;\;\; T'<J_0$
\hspace{1cm} $Q_{ab} = (1-q_{\rm in}) \delta_{ab} + q_{\rm in} \; .$ \\
with $q_{\rm in}$ the Edwards-Anderson parameter that is equal to
\begin{equation}
q_{\rm in}=1-T'/J_0
\; . 
\end{equation}  
This model
is particularly simple since it is solved by a replica symmetric {\it Ansatz}.
The two cases listed above are encoded by the second form if one simply
keeps in mind that the $T'>T_c$ corresponds to setting $q_{\rm in}=0$.

The Lagrange multiplier varies with temperature in the disordered phase
while it gets fixed to the highest eigenvalue of the interaction matrix
at low temperature:

Disordered initial conditions $\;\;\; T'>J_0$
\hspace{1cm} $z=T'+J_0^2/T' \; .$

Ordered initial conditions $\;\;\;\;\;\;\; T'<J_0$
\hspace{1cm} $z=2J_0 \; .$

Equilibrium below $T_c$ corresponds to the condensation of the system in the direction of the 
largest eigenvector of the random matrix~\cite{KoThJo76}; for this reason we will refer to this case as the 
condensed one.

\subsection{Replica structure in the course of time}

Although we change the Hamiltonian from $H_0$ to $H$ at the initial time $t=0$,
the dynamic equations keep the initial replica structure. There will be two kinds of correlations with the initial condition
\begin{equation}
C_1(t,0) \equiv C_{a=1 \, b=1}(t,0)  \qquad\mbox{and}\qquad C_{b\neq 1}(t,0)  \equiv C_{a=1 \, b\neq1}(t,0)
\; ,
\end{equation}
where we singled out the replica labeled one and we shortened the
notation  to keep only one subscript.
The interpretation of the correlations $C_1(t,t')$ and $C_{b\neq 1}(t,t')$
is similar to the one of real replicas. The former is the self-correlation between
one replica $\{s_i\}(t')$ and the same one further evolved until a later time $t$,
$\{s_i\}(t)$. For this reason, $C_1(t,t') \mapsto C(t,t')$. The latter is the correlation between replica
labeled $b$, let us say $\{\sigma_i\}(t')$, at time $t'$ and the
singled out replica one evolved until
time $t$ and represented by $\{s_i\}(t)$.
It turns out that only $C_b(t,0)$, the correlation with replica $b$
evaluated at the initial time, intervenes in the other equations.
This is the only correlation between different replicas that
appears in the calculations.

The border conditions on the correlation functions then are
\begin{eqnarray}
\begin{array}{ll}
&C_1(t,0)= C(t,0)
\qquad
\mbox{implying} \qquad
C_1(0,0)= C(0,0) =1
\; ,
\\
&
C_{b\neq 1}(0,0) = Q_{1,2} \qquad
\mbox{implying} \qquad C_{b\neq 1}(0,0) = q_{\rm in}
\; .
\end{array}
\end{eqnarray}

\subsection{Equal time values}

The equal-time conditions are
\begin{eqnarray}
&&
C(t,t)  =  1 \; ,
\qquad\qquad\qquad\qquad
R(t,t) =  0 \; ,
\\
&&
\partial_{t}C(t,t')\vert_{t'\rightarrow t^{-}}=\partial_{t}C(t,t')\vert_{t'\rightarrow t^{+}}  =  0 \; ,
\\
&& \partial_{t}C_{b\neq 1}(t,t')\vert_{t'\rightarrow t^{-}}=
\partial_{t}C_{b\neq 1}(t,t')\vert_{t'\rightarrow t^{+}}  =  0 \; ,
\\
&& \partial_{t}C_{b\neq 1}(t,0)\vert_{t\rightarrow 0^+}  =  0 \; ,
\qquad\qquad
\partial_{t}R(t,t')\vert_{t'\rightarrow t^{-}}  = \frac{1}{m} \; ,
\end{eqnarray}
for all times $t, t'$ larger than or equal to  $0^+$, when the dynamics start.

\subsection{The final set of equations}

The set of equations that fully determine the evolution of the system from an initial condition in
canonical Boltzmann equilibrium at any temperature $T'$ are
\begin{eqnarray}
\left(m\partial_{t}^{2}+z(t)\right)R(t,t') \! &  \!\! = \!\! & \!
\delta(t-t')
+
J^2 \int_{t'}^{t}dt''  R(t,t'') R(t'',t')
\; ,
\label{eq:dyn-eqs-R}\\
\left(m\partial_{t}^{2}+z(t)\right)C(t,t') \!\! &  \!\! = \!\! & \!\!
J^2\int_{0}^{t}dt'' R(t,t'') C(t'',t')
+ J^2\int_{0}^{t'}dt'' R(t',t'')C(t,t'')
\;
\nonumber\\
&&
+\frac{J J_0}{T'} \; [C(t,0) C(t',0) - C_{b\neq 1}(t,0) C_{b\neq 1}(t',0)]
\; ,
\label{eq:dyn-eqs-C}
\\
\left(m\partial_{t}^{2}+z(t)\right)C_{b\neq 1}(t,0) \!\! &  \!\! = \!\! & \!\!
J^2\int_{0}^{t}dt'' R(t,t'') C_{b\neq 1}(t'',0)
+ \frac{J J_0}{T'} \left[  q_{\rm in} C(t,0)  + (1-2q_{\rm in})  C_{b\neq 1}(t,0)  \right]
\; ,
\label{eq:dyn-eqs-C1}
\\
z(t) \!\! &  \!\! = \!\! & \!\!
\left. -m\partial_{t}^{2}C(t,t')\right|_{t'\rightarrow t^{-}}
\!\!\!\! +
2J^2 \!\! \int_{0}^{t}dt'' R(t,t'')C(t,t'')
 +
\frac{J J_0}{T'} \, [C^2(t,0) - C^2_{b\neq 1}(t,0)]
\label{eq:dyn-eqs-z}
\; .
\end{eqnarray}
High and low temperature initial states are distinguished by $q_{\rm in}=0$ for $T'>J_0$, and $q_{\rm in} =1-T'/J_0$
for $T'<J_0$, respectively.
The equation for $C(t,0)$ is just the one for $C(t,t')$ evaluated at $t'=0$ so we do not write it explicitly.

\subsection{Asymptotic analysis}

We have shown in~\cite{CuLoNePiTa18} that the dynamics approach an asymptotic limit
such that $z(t)$ reaches a constant
$z(t) \to z_f$ and $R(t,t')$ a stationary function of the two times, $R(t-t')$.
In this limit the response equation can be Fourier transformed and Eq.~(\ref{eq:dyn-eqs-R}) determines
its  frequency dependence
\begin{equation}
\hat R(\omega)
=
\frac{1}{2 J^2} \left[ -m\omega^2 + z_f \pm m \sqrt{(\omega_-^2-\omega^2) (\omega_+^2-\omega^2)} \right]
\label{eq:response_fourier_transform-text1}
\end{equation}
with
\begin{equation}
m\omega_{\pm}^2 = z_f\pm 2J
\label{eq:omega_pm}
\end{equation}
(note the unusual choice of sign for the imaginary part that we adopted in~\cite{CuLoNePiTa18}).
The numerical solution of the full dynamical equations show that the relevant sign is the
minus one.
$\hat R(\omega)$, and also $R(t)$,  are independent of the initial
temperature for $T'<T_c=J$ while they depend on temperature
through $z_f$ for $T'>T_c=J$.

In terms of the physical parameters,
$\hat R(\omega)$ is real for $|-m\omega^2+z_f|> 2J$.
In two of the dynamical phases that we identified in Ref.~\cite{CuLoNePiTa18},
$z_f=2J$, and this implies that the lower characteristic frequency vanishes
$\omega_-=0$ and the imaginary part
of the linear response is gapless.

\subsection{Phase diagram}

\begin{figure}[h!]
\begin{center}
\includegraphics[scale=0.6]{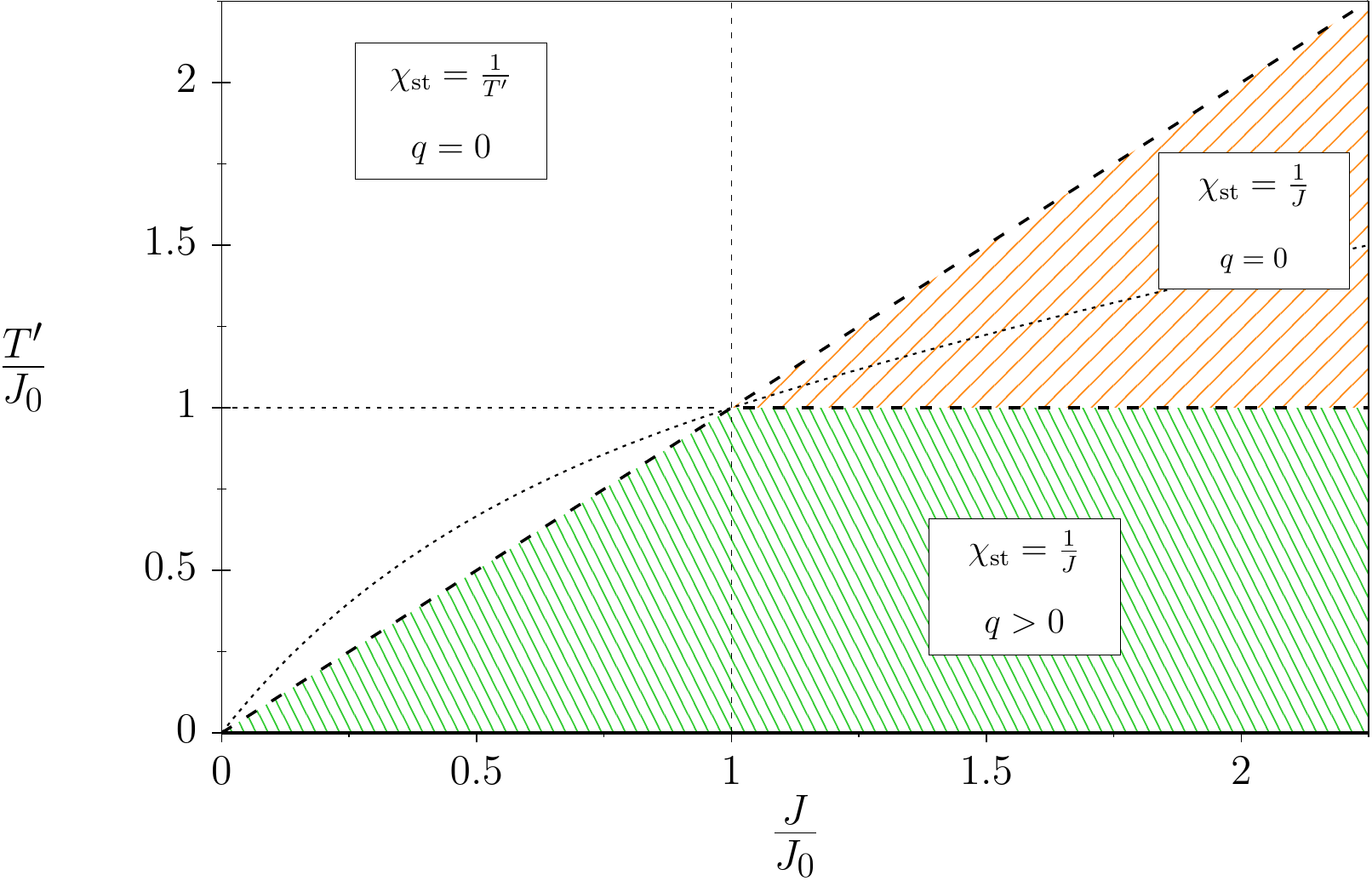}
\end{center}
\caption{\small
The dynamic phase diagram.
}
\label{fig:phase-diagram}
\end{figure}

The phase diagram in Fig.~\ref{fig:phase-diagram} has three phases that are
distinguished by the asymptotic value of the Lagrange multiplier, $z_f$,
the static susceptibility, $\chi_{\rm st} = \hat R(\omega=0)$,
 the long-time limits of the delayed
self-overlap, 
\begin{equation}
q \equiv \lim_{t-t' \to\infty} \lim_{t' \to\infty} C(t,t')
\; , 
\label{eq:def-q0}
\end{equation}
and the overlap with the initial condition, 
\begin{equation}
q_0 \equiv  \lim_{t \to\infty} C(t,0)
\; . 
\label{eq:def-q}
\end{equation}
The first four columns in Table 1 summarise the values that these
quantities take in the three phases.

\begin{table}[h!]
\vspace{0.5cm}
\begin{center}
\begin{tabular}{|c||c|c|c|c|c|c|}
\hline
 & $z_f$ & $\chi_{\rm st}$ & $q_0$ & $q$ & $m\omega^2_-=z_f - 2J$ & $m\omega^2_+=z_f + 2J$
\\
\hline
\hline
I & $T'+J^2/T'$ & $1/T'$ & 0 & 0 & $(\sqrt{T'}-J/\sqrt{T'})^2$ & $(\sqrt{T'}+J/\sqrt{T'})^2$
\\
\hline
\textcolor{orange}{II} & $\displaystyle{2J}$ & $1/J$ & 0 & 0 & 0& $4J$
\\
\hline
\textcolor{green}{III} & $\displaystyle{2J}$ & $1/J$ & $\neq 0$ & $\neq 0$ & 0 & $4J$
\\
\hline
\end{tabular}
\end{center}
\label{table1}
\caption{\small This table summarises the asymptotic values of the Lagrange multiplier,
the correlation with the initial condition and the infinitely long time delay of the
self correlation in the three first columns. The last two columns show the parameter
dependence of the two characteristic frequencies $\omega_\pm$.}
\end{table}

\subsection{The various overlaps}

\begin{figure}[h!]
\begin{center}
\includegraphics[scale=0.65]{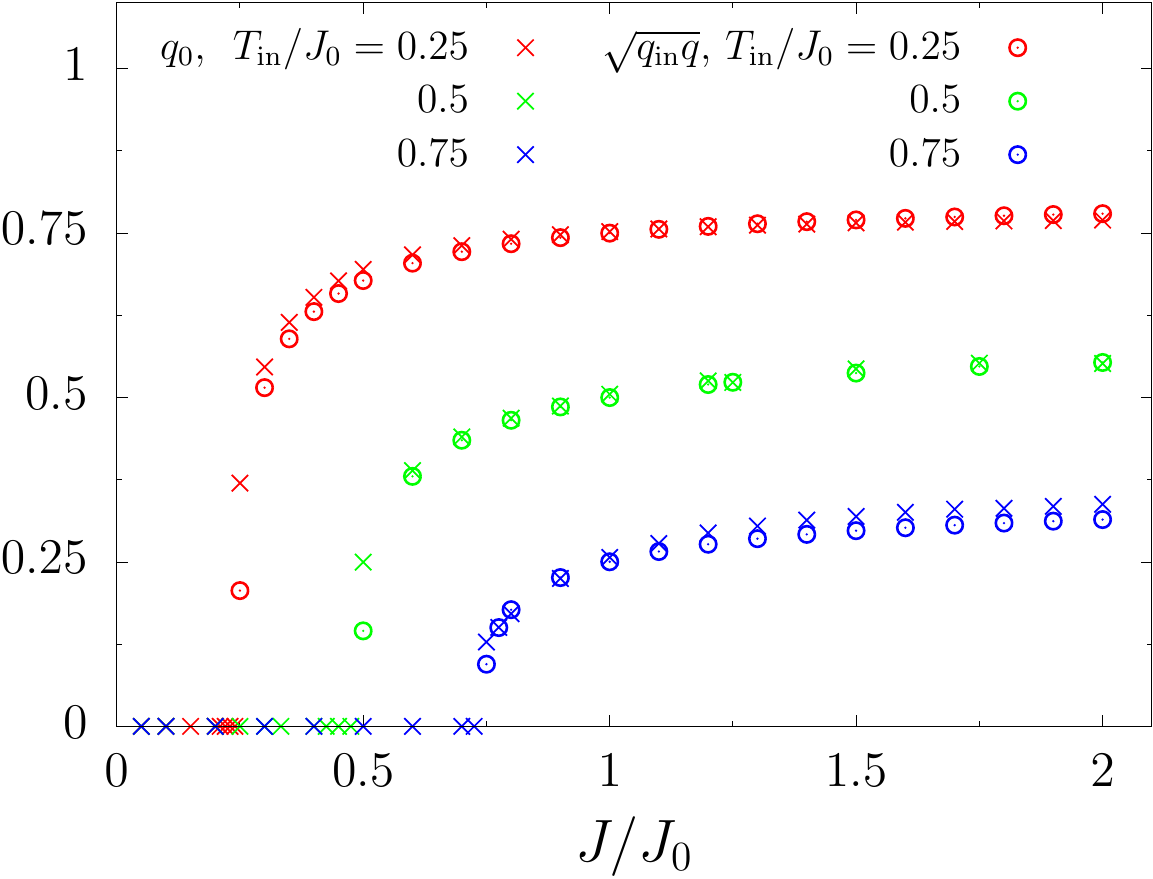}
\end{center}
\caption{\small
The asymptotic value $q_0 = \lim_{t \to +\infty} C(t,0)$ estimated from the numerics (crosses)
and the quantity $ \sqrt{q_{\rm in} q}$ (circles), where 
$q = \lim_{\tau \to +\infty} C(t_2 + \tau,t_2)$, for $t_2 \gg 1$, is also estimated from the numerics.
The two quantities are plotted against $J/J_0$,
for three choices of the initial condition ($T'/J_0$) given in the key.
}
\label{fig:asympt-qs}
\end{figure}

In Fig.~\ref{fig:asympt-qs} we show the numerical value of  the overlap with the initial condition, 
$
q_0 \equiv  \lim_{t \to\infty} C(t,0)
$,
and we compare it to $\sqrt{q_{\rm in} q}$, where $q$ is the one of the self-overlap, 
$
q \equiv \lim_{t-t' \to\infty} \lim_{t' \to\infty} C(t,t')
$,
 and $q_{\rm in} = 1-T'/J_0$ is the overlap of the initial condition. 
 The data are plotted as a function of the parameter $J/J_0$ that controls the energy injection or extraction. 
 The agreement of the numerical data points is very good for the three initial conditions considered, 
 that are made explicit in the key (see App.~\ref{app1} and Ref.~\cite{Ba97}).

\vspace{0.3cm}

\subsection{Approach to the asymptotic limit}

In this Section we present the rest of the original results in this contribution. They concern the
way in which the functions reach the asymptotic values $z_f$, $q_0$ and $q$,
recalled in Table~1 and confronted in Fig.~\ref{fig:asympt-qs}.
We will find that the approach is always algebraic with oscillations, and we will study
the exponent as well as the relevant frequencies involved in the periodic behaviour.
The dependence of the exponents for $|z(t)-z_f|$, $|C(t,0)-q_0|$ and $|C(t,t')-q|$
on the control parameters $T'/J_0$ and $J/J_0$
is given in Figs.~\ref{fig:exp_decay_z}~(a) and (b), and Fig.~\ref{fig:exp_decay_corr_lt}, respectively.

\subsubsection{Pre-asymptotic behaviour of $z(t)$}
\label{sec:Lagrange-multiplier}

In~\cite{CuLoNePiTa18} we established that the long-time limit of $z(t)$ is given by the
two values in the second column in Table~1 at the two sides of the diagonal
$T'/J_0 = J/J_0$ in the phase diagram. Let us now study the approach to this value.

\begin{figure}[h!]
 \vspace{0.5cm}
\begin{center}
$\;$ \hspace{1.75cm} (a) \hspace{5.5cm} (b) \hspace{4cm} $\;$
\vspace{0.2cm}
 \includegraphics[scale=0.4]{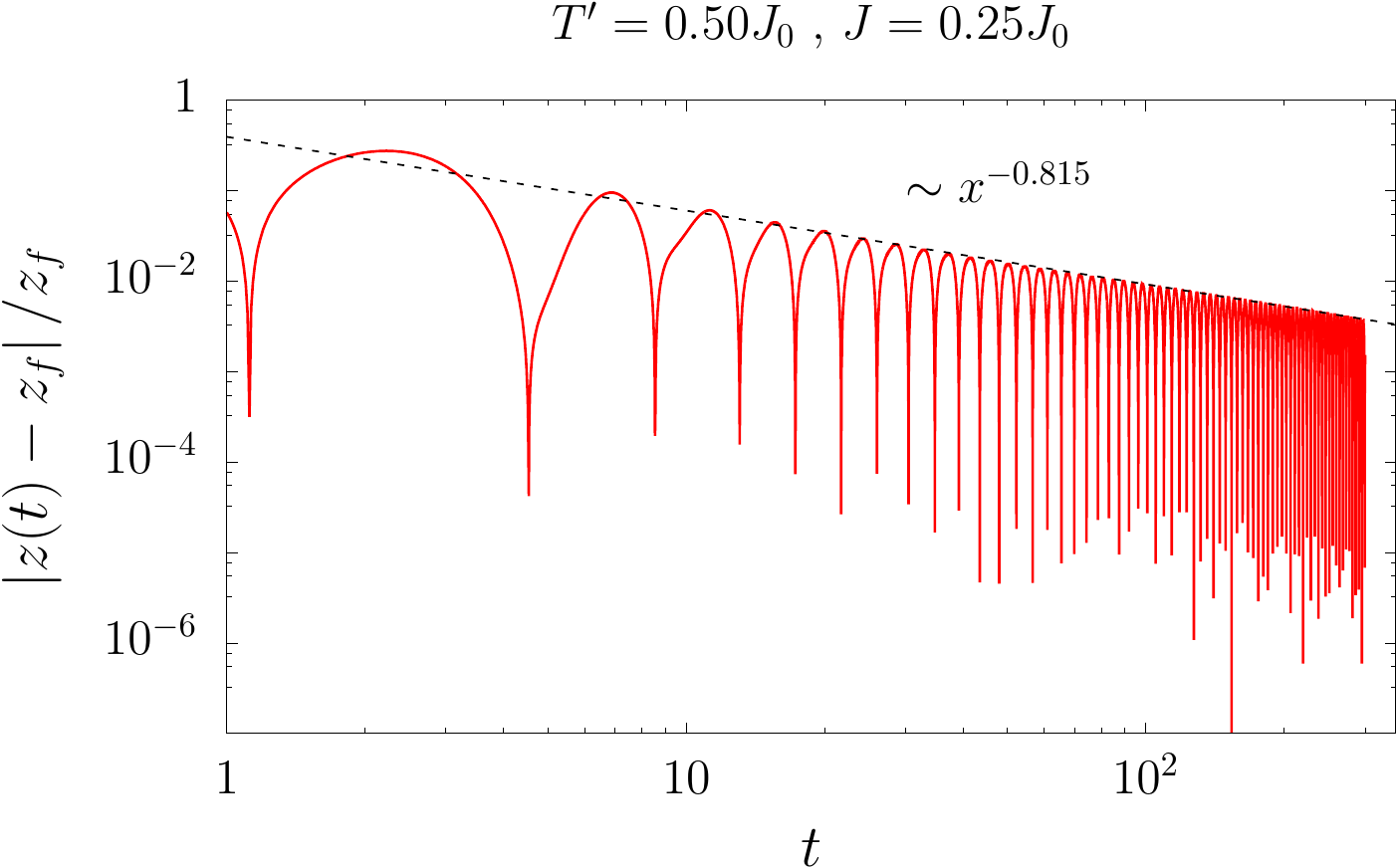}\quad%
 \includegraphics[scale=0.4]{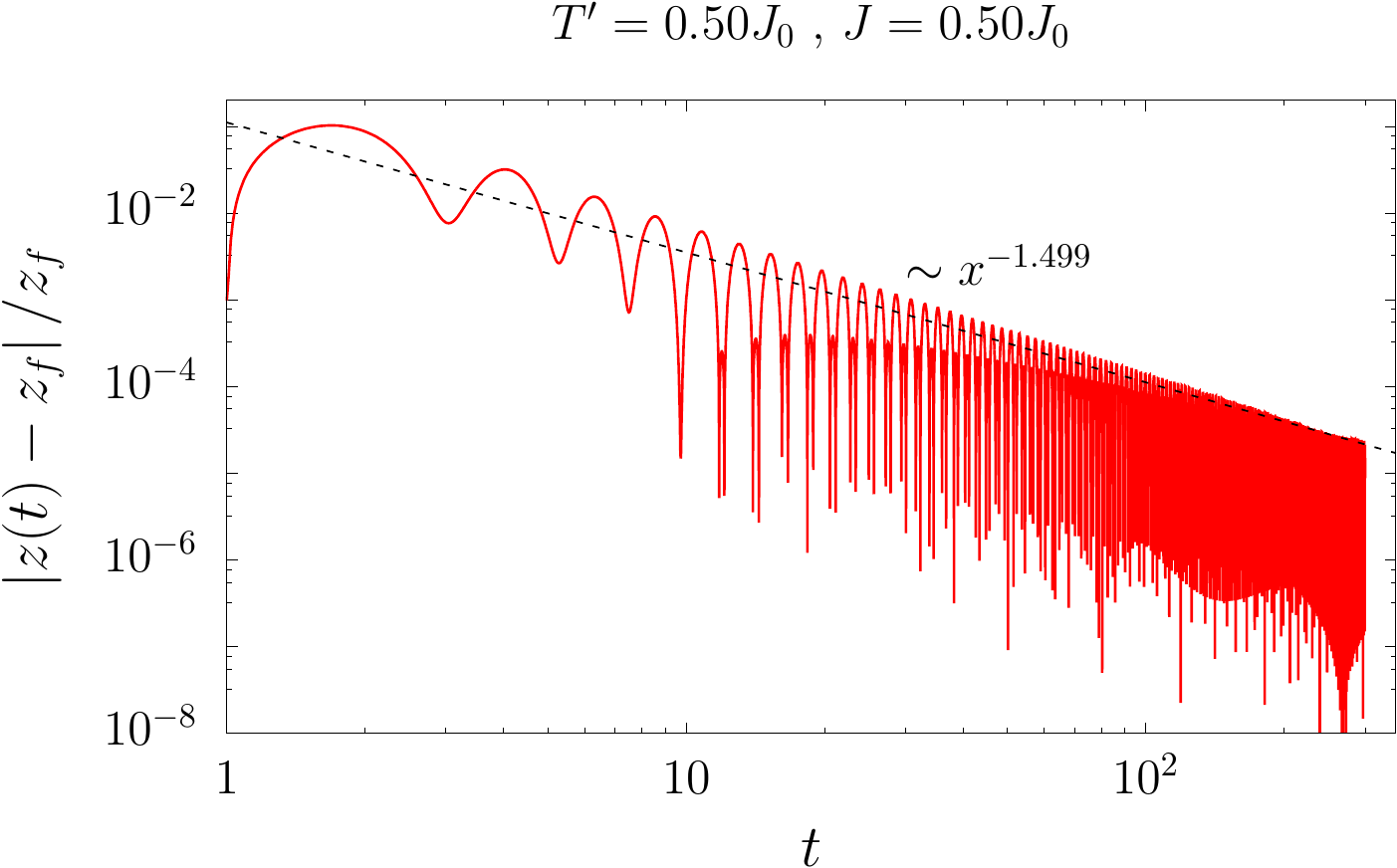}%

 \vspace{0.5cm}
 $\;$ \hspace{1.75cm} (c) \hspace{5.5cm} (d) \hspace{4cm} $\;$
\vspace{0.2cm}
 \includegraphics[scale=0.4]{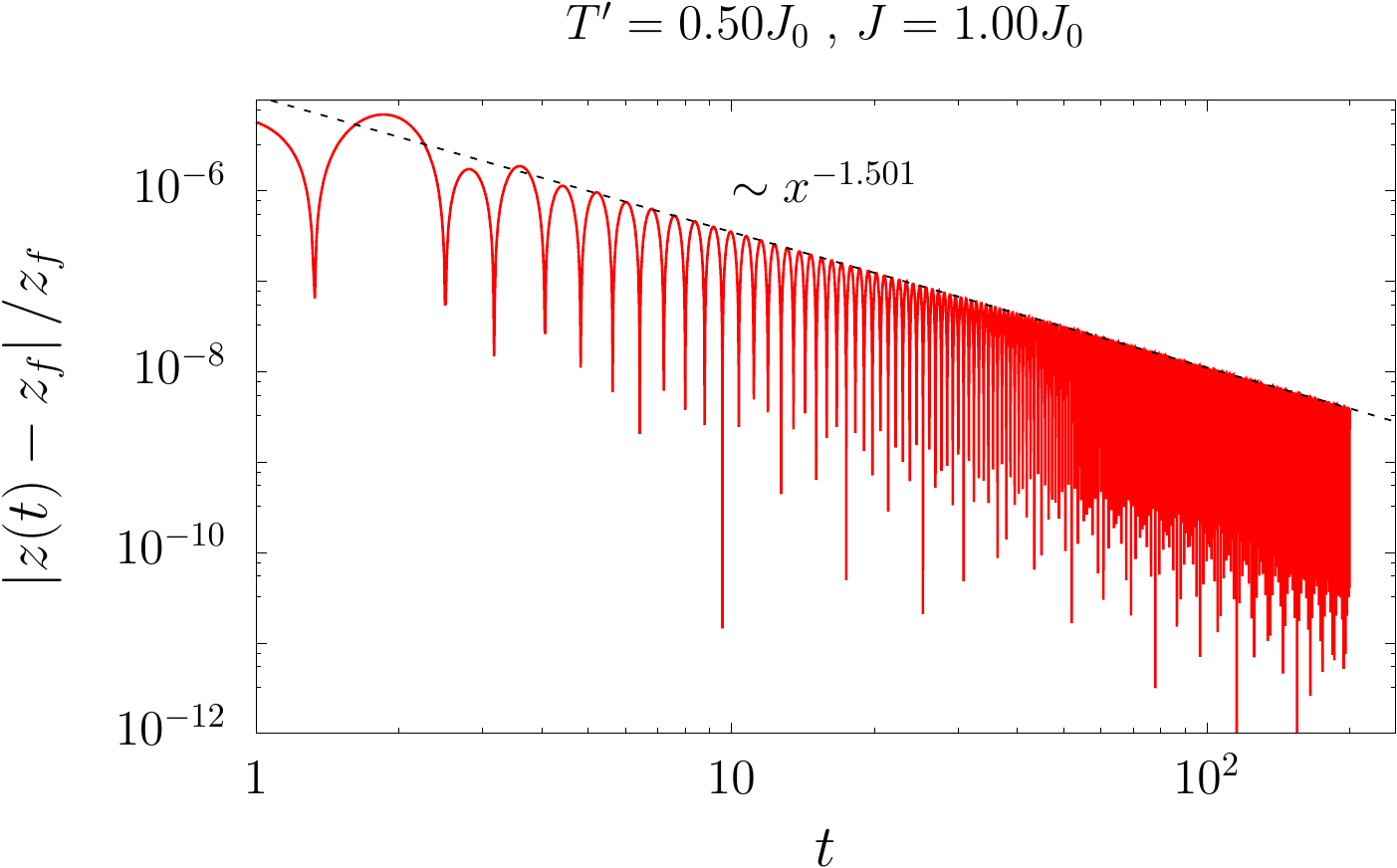}\quad%
 \includegraphics[scale=0.4]{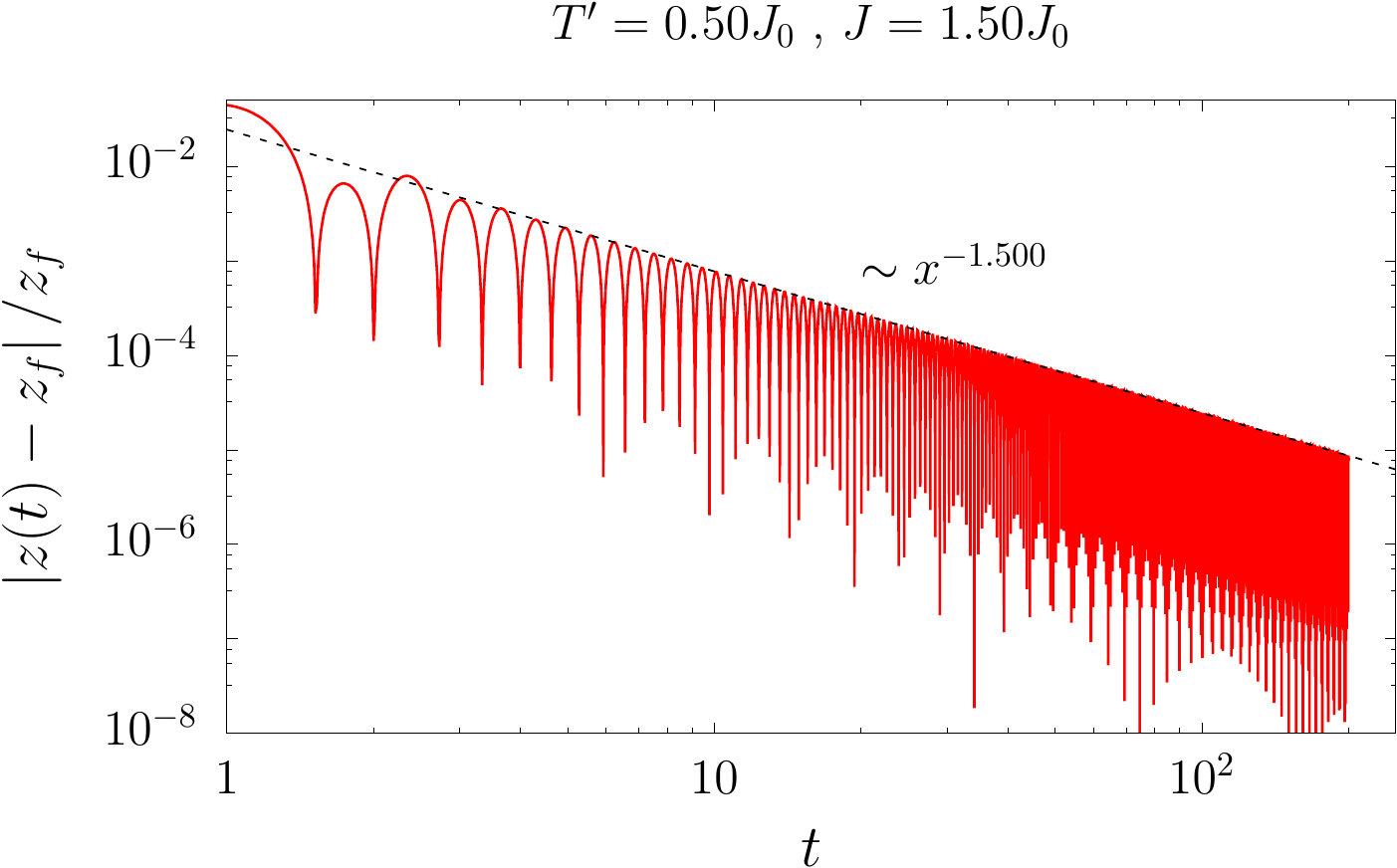}%
\end{center}
\caption{\small
Pre-asymptotic behaviour of $z(t)$.
The quantity $ \left| z(t) - z_f \right| / z_f $, with $ z_f = \lim_{t \to +\infty} z(t) $ is plotted against $t$
in the case of quenches with $T' = 0.5 \ J_0 $, that is to say, from a
condensed state. The parameter controlling the energy injection or
extraction is given above each plot.
The function $f(x) = a \, x^{-b}$ has been fitted to the upper envelope of
$ \left| z(t) - z_f \right|/ z_f$ for $t \gg 1$
(dashed black line). The value of $z_f$ was fixed to the one that is  expected,
see Table~1. The numerical values of $b$ are indicated close to the fits.
The reason why the decay in the critical quench seems to be steeper than what is shown with
the dashed line is explained in the text.
}
\label{fig:z_T0p50}
\end{figure}

We first use the algebraic decay proposal
\begin{equation}
   \mathrm{env} \left[ z(t) - z_f \right] \ \sim \ t^{-\alpha_{ \small z }} \,
   \; ,
 \label{eq:z-asymptotics}
\end{equation}
where env stands for ``the envelope of'',
with the values of $z_f$ in Table~1, to extract the exponent
$\alpha_z$. In practice, we will use the upper envelope of the
data that, in all cases excluding the critical quenches,
coincides with the results obtained from using the lower envelope. The critical quench
is special and we will discuss it in detail below.

In Fig.~\ref{fig:z_T0p50}
we display four panels with parameters such that $T' $ is fixed to
$0.5 = T'< J_0 = T^0_c$ and $J/J_0$ changes
in such a way that:  there is injection of energy (upper left), the parameters are on the
critical line (upper right),  and there is extraction of energy (the two lower panels). A detailed
study of these and other cases  with $T'<J_0$
indicates that the exponent $\alpha_z$ varies on the left of
$T'=J$, it takes the value $\alpha_z=3/2$ at the dynamic transition at $T'=J$, and it remains fixed to
$3/2$ on the  right of it.

The decay in the critical quench in panel (b) seems to be steeper than
what is shown with the dashed line that represents the fit.
The reason for this is that the upper ($z(t) - z_f > 0$) and lower ($z(t) - z_f < 0$)
envelopes of the function to fit do not have the same decay,
at least up to the times $t\approx 300$ that we are able to explore with the numerical computation.
The dashed line shown in the plot is an ``average'' between the fits of the upper and lower
envelopes over the available times. In fact, a fit to the  upper envelope only yields
$  \alpha_z > 1.5$, while a fit to the lower envelope yields
$ \alpha_z < 1.5$ (the lower envelope is not visible in the plot).
Much longer times would be needed to reach a regime in which the envelopes have the same algebraic decay
during the computational times and thus
make a clean fit in this case.

A summary of the  $\alpha_z$  values for $T'/J_0 < 1$ are shown in Fig.~\ref{fig:exp_decay_z} (a)
with  red points.

\begin{figure}[h!]
 \vspace{0.5cm}
\begin{center}
$\;$ \hspace{1.75cm} (a) \hspace{5.5cm} (b) \hspace{4cm} $\;$
\vspace{0.2cm}
 \includegraphics[scale=0.4]{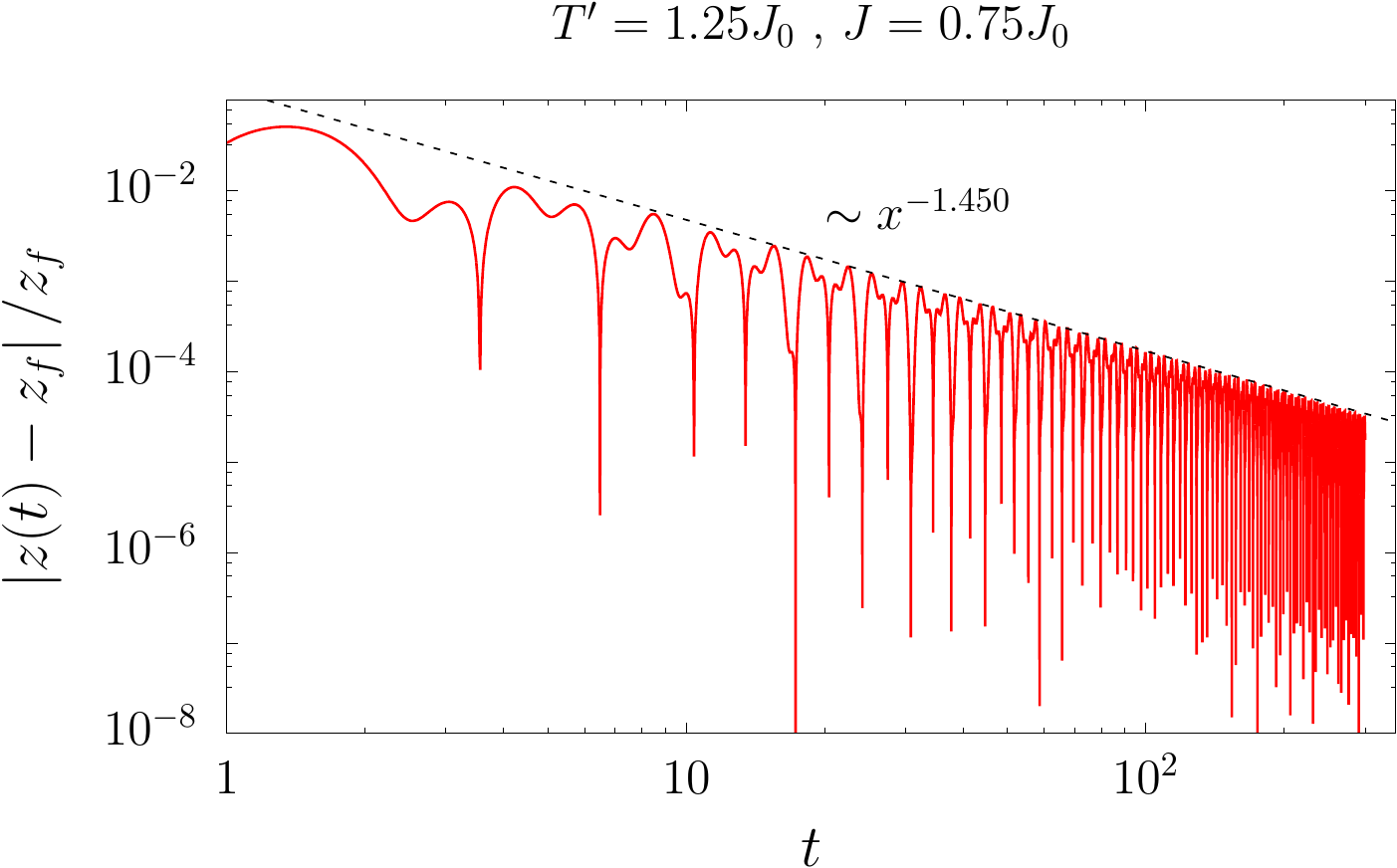}\quad%
 \includegraphics[scale=0.4]{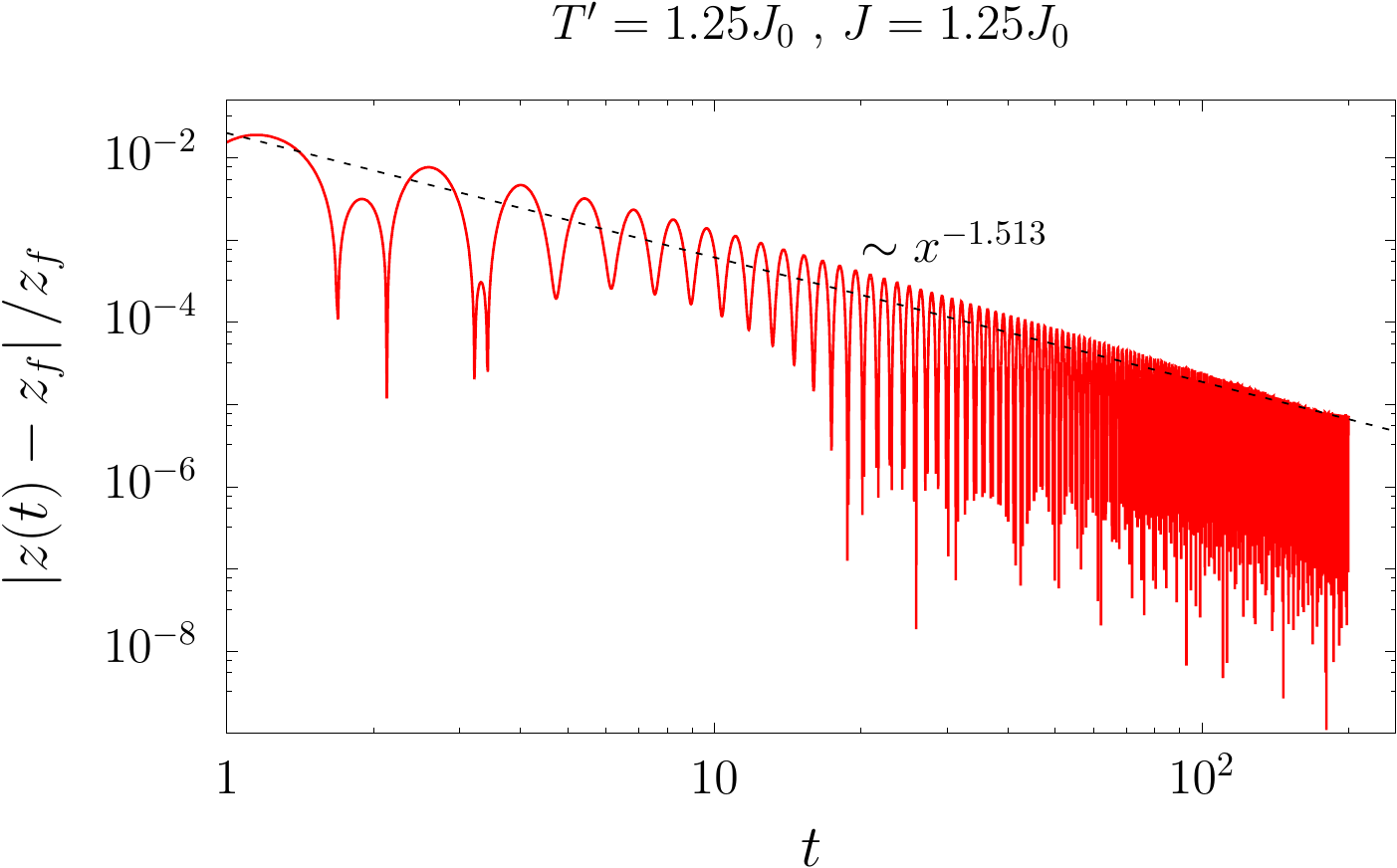}%

  \vspace{0.5cm}
  $\;$ \hspace{1.75cm} (c) \hspace{5.5cm} (d) \hspace{4cm} $\;$
\vspace{0.2cm}
 \includegraphics[scale=0.4]{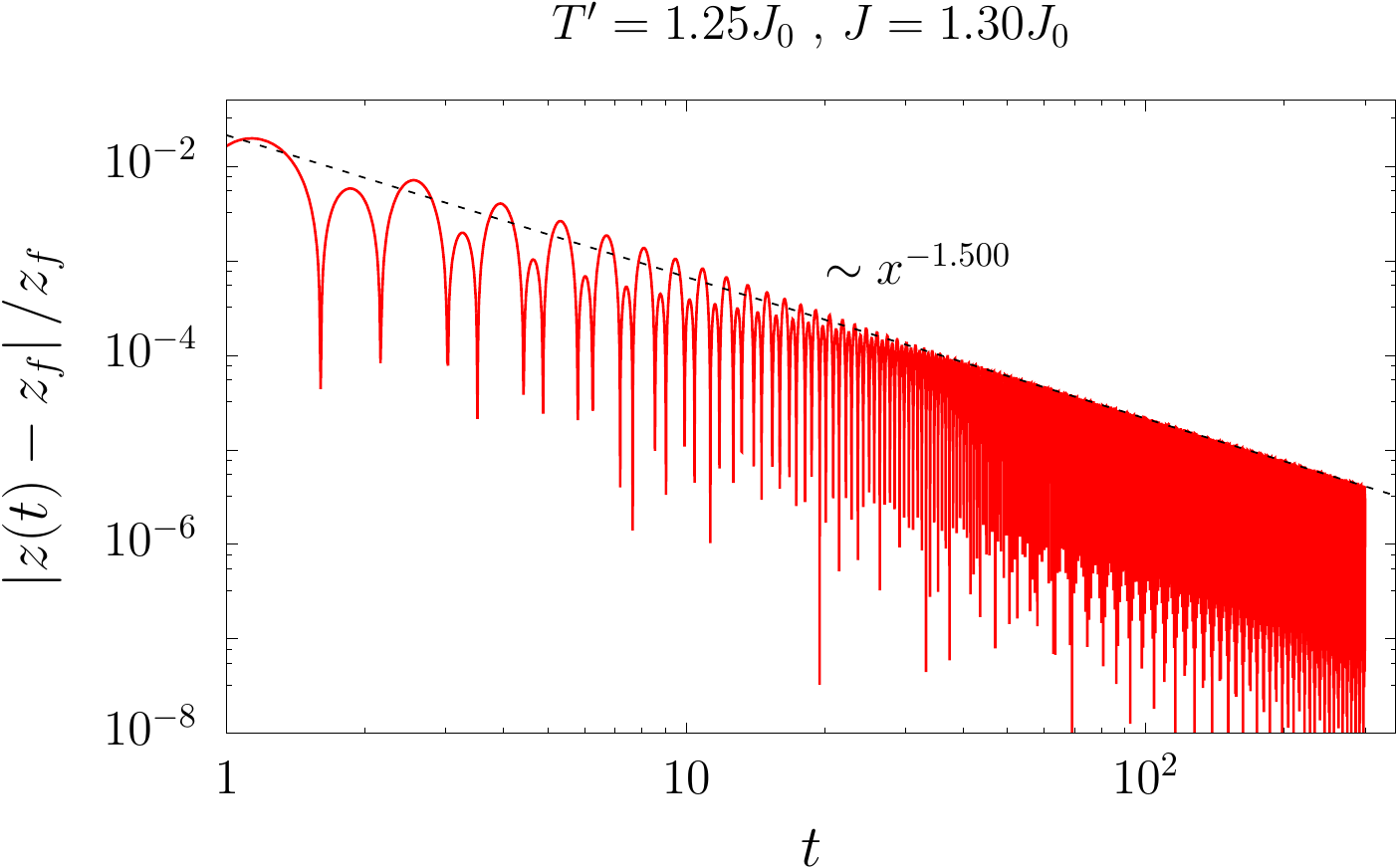}\quad%
 \includegraphics[scale=0.4]{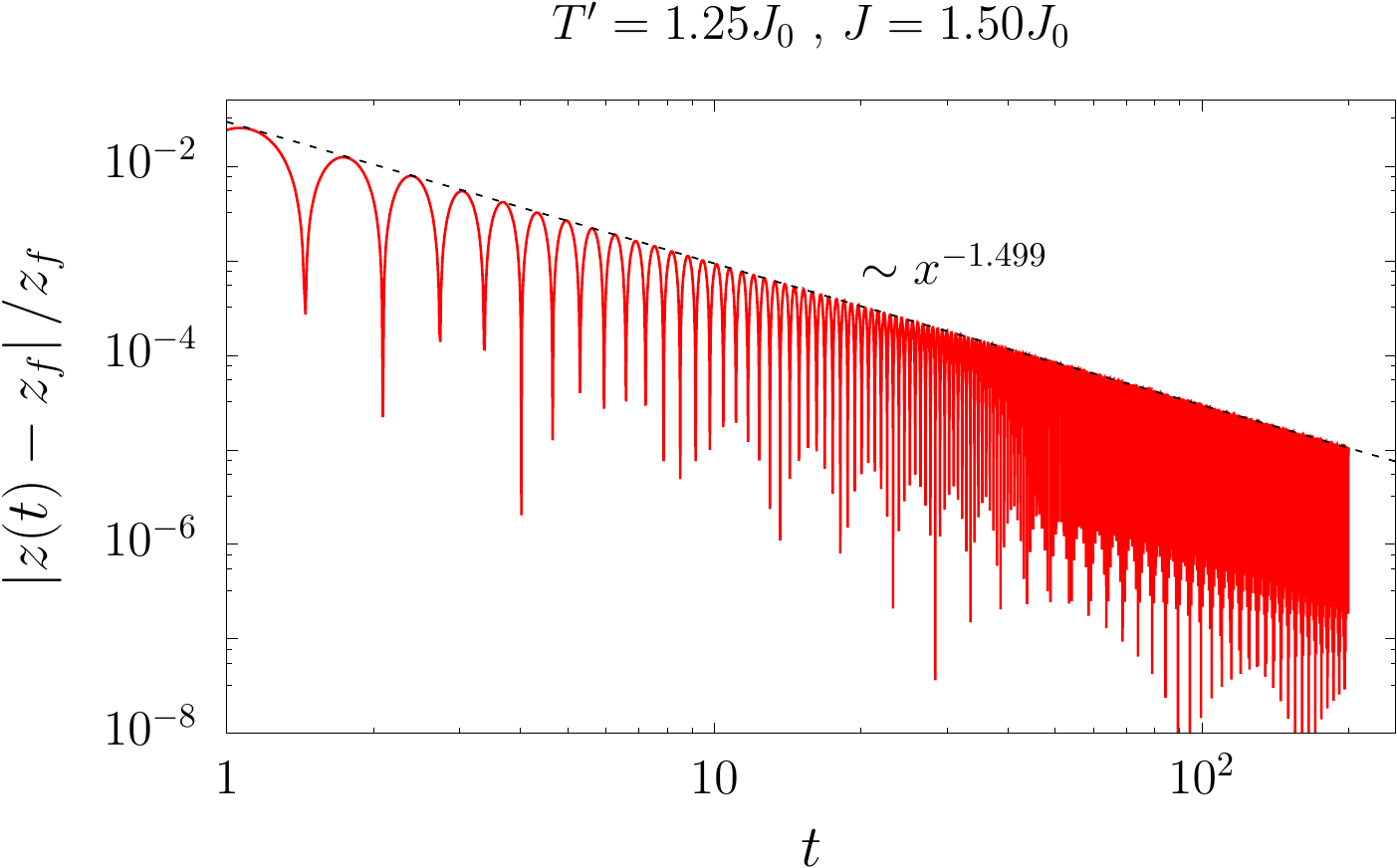}%
\end{center}
\caption{\small
Pre-asymptotic behaviour of $z(t)$.
The quantity $ \left| z(t) - z_f \right| / z_f $, with $ z_f = \lim_{t \to +\infty} z(t) $, is plotted against $t$ in the case of quenches with $T' = 1.25 \ J_0 $, that is to say, from a disordered
state. The parameter controlling the energy injection or
extraction is given above each plot.
The function $f(x) = a \, x^{-b}$ has been fitted to the upper envelope of
$ \left| z(t) - z_f \right|/ z_f$ for $t \gg 1$
(dashed black line). The value of $z_f$ was fixed to the one that is  expected,
see Table~1.
The numerical values of $b$ are indicated close to the fits.
The same explanation for the difference between the data and the dashed line in the
critical quench given close to Fig.~\ref{fig:z_T0p50} applies to this case as well.
}
\label{fig:z_T1p25}
\end{figure}

In Fig.~\ref{fig:z_T1p25} we repeat this analysis for $1.25 = T'>J_0 = T^0_c$ and we see that,
also for this kind of initial states, the exponent $\alpha_z$ depends on the parameters
on the left of $T'=J_0$ while it sticks to the value $3/2$ at $T'=J$ and beyond it.
The behaviour of $\alpha_z$ in between $J=J_0$ and
$T'=J$ is harder to determine from the numerical data. The exponent seems to reach the
value $3/2$ at $J=J_0$ and it either remains constant beyond this (no quench) value or,
if it varies, it does very smoothly, increasing slightly beyond $3/2$, reaching a shallow maximum and
then decreasing again towards $3/2$ at $T'=J$. A deeper analysis of this intermediate regime
is easier to perform using a larger $T'/J_0$ value that makes the interval $J_0 < J < T'$
wider, and we discuss these cases below.

A summary of the $\alpha_z$ values for $T'/J_0 = 1.25$ are shown
in Fig.~\ref{fig:exp_decay_z} (a) with blue points.

\begin{figure}[h!]
\vspace{0.5cm}
\begin{center}
(a) \hspace{8cm} (b) \hspace{4cm} $\;$
\\
\vspace{0.25cm}
 \includegraphics[scale=0.6]{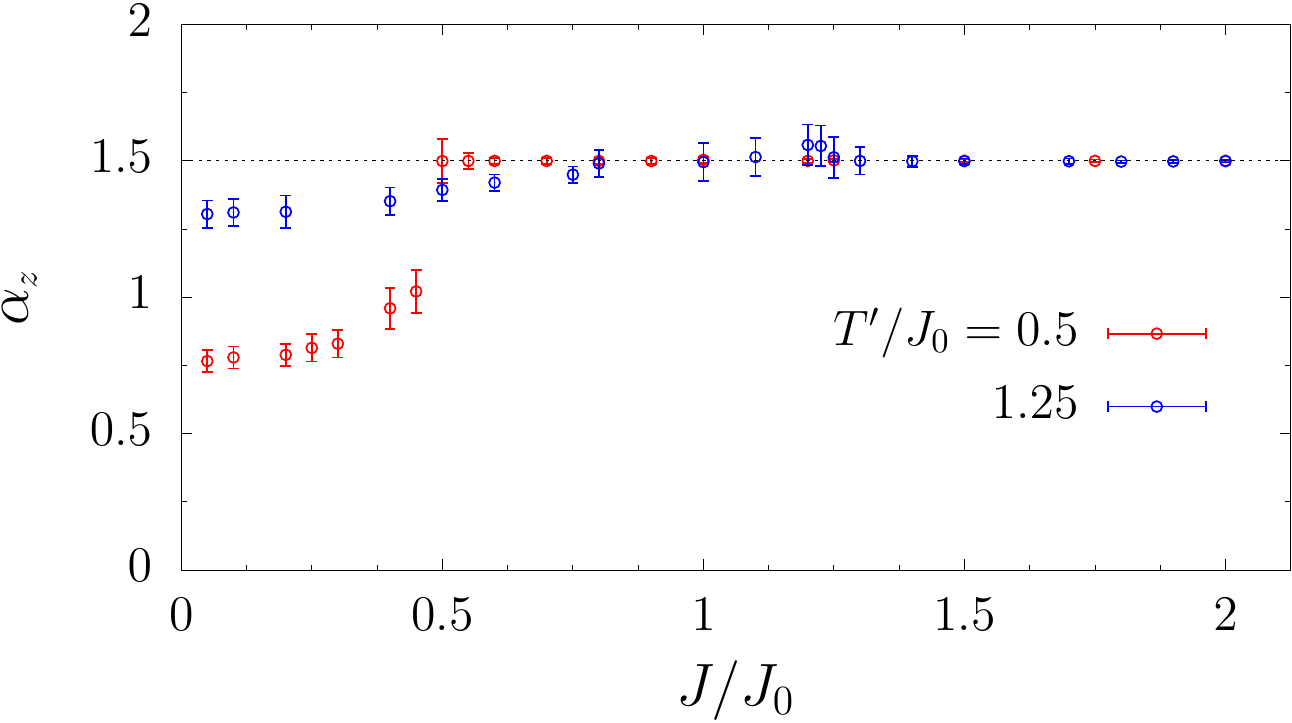}\quad
  \includegraphics[scale=0.6]{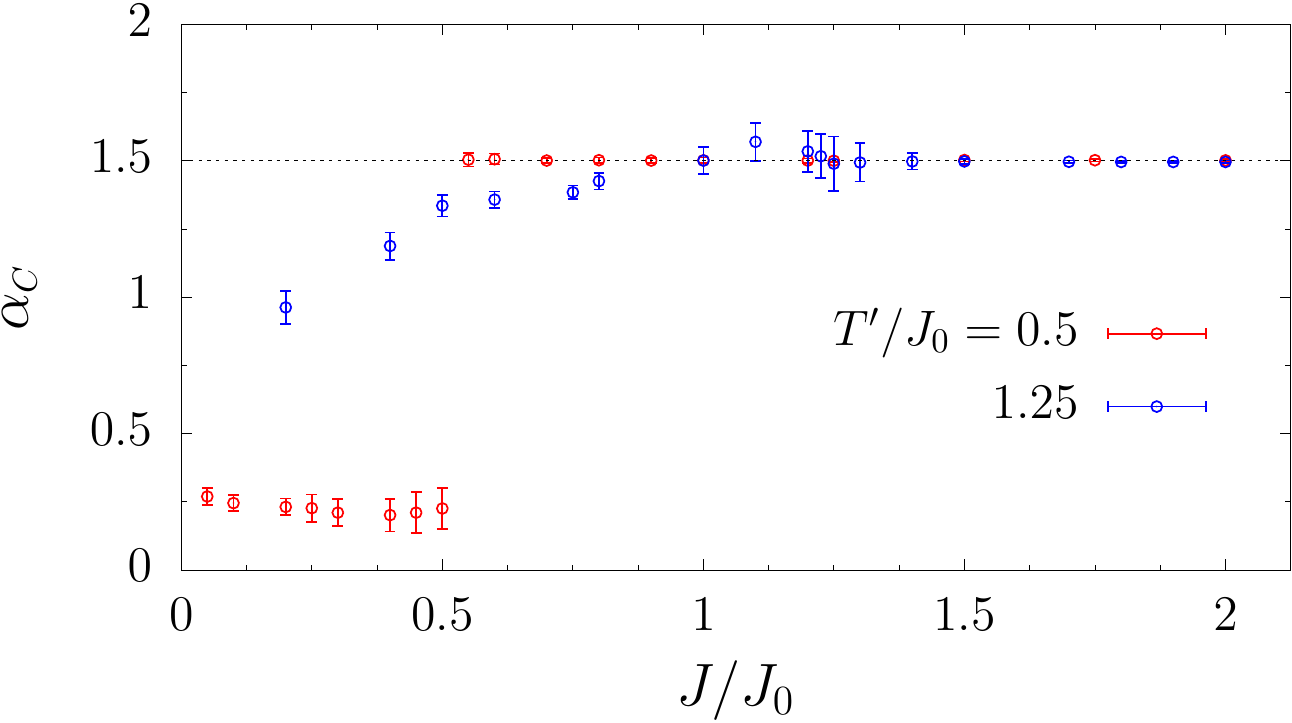}
\end{center}
\caption{\small
The exponents $\alpha_z$ (a) and $\alpha_C$ (b) defined in  Eqs.~(\ref{eq:z-asymptotics})
and (\ref{eq:corr-t0-asymptotics}), obtained from
fitting the upper envelopes of
$ \left| z(t) - z_f \right| / z_f $ and  $ \left| C(t,0) - q_0 \right|$ as functions of the parameter $J/J_0$, for two different
values of the temperature with which one draws the initial states,
$T'/J_0 = 0.5$ (red points) and $T'/J_0 = 1.25$ (blue points). Cases with
larger $T'/J_0$ are treated below to try to establish the behaviour of the
exponent in the region $1\leq T'/J_0 \leq J/J_0$, see
Fig.~\ref{fig:exp_decay_z_ht} where we report the exponents
$ \alpha_z $ and $\alpha_C $ vs. $J/J_0$ for these $T'/J_0$.
}
\label{fig:exp_decay_z}
\end{figure}

\begin{figure}[h!]
\vspace{0.5cm}
\begin{center}
(a) \hspace{8cm} (b) \hspace{4cm} $\;$
\\
\vspace{0.25cm}
  \includegraphics[scale=0.6]{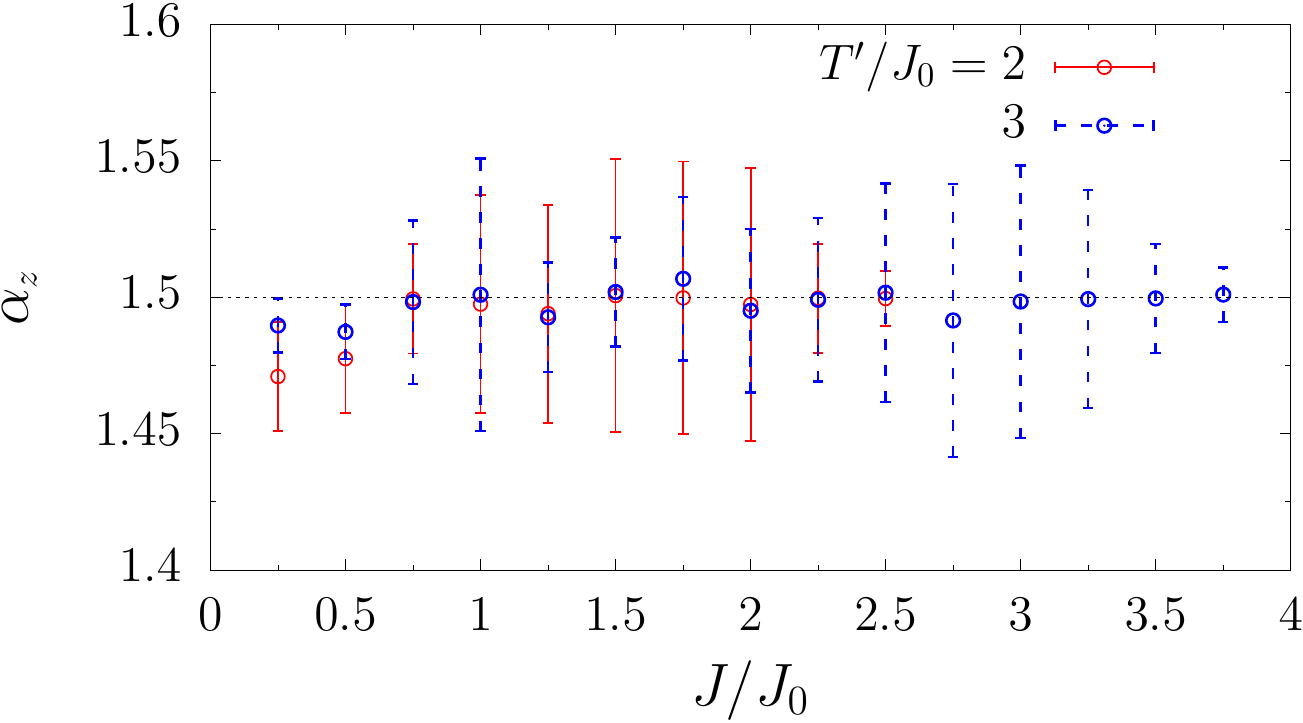}\quad
  \includegraphics[scale=0.6]{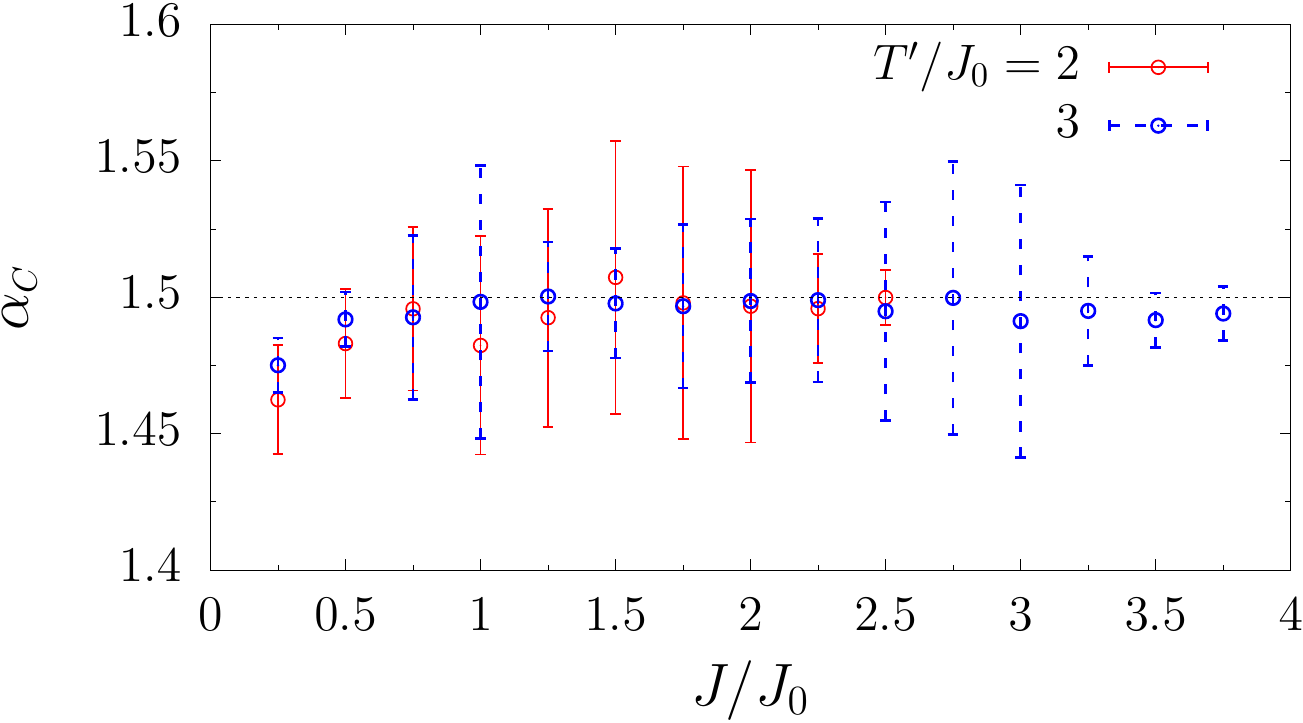}
\end{center}
\caption{\small
The exponents $\alpha_z$ (a) and $\alpha_C$ (b) defined in  Eqs.~(\ref{eq:z-asymptotics})
and (\ref{eq:corr-t0-asymptotics}), obtained from
fitting the upper envelopes of
$ \left| z(t) - z_f \right| / z_f $ and  $ \left| C(t,0) - q_0 \right|$, as functions of the parameter $J/J_0$,
for two different values of $T'/J_0 > 1$.
}
\label{fig:exp_decay_z_ht}
\end{figure}

With the aim of determining the actual behaviour of $\alpha_z$ in the uncertain region, we
repeated this analysis for much higher values of the initial temperature, $T'/J_0=2, \, 3$,
in such a way that the interval between $J=J_0$ and $T'=J$ widens.
We found that for these $ T'/J_0$, the exponent $\alpha_z$  is equal to $3/2$ or very
close to it also for $ T' < J$, that is to say, on the left of the critical line.
It is possible that the change in behaviour occurs at $J=J_0$
(the line separating quenches with injection of energy from those with extraction)
since we detect  a very small deviation
from $3/2$ for $J<J_0$. However, this is hard to establish beyond doubt. Although the
error bars that we estimate are smaller for $J<J_0$ than for $J>J_0$, the actual values
of the exponent are pretty close to $3/2$ even for $J< J_0$. The summary of our
findings for $\alpha_z $ at $T'/J_0=2$ are shown with red points and the ones at $T'/J_0 = 3$
are shown with blue points in Fig.~\ref{fig:exp_decay_z_ht} (a),

It is quite clear that the algebraic envelope captures part of the approach of $z(t)$ to its
asymptotic value $z_f$ but not everything: the curves in
Figs.~\ref{fig:z_T0p50} and \ref{fig:z_T1p25} show strong oscillations. We now study
the periodicity of the approach to the asymptotic value
by focusing on
\begin{equation}
  Z_{\rm osc}(\omega) \equiv
  \int \ \mathrm{d} t \ \mathrm{e}^{-i \omega t}
  \ t^{\alpha_{ \small z  }} \ \left( \frac{z(t) - z_f}{z_f} \right)
  \; ,
 \label{eq:z-osc-ft}
\end{equation}
with $ \alpha_{ \small z}$ the exponent extracted from the analysis in Eq.~(\ref{eq:z-asymptotics}) and
$z_f$ fixed to the values given in Table~1. This equation is the Fourier transform of the pure
oscillating part of $z(t)$,
in the long time limit. It is clear that our data will be a bit corrupted by the fact that the
numerical time interval is forcefully rather short, going from a minimal time $t_{\rm min}$
where we expect to have reached the preasymptotic state to the maximal
time in the simulation.

\begin{figure}[h!]
\vspace{0.5cm}
\hspace{2.5cm} (a)  \hspace{7cm} (b)
\begin{center}
 \includegraphics[scale=0.3]{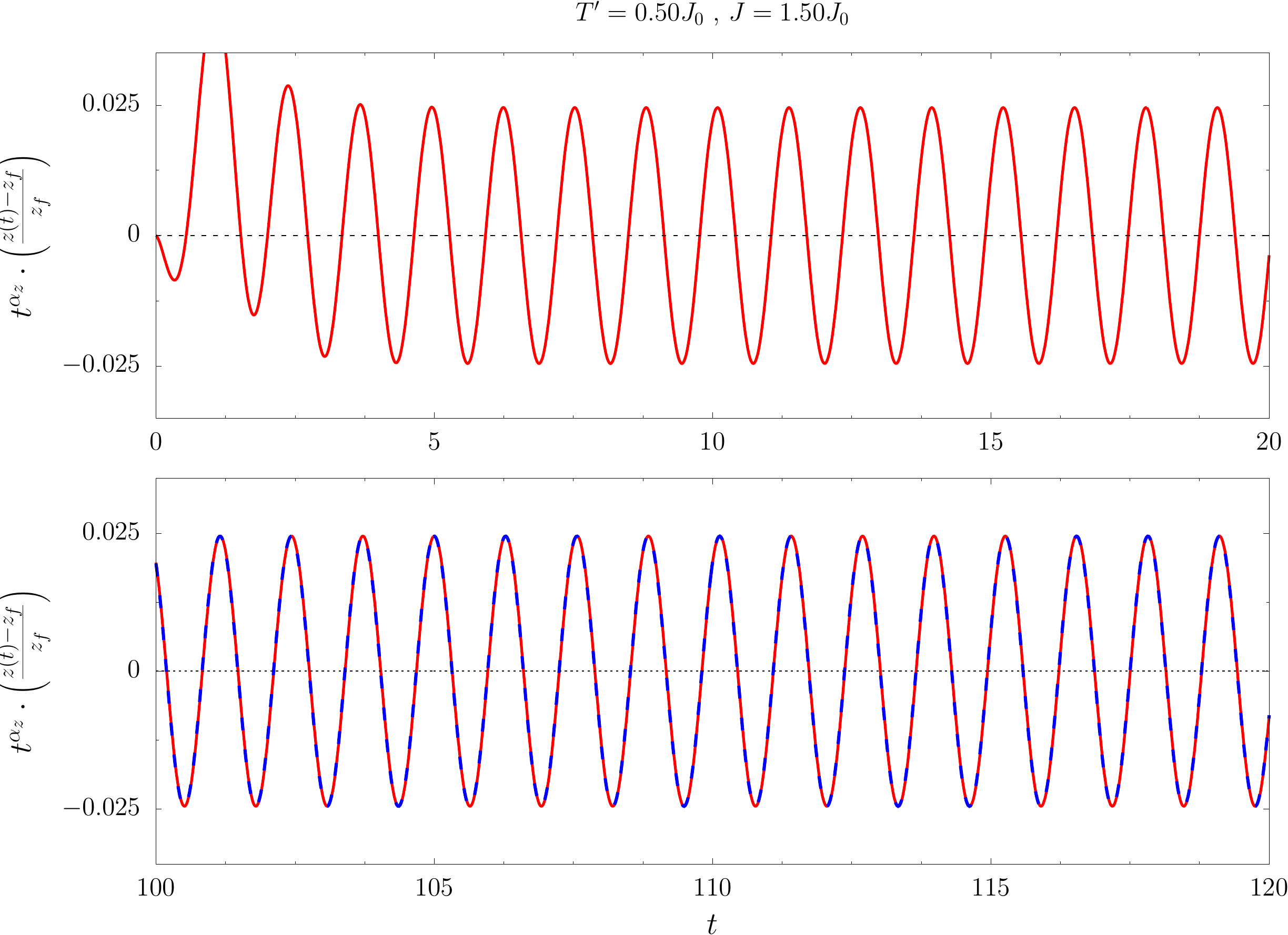}%
\hspace{0.25cm}
 \includegraphics[scale=0.58]{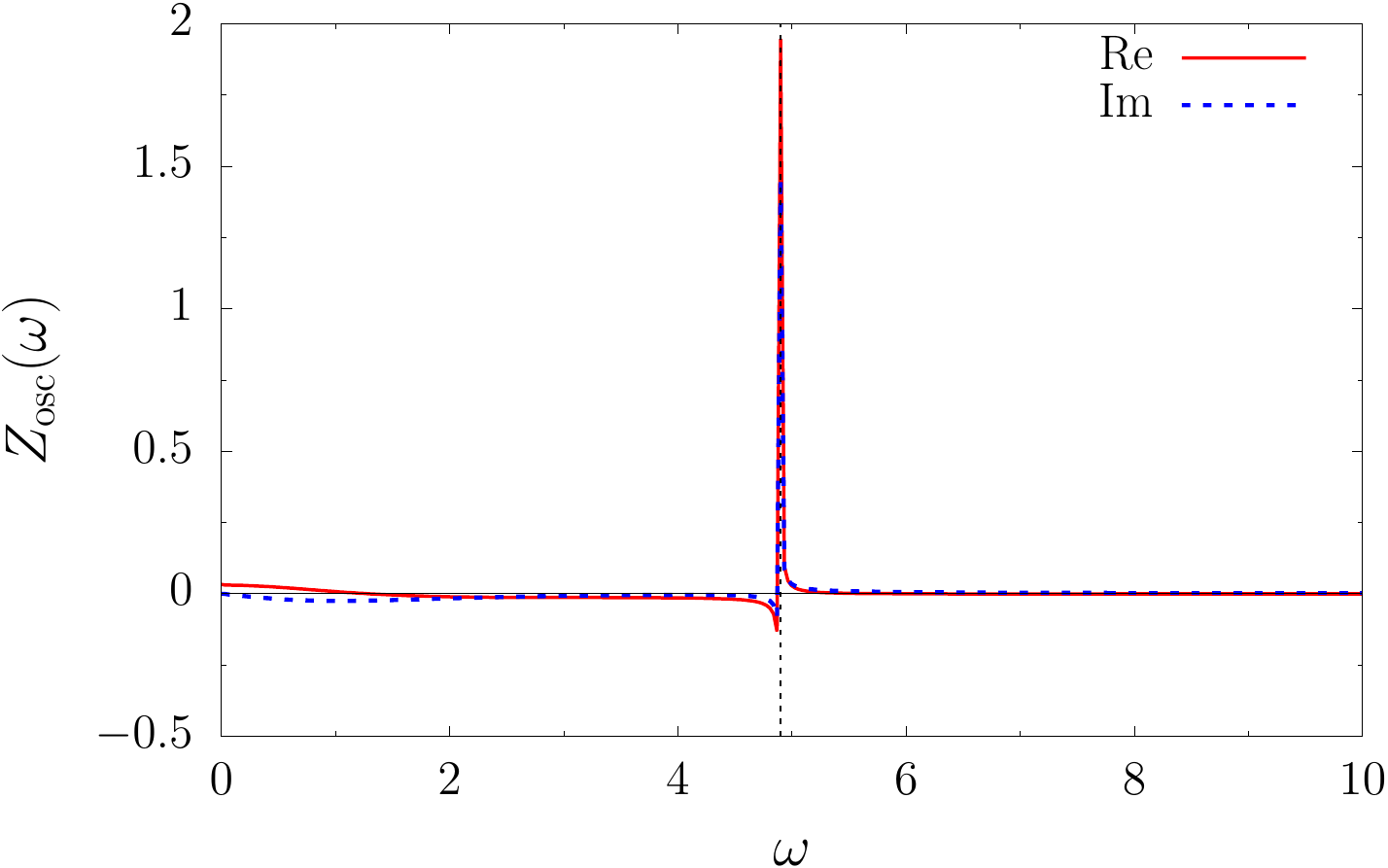}%
\end{center}
\caption{\small
Energy extraction from a condensed state ($T' = 0.5 \ J_0$, \ $J= 1.5 \ J_0$).
(a) The oscillating part of $z(t)$ at short (above) and long (below) times with the exponent fixed to the
value $\alpha_{\small z} = 1.500$ obtained from fitting the envelope,
(b) Real and imaginary parts of the
Fourier transform of the oscillating part, $Z_{\rm osc}(\omega)$,
see Eq.~(\ref{eq:z-osc-ft}).
The vertical dashed line is at $2 \omega_{+}$, where $\omega_{+}=\sqrt{(z_f + 2 J)/m}= 2\sqrt{J/m}$.
In the lower panel in (a), we also plotted the
function $h(t) = A \cos(2 \omega_{+} +\phi)$ (blue dashed line) with
$ A \simeq 0.0245$ and $ \phi \simeq 0.8399$, that
matches the data almost perfectly in the given time window.
}
\label{fig:z_ft_T0p50_J1p50}
\end{figure}

Figure~\ref{fig:z_ft_T0p50_J1p50} shows the oscillating part of $z(t)$, that is
$t^{\alpha_{ \small z  }}  [z(t) - z_f]/z_f  $, with the exponent
$\alpha_{\small z}$ defined in Eq.~(\ref{eq:z-asymptotics}) and fixed to the value
$\alpha_{\small z} = 1.500$ obtained from fitting the envelope. The parameters are
$T' = 0.5 \ J_0$, \ $J= 1.5 \ J_0$, energy is extracted from the sample with this quench,
and $z_f=2J$.
The panels on the left display the time dependence in two time intervals: at relatively short times when the
decay is still far from the asymptotic limit, and at sufficiently long times to have reached it.
The right panel is the Fourier transform in Eq.~(\ref{eq:z-osc-ft}) with a clear peak at the
frequency
\begin{equation}
\omega=2\omega_+=2\sqrt{(z_f+2J)/m} = 4\sqrt{J/m}
\; .
\end{equation}
This figure shows that the characteristic frequency of the linear response function,
$\omega_+$, the only non-vanishing one for this set of parameters, also determines the
periodic dependence of the approach to the constant long-time limit of $z_f$. Once we have
found the characteristic frequency $2\omega_+$ from the Fourier analysis, we
plotted the form $A\cos(2\omega_+ t + \phi)$ together with the  numerical data
in the lower panel on the left. The agreement between the two is perfect within the scale
of the figure, see Fig.~\ref{fig:z_ft_T0p50_J1p50} (a).

\begin{figure}[h!]
\vspace{0.5cm}
\hspace{2.5cm} (a)  \hspace{7cm} (b)
\begin{center}
 \includegraphics[scale=0.42]{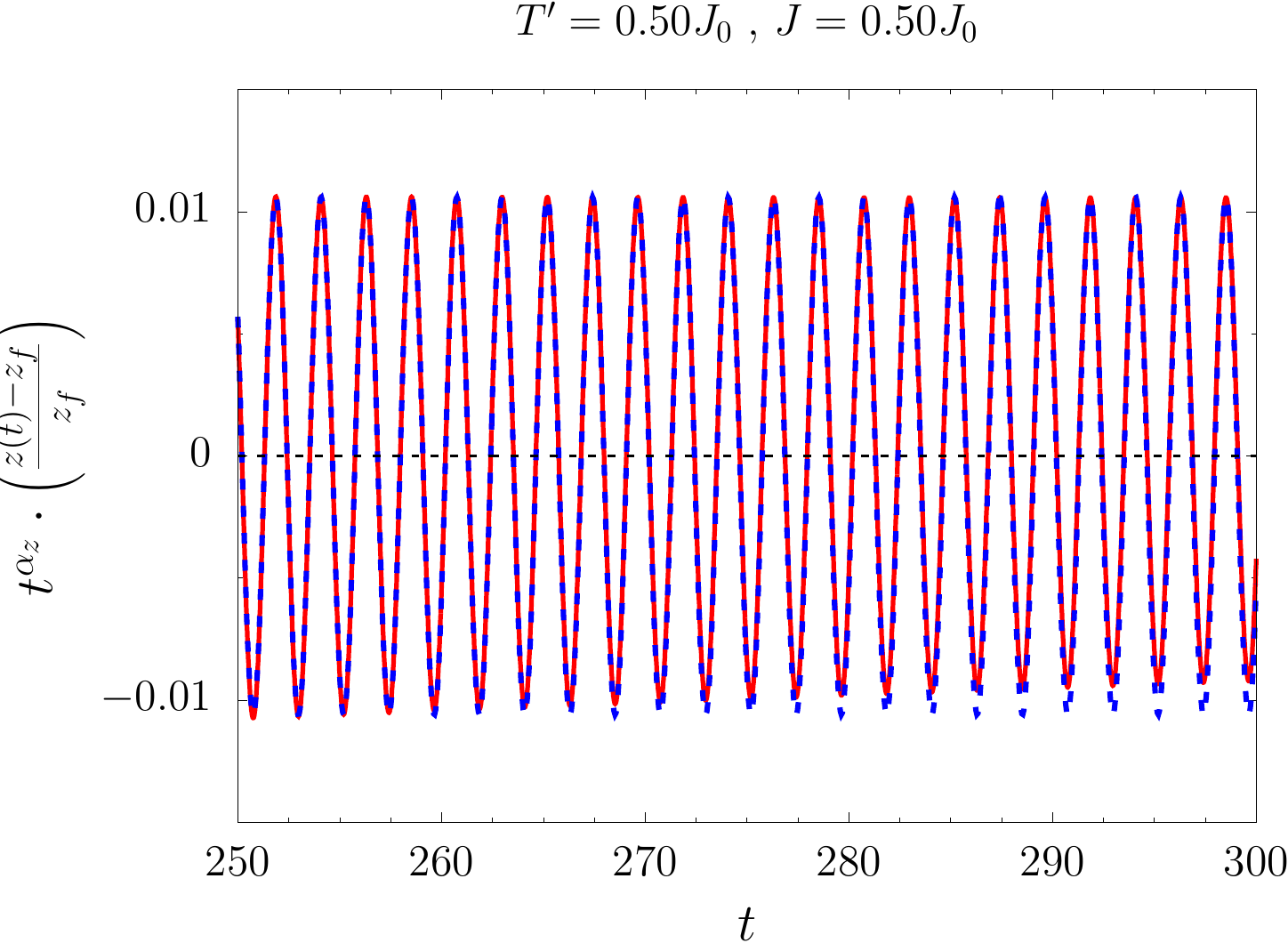}%
\hspace{0.25cm}
 \includegraphics[scale=0.46]{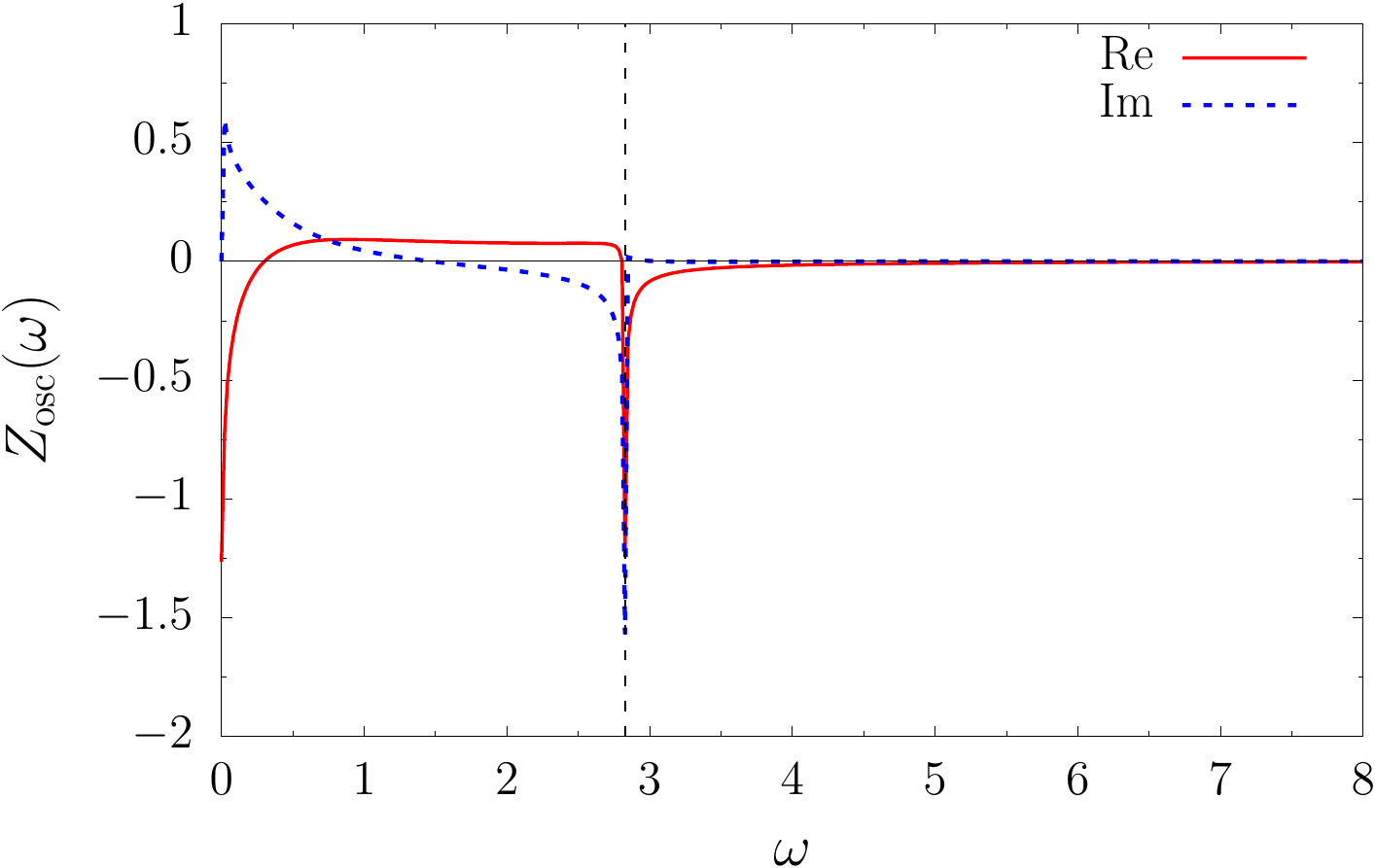}%
\end{center}
\caption{\small
Quench from a condensed state to the critical line ($T' = 0.5 \ J_0$, \ $J= 0.5 \ J_0$).
(a) The oscillating part of $z(t)$, with $\alpha_{\small z} = 1.500$ as obtained from fitting the upper envelope,  in a selected time window ($250 < t < 300$) together with the
function $h(t) = A \cos(2 \omega_{+} +\phi)$ (blue dashed line)
with $ A \simeq 0.0106$ and $ \phi \simeq 0.7552$. Notice that the lower envelope
has a remanent weak time dependence.
(b) Fourier transform of the oscillating part, $Z_{\rm osc}(\omega)$, see Eq.~(\ref{eq:z-osc-ft}).
The vertical dashed line is at $2 \omega_{+}$, where $\omega_{+}=\sqrt{(z_f + 2 J)/m}=2\sqrt{J/m}$.
Since the lower frequency $\omega_{-}$ is zero in this case, a single harmonic at
$2 \omega_{+}$ is enough to fit the data in (a).
}
\label{fig:z_ft_T0p50_J0p50}
\end{figure}

In the next figure, Fig.~\ref{fig:z_ft_T0p50_J0p50}, we investigate a quench to the critical line. The oscillating part of $z(t)$ displayed in the left panel proves that  this quantity has not yet reached a limit in which all the decaying time-dependence is captured by the power law. More precisely, we determined the exponent $\alpha_z$ from the upper envelope and the data show that the lower part of the time-dependence is still varying quite a lot. We believe that this is due to the very slow dynamics in a quench to the critical line and that, basically, for this reason the system has not yet reached the pre-asymptotic regime. Still, the Fourier analysis points towards the special role played by the frequency $2\omega_+$ and the appearance of a (small) peak at $\omega_-=0$, signalling that this frequency will detach from zero beyond criticality and will play a role in the pre-asymptotic behaviour of $z(t)$ as well. The comparison between the numerical data and
the curve $A\cos(2\omega_+ t + \phi)+B$ is
shown in Fig.~\ref{fig:z_ft_T0p50_J0p50} and the agreement is rather good, apart from the
remanent time dependence in the lower part of the curve, that is seen as a deviation between the
blue (fit) and red (data).

\begin{figure}[h!]
\vspace{0.5cm}
\hspace{2cm} (a)  \hspace{8cm} (b)
\vspace{-0.25cm}
\begin{center}
\vspace{0.5cm}
 \includegraphics[scale=0.28]{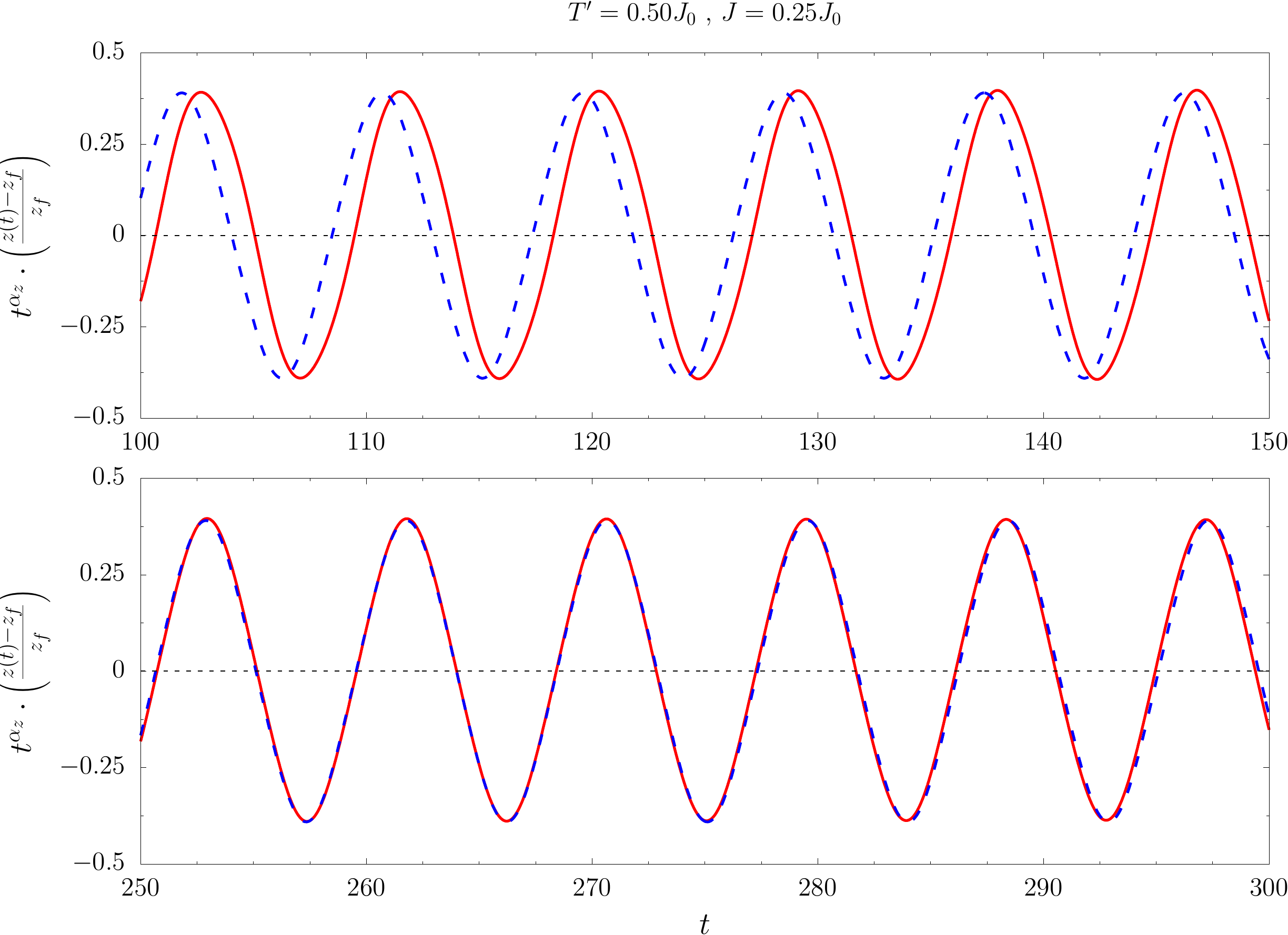}%
 \hspace{0.5cm}
 \includegraphics[scale=0.56]{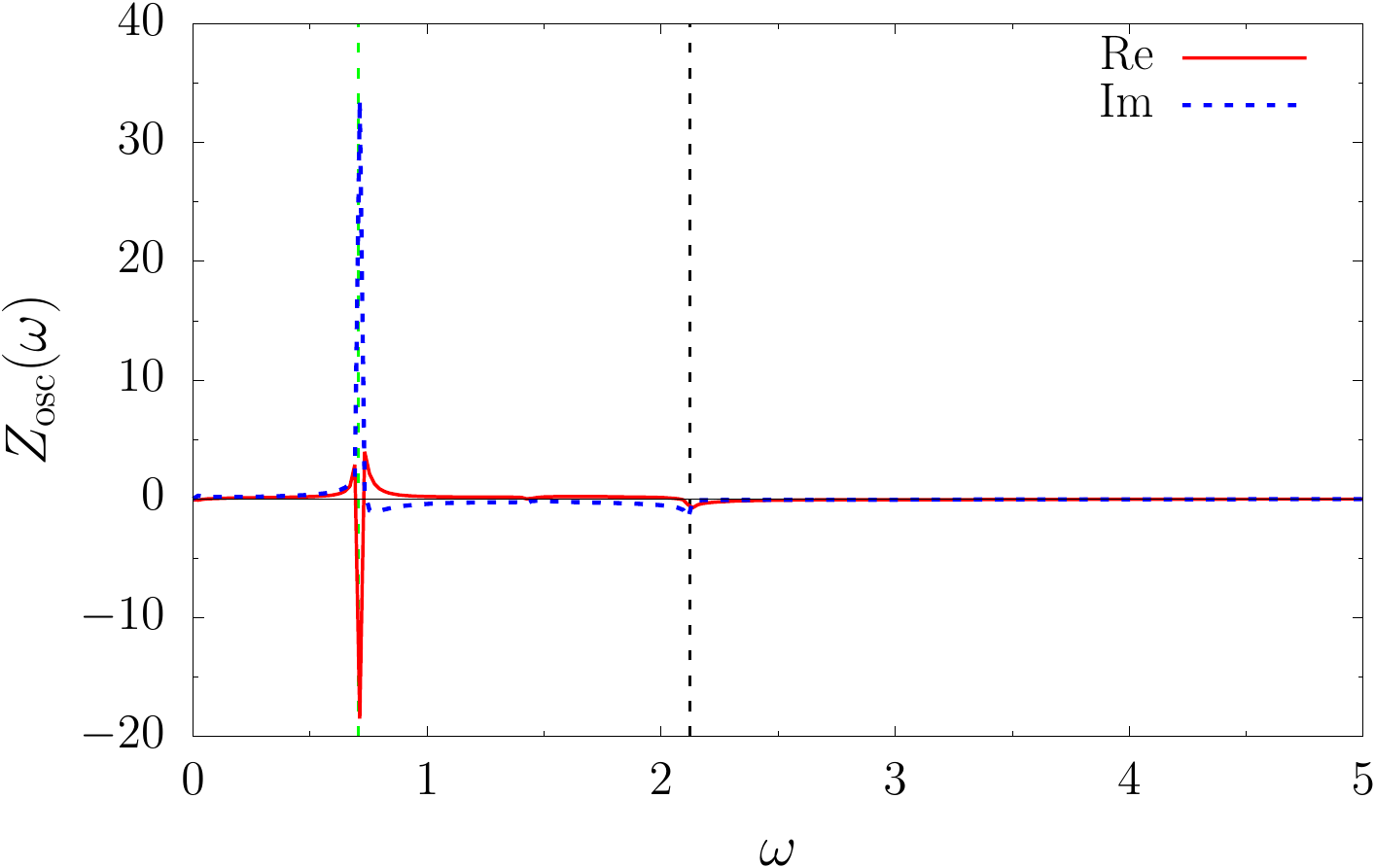}%
\end{center}
\caption{\small
Energy injection on a condensed state ($T' = 0.5 \, J_0$, $J= 0.25 \, J_0$).
(a) The oscillating part of $z(t)$ with the exponent $\alpha_{\small z} \simeq 0.815$ obtained from fitting the long-time envelope, at short (above) and long (below) times
after the quench.
(b) Real and imaginary parts of the
Fourier transform of the oscillating part, $Z_{\rm osc}(\omega)$,
see Eq.~(\ref{eq:z-osc-ft}).
The vertical dashed lines are at $2 \omega_{-}$ and
$2 \omega_{+}$, where $\omega_{\pm}=\sqrt{(z_f \pm 2 J)/m}$ and $z_f=T'+J^2/T'$.
In (a) we also  plotted the
function $h(t) = A_{+} \cos(2 \omega_{+} +\phi_{+}) + A_{-} \cos(2 \omega_{-} +\phi_{-}) $
(blue dashed line) with $ A_{+} \simeq 0.0076$, $ \phi_{+} \simeq 0.7029$,
$ A_{-} \simeq 0.3832$ and $ \phi_{-} \simeq 0.2675$,
that fails to match the data at short times but does almost perfectly at long times.
In this case, the contribution of the higher frequency, $2 \omega_{+}$, is negligible
 ($ A_{+} \simeq 0.0076$) and we would have obtained an equally good numerical agreement
keeping only the frequency $2 \omega_{-}$.
}
\label{fig:z_ft_T0p50_J0p25}
\end{figure}

\begin{figure}[h!]
\vspace{0.5cm}
\hspace{2cm} (a)  \hspace{8cm} (b)
\vspace{-0.25cm}
\begin{center}
\vspace{0.5cm}
 \includegraphics[scale=0.28]{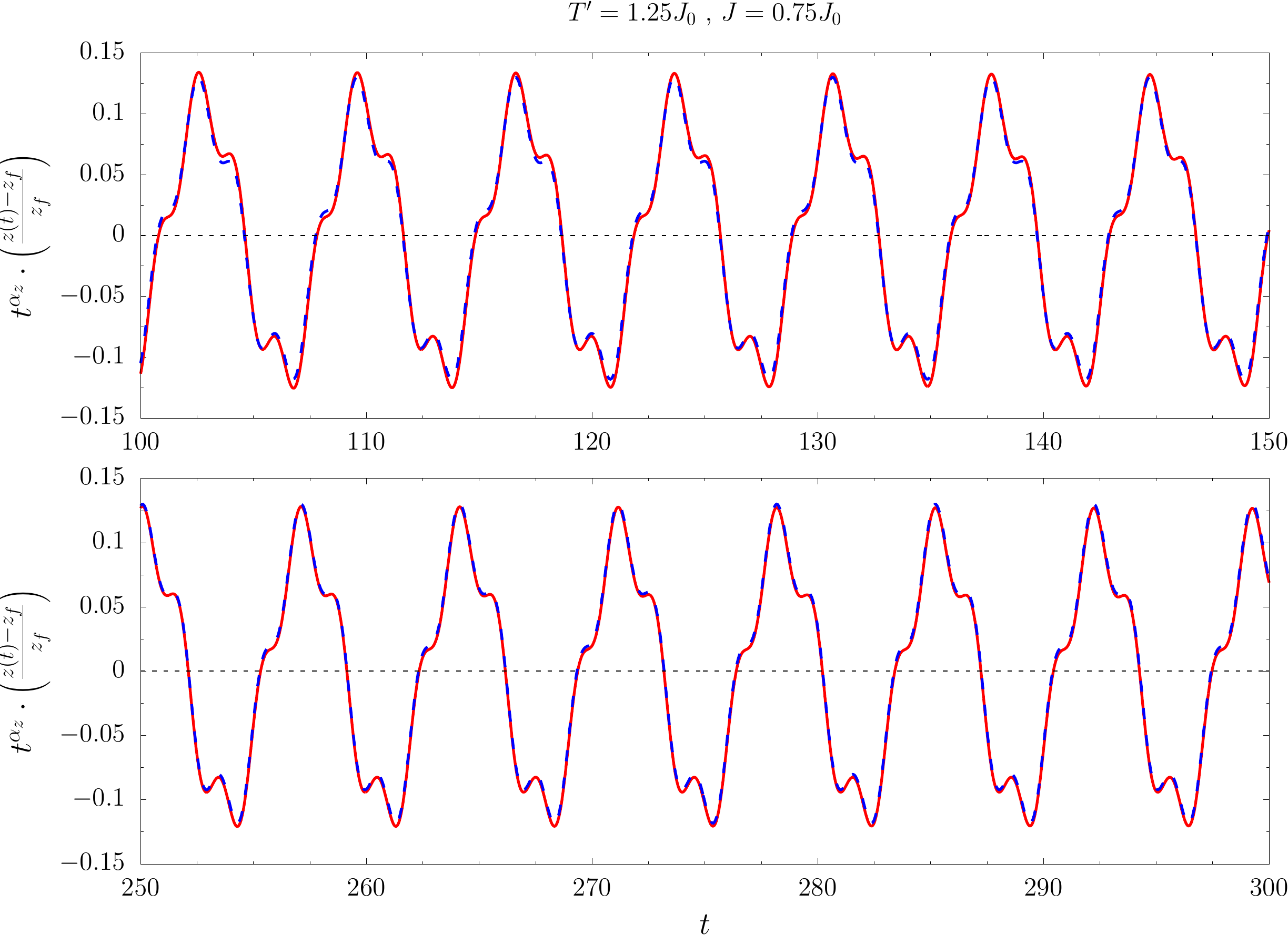}%
 \hspace{0.25cm}
  \includegraphics[scale=0.56]{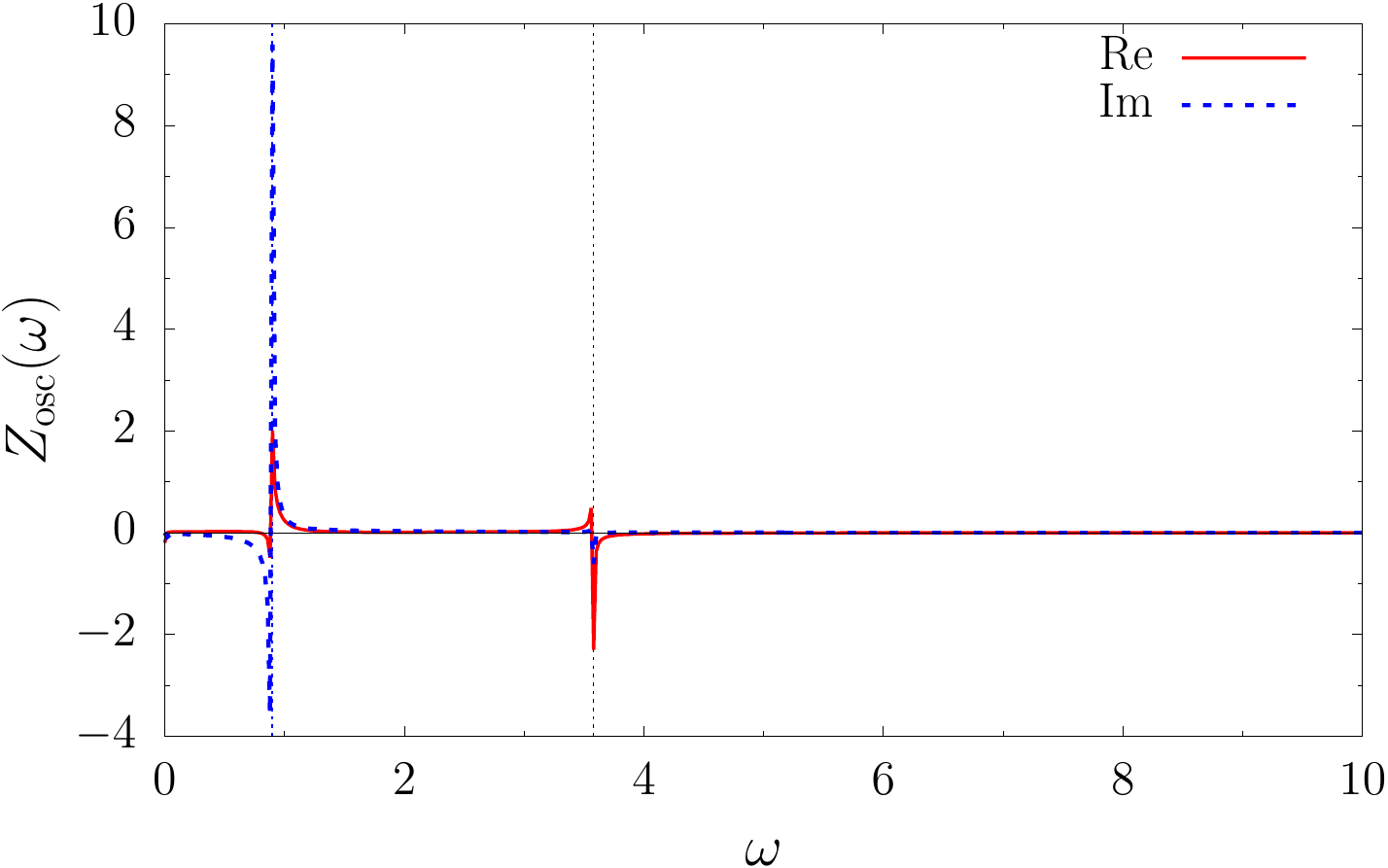}%
\end{center}
\caption{\small
Energy injection on  the disordered state ($T' = 1.25 \, J_0$, $J= 0.75 \, J_0$).
(a) The oscillating part of $z(t)$ with the exponent $\alpha_{\small z} \simeq 1.450$ obtained from fitting the envelope at long times.
(b) Real and imaginary parts of the Fourier transform of
the oscillating part, $Z_{\rm osc}(\omega)$, see Eq.~(\ref{eq:z-osc-ft}).
The vertical dashed lines are at $2 \omega_{-}$ and
$2 \omega_{+}$, where $\omega_{\pm}=\sqrt{(z_f \pm 2 J)/m}$ and $z_f = T'+J^2/T'$.
On the left, we also plotted the
function $h(t) = A_{+} \cos(2 \omega_{+} t +\phi_{+}) + A_{-} \cos(2 \omega_{-} t +\phi_{-}) $
(blue dashed line) with $ A_{+} \simeq -0.0242$ and $ \phi_{+} \simeq -2.3246$, and
$ A_{-} \simeq 0.1077$ and $ \phi_{-} \simeq -0.7846$.
This function matches the data almost perfectly at long times.
}
\label{fig:z_ft_T1p25_J0p75}
\end{figure}

Finally, in Figs.~\ref{fig:z_ft_T0p50_J0p25} and \ref{fig:z_ft_T1p25_J0p75}
 we treat cases with large energy injection taking the
 system from a condensed state or the disordered state
 into the phase in which $z_f$ depends on the
 two control parameters, that is to say, beyond the critical diagonal to the
 left of it.  On this side of the transition both frequencies $\omega_-$ and
 $\omega_+$ are non zero and they influence the pre-asymptotic
 behaviour. This is not so clear in Fig.~\ref{fig:z_ft_T0p50_J0p25} where the
 system is taken out of a condensed state, but it is quite remarkable in
 Fig.~\ref{fig:z_ft_T1p25_J0p75}.
 In the case $T'/J_0 = 0.5$ and $J/J_0= 0.25$ the peak at $2 \omega_{+}$ is negligible
 with respect to the one at $2 \omega_{-}$
 (but should be real, not an effect of numerical error), see Fig.~\ref{fig:z_ft_T0p50_J0p25}.
 Instead, in the case $T'/J_0 = 1.25$ and $J/J_0= 0.75$ (see Fig.~\ref{fig:z_ft_T1p25_J0p75})
 the higher frequency $2 \omega_{+}$ is clearly present.

We summarise the results in this section in the equation
\begin{equation}
z(t) \simeq z_f + t^{-\alpha_z} [ A_+ \cos(\omega_+ t + \phi_+) + A_- \cos(\omega_- t + \phi_-) ]
\end{equation}
that condenses in a single equation the time dependence of the Lagrange multiplier in the
first correction to the asymptotic constant value.

\subsubsection{The susceptibility, and the kinetic and potential energies}

\begin{figure}[h!]
\vspace{0.5cm}
 \hspace{1cm} (a) \hspace{5cm} (b) \hspace{5cm} (c)
\begin{center}
\vspace{-0.2cm}
  \includegraphics[scale=0.375]{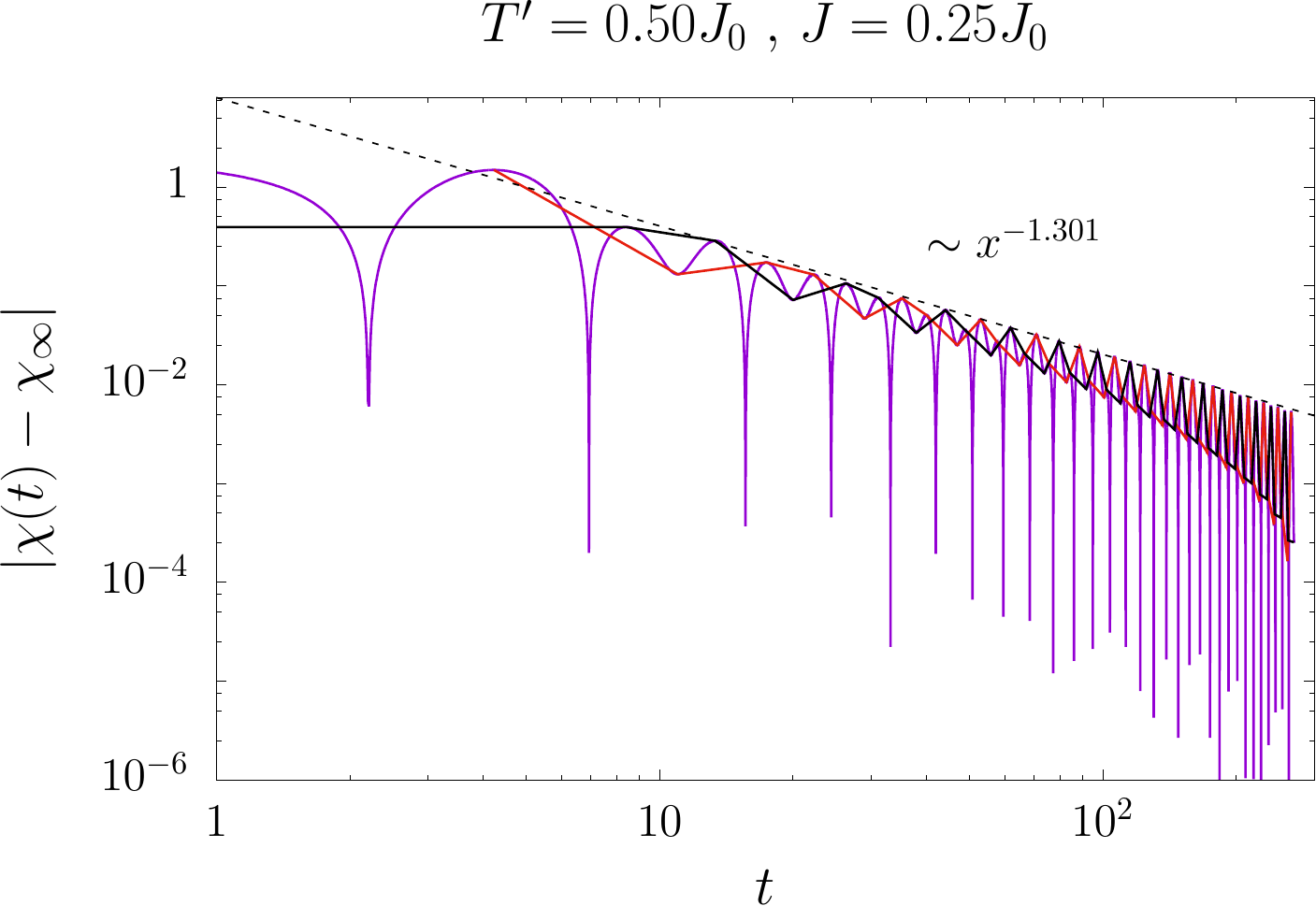}\quad%
  \includegraphics[scale=0.375]{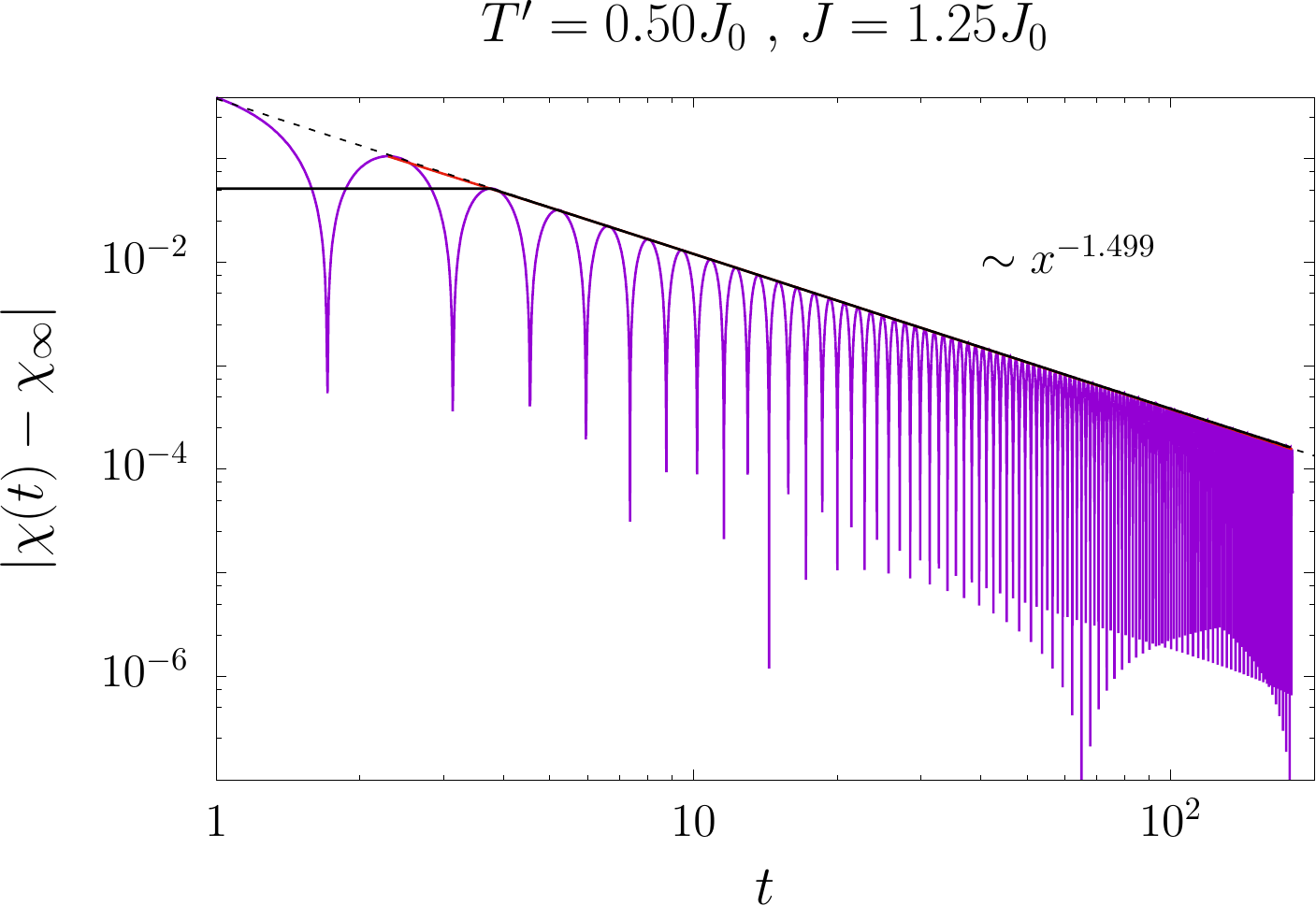}\quad%
 \includegraphics[scale=0.375]{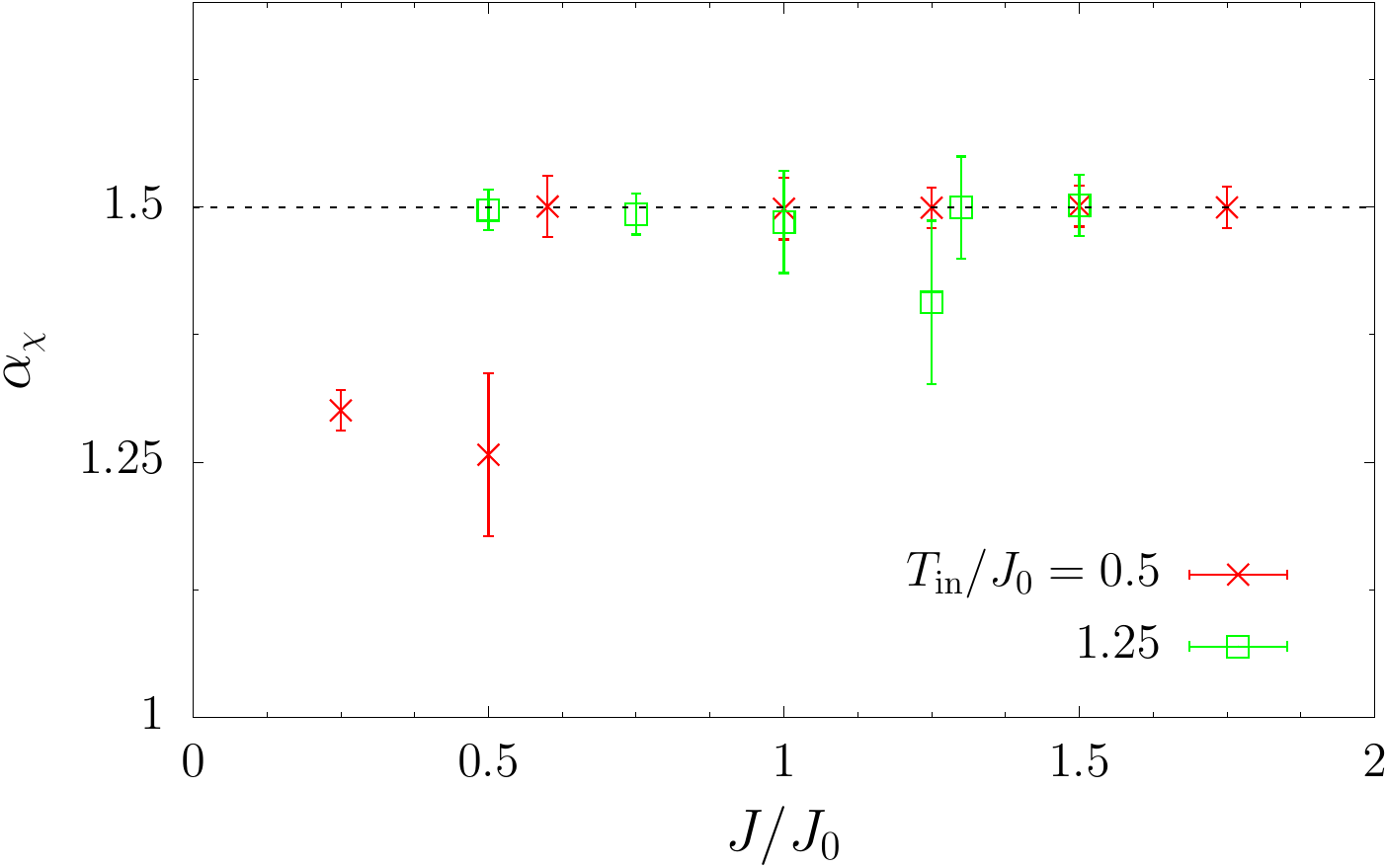}
\end{center}
\caption{\small
(a) and (b) Asymptotic behaviour of the dynamical susceptibility, $\chi(t)$.
The quantity $ \left| \chi(t) - \chi_{\infty} \right|$,
with $\chi_{\infty} = \lim_{t \to +\infty} \chi(t) $, is plotted against $t$,
in the case of quenches with $T' = 0.5 \ J_0 $ (purple curve).
The black and red continuous lines correspond to the upper and lower envelopes, respectively.
The black dashed line represents a fit of  the upper envelope to the function
$f(x) = a \, x^{-b}$. The numerical values of $a$ and $b$ are indicated in the key.
(c) The exponent $b$ obtained from fitting the envelope of $ \chi(t) - \chi_{\infty}$,
plotted against the parameter $J/J_0$, for two different values of $T'/J_0$.
}
\label{fig:susc_T0p50}
\end{figure}

The dynamic susceptibility also approaches a constant asymptotic value
with an algebraic decay. The first two panels in Fig.~\ref{fig:susc_T0p50}
show examples of this behaviour for $T'<J_0$. The black and red continuous
curves follow the upper and lower envelopes. The dotted black line is a
fit to the power law decay. For higher values of $J/J_0$ we still find
$b=1.5$. The summary of exponent values found is given in panel (c) in the same
figure both for $T'/J_0<1$ and $T'/J_0>1$.

The ``oscillatory'' part of $ \chi(t) - \chi_f$, defined as
$\chi_{\rm osc}(t) = ( \chi(t) - \chi_f ) \cdot t^{\alpha_{\chi}}$
where $\alpha_{\chi}$ is the exponent obtained by fitting
the envelope of $ \chi(t) - \chi_f $ with a power law, as done in  Fig.~\ref{fig:susc_T0p50}
(a) and (b), follows the same time dependence as the one of $z(t)$. More precisely,
\begin{equation}
\chi(t) \simeq \chi_f +  t^{-\alpha_\chi} \,
[ B_{-} \cos(\omega_{-} t + \varphi_{-}) + B_{+} \cos(\omega_{+} t + \varphi_{+}) ]
\; ,
\end{equation}
with $ B_{-}, B_{+}, \varphi_{-}, \varphi_{+}$ fitting parameters and
$\omega_{\pm} = \sqrt{(z_f \pm 2 J)/m }$,
describes the data very accurately. As beforehand,
when $z_f=2 J$, $\omega_{-} = 0$, and thus $\chi(t)$ reduces to a single harmonic.
We also note that this form implies
\begin{eqnarray}
R(t)
\!\! &\!\!  \simeq \!\! &  \!\!
-\alpha_\chi t^{-\alpha_\chi-1} \,
[ B_{-} \cos(\omega_{-} t + \varphi_{-}) + B_{+} \cos(\omega_{+} t + \varphi_{+}) ]
\nonumber\\
&&
-
t^{-\alpha_\chi} \,
[ B_{-} \omega_- \sin(\omega_{-} t + \varphi_{-}) + B_{+} \omega_+ \sin(\omega_{+} t + \varphi_{+}) ]
\; ,
\end{eqnarray}

The kinetic and potential energies, $e_{\rm kin}(t)$ and $e_{\rm pot}(t)$, depend
 linearly on $z(t)$,
\begin{eqnarray}
 e_{\rm kin}(t) = \frac{1}{2} e_{\rm tot} + \frac{1}{4} z(t) \; , & \qquad\qquad &
 e_{\rm pot}(t) = \frac{1}{2} e_{\rm tot} - \frac{1}{4} z(t) \; ,
\end{eqnarray}
with $e_{\rm tot}$ the conserved total energy density and, therefore, they have the
same time dependence as $z(t)$ once the factor and sign are taken into account.

\subsubsection{Correlation with  the initial state}
\label{sec:corr-t0}

A very similar analysis to the one described in the previous Subsection
can be applied to the correlation between the instantaneous configuration
and the initial one. The envelope to be studied is in this case
\begin{equation}
   \mathrm{env} \left[ C(t,0) - q_0 \right] \ \sim \ t^{-\alpha_{ \small C }} \,
 \label{eq:corr-t0-asymptotics}
\end{equation}
with $q_0 = \lim_{t \to +\infty} C(t,0) $ being 0 for $T'/J_0>1$
and different from zero for $T'/J_0<1$ and $T'<J$ (sector III, dashed in green in dynamic phase diagram
displayed in Fig.~\ref{fig:phase-diagram}).
We have to note an important difference
with the analysis of the pre-asymptotic behaviour of $z(t)$ performed in the previous
section. While $z_f$ is continuous, $q_0$ is discontinuous at the transition $T'=J$. We do not have an analytic expression
for $q_0$ but we have clear numerical evidence for this claim, that was shown in Fig.~18 (b) in~\cite{CuLoNePiTa18}.
This implies that the behaviour on the left of the critical line may not be the same as the one
on the right of it entailing, as we will see, a discontinuity of the exponent $\alpha_C$ itself.

\begin{figure}[h!]
\vspace{0.5cm}
\begin{center}
 (a) \hspace{5cm} (b) \hspace{5cm} (c) \hspace{3cm} $\;$
\\
\vspace{0.2cm}
 \includegraphics[scale=0.36]{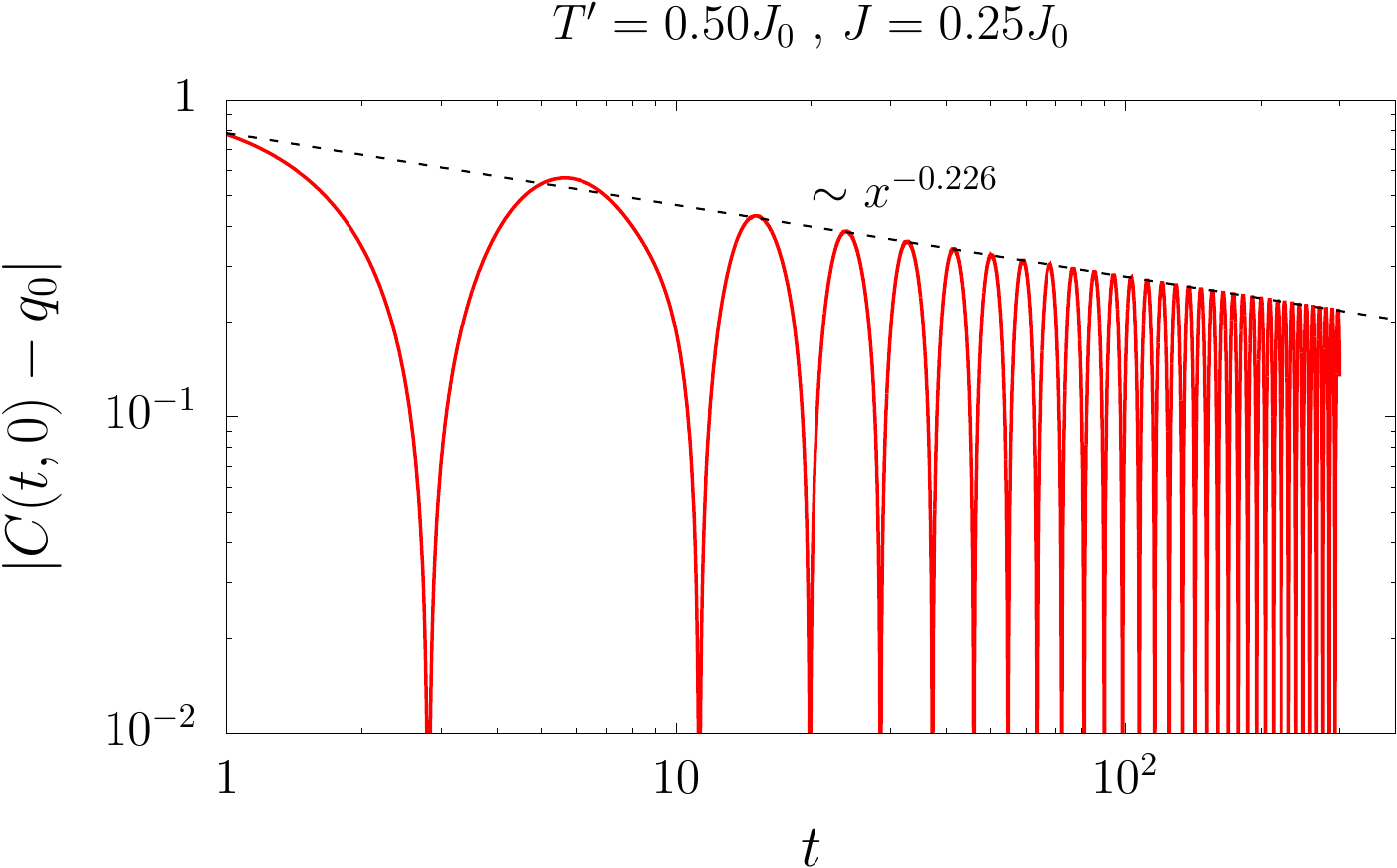}\quad%
 \includegraphics[scale=0.36]{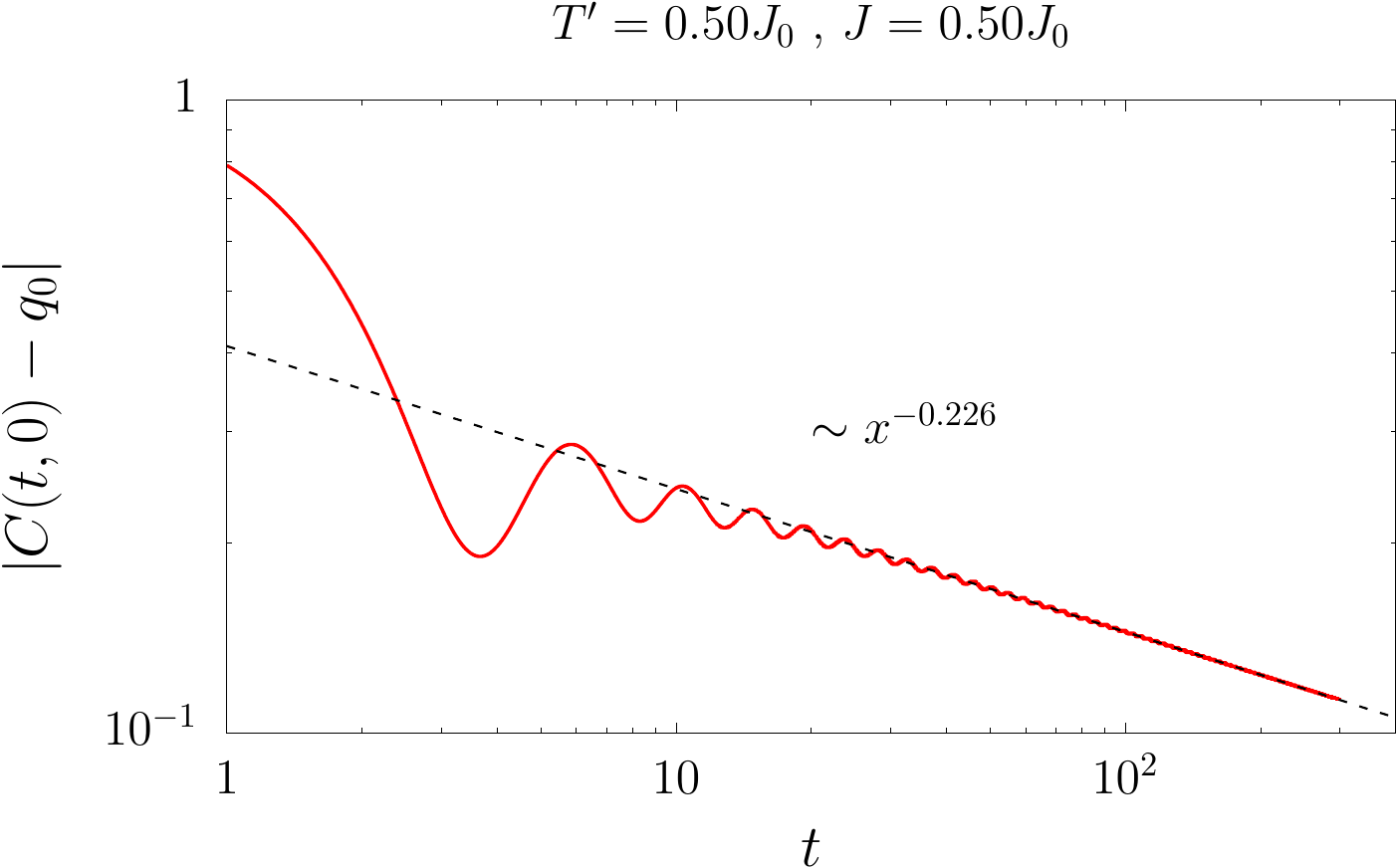}\quad%
 \includegraphics[scale=0.36]{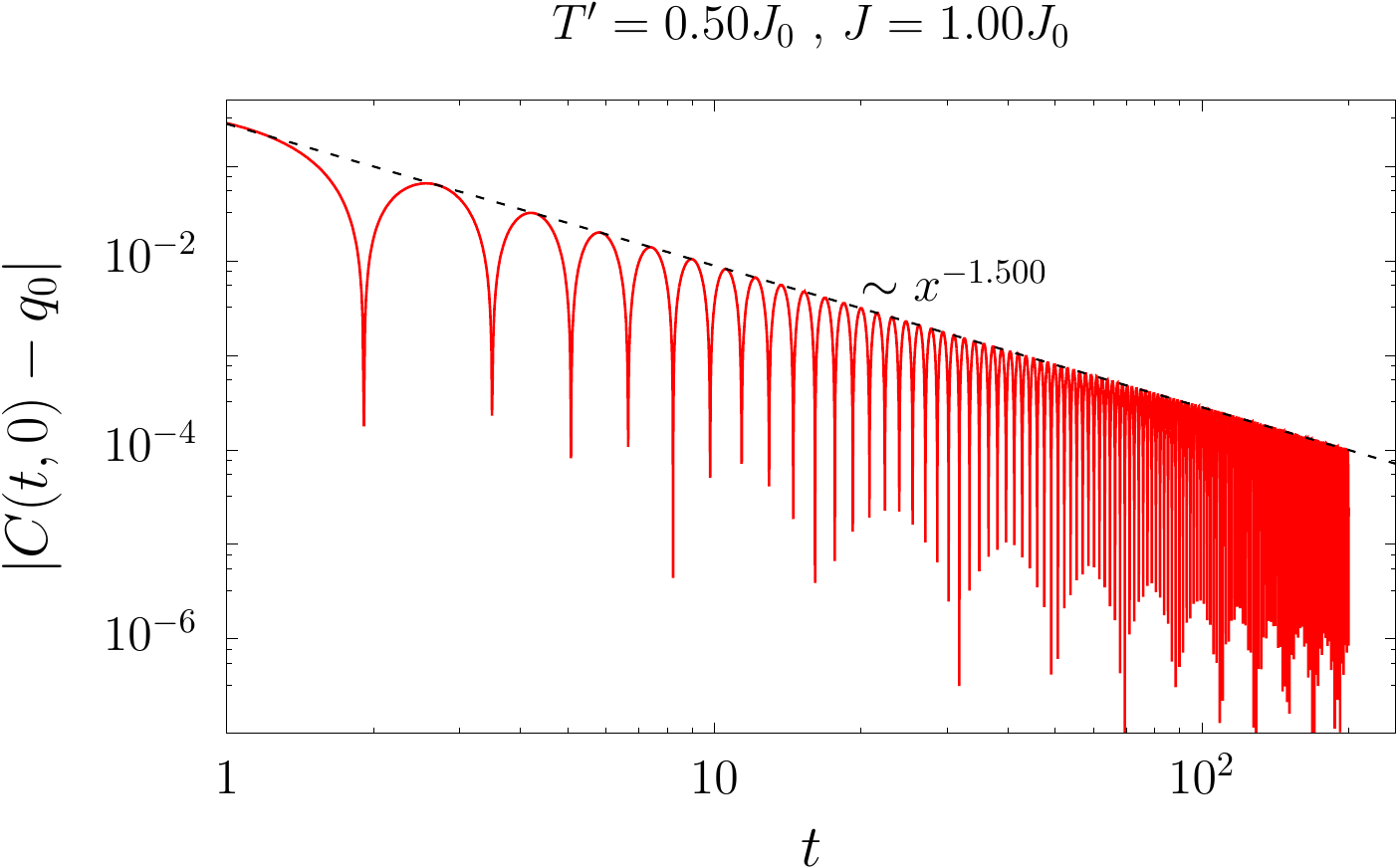}\quad%
\\
\vspace{0.5cm}
 (d) \hspace{5cm} (e) \hspace{5cm} (f) \hspace{3cm} $\;$
\\
\vspace{0.2cm}
 \includegraphics[scale=0.36]{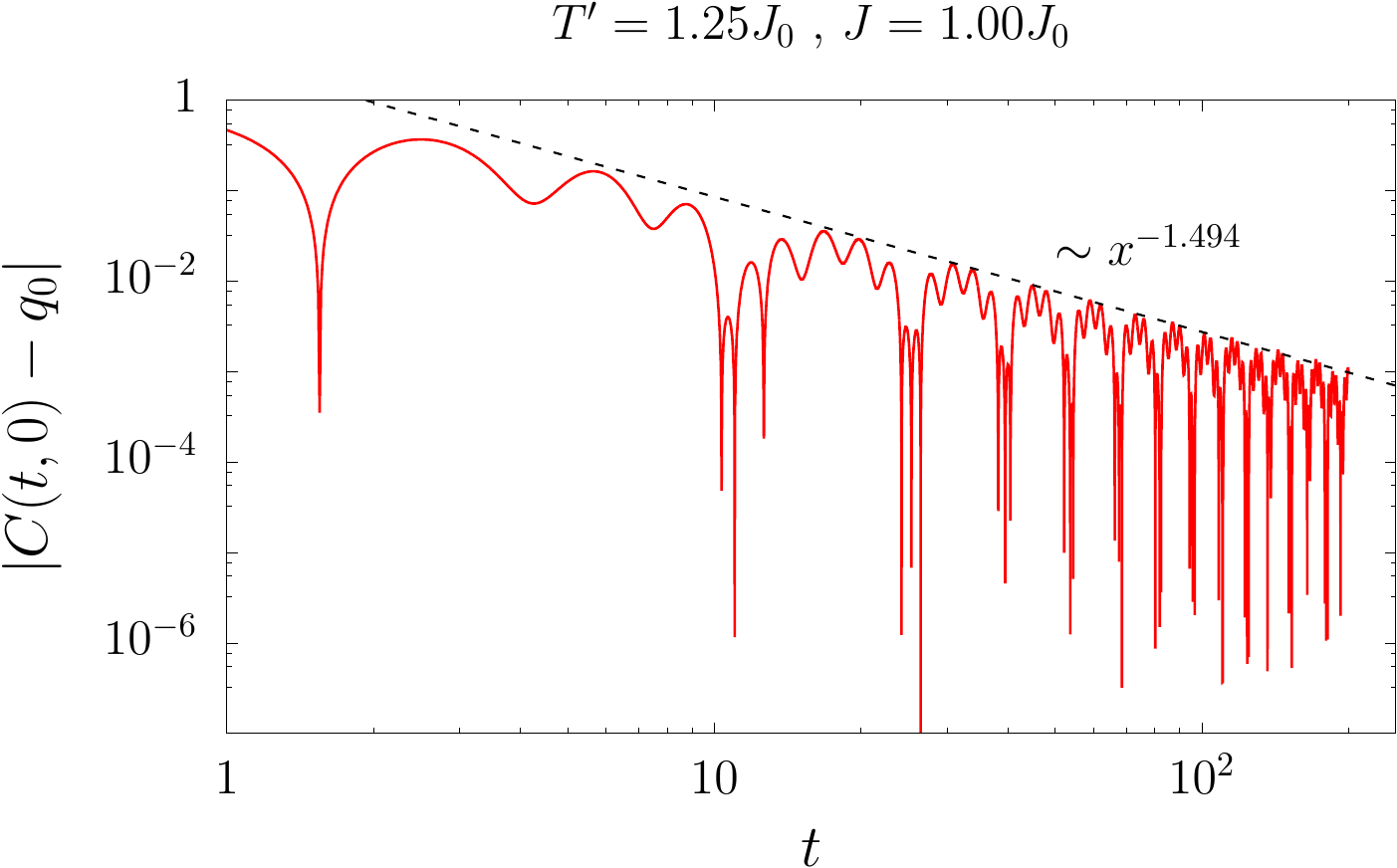}\quad%
 \includegraphics[scale=0.36]{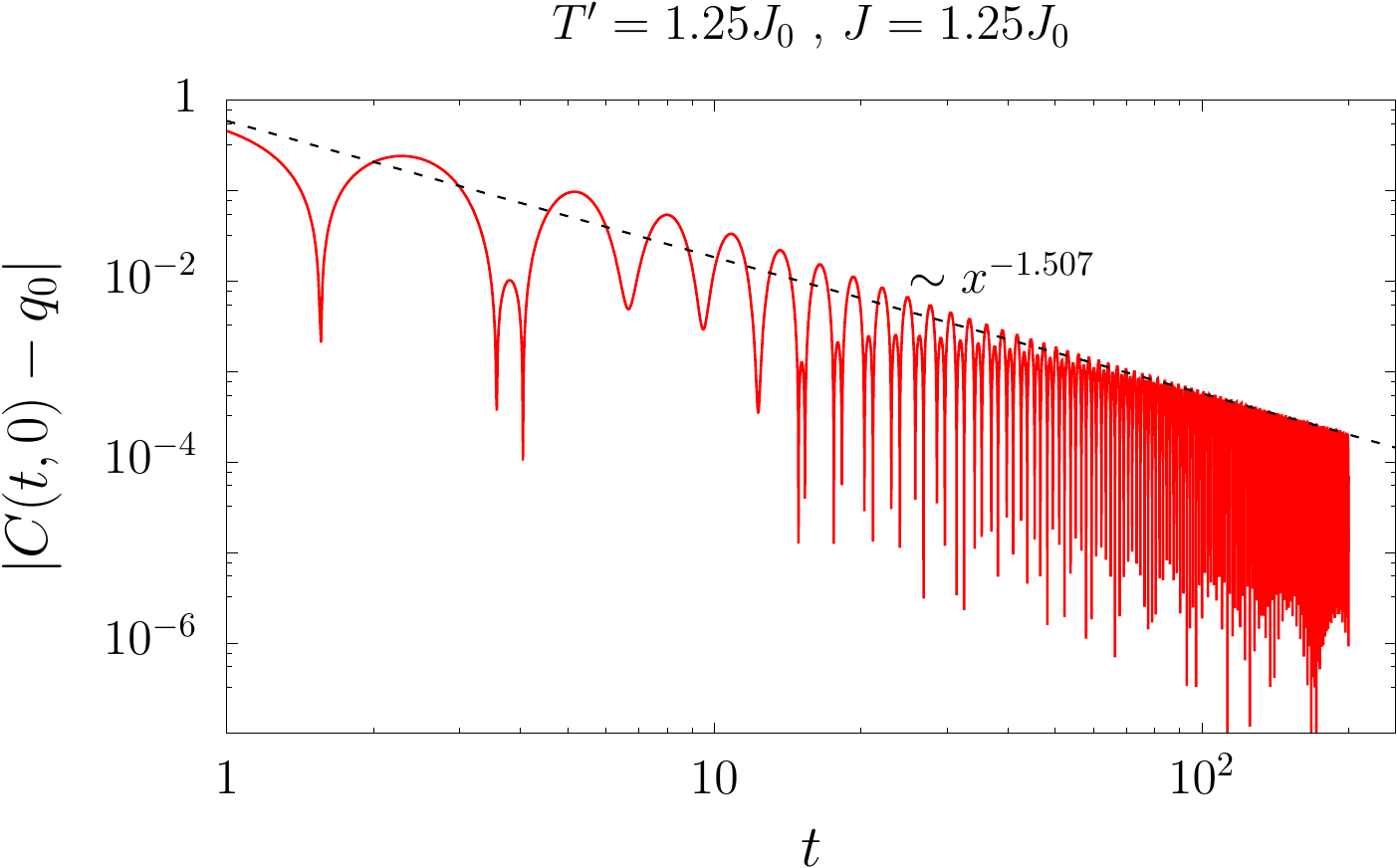}\quad%
 \includegraphics[scale=0.36]{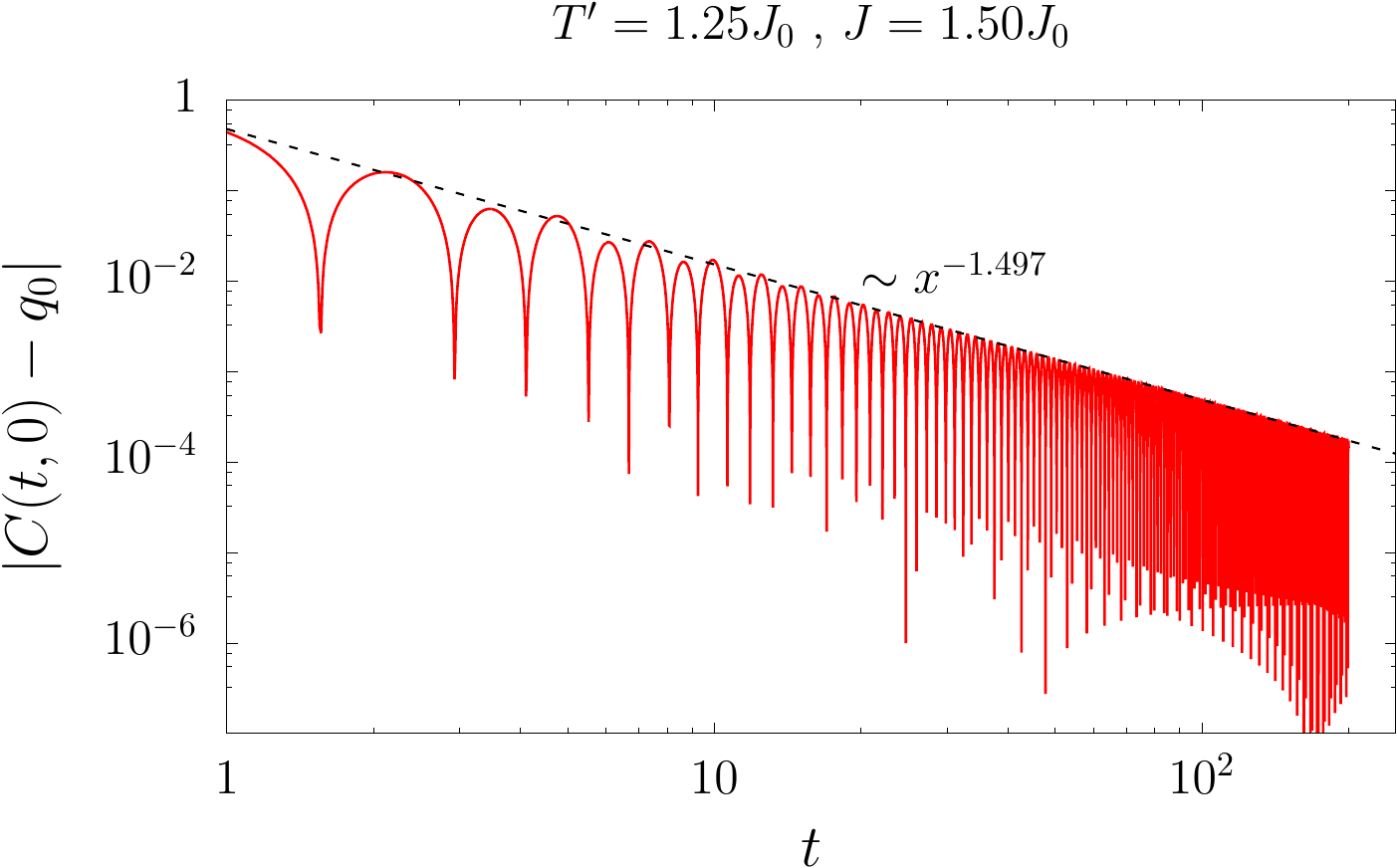}%
\end{center}
\caption{\small
Pre-asymptotic behaviour of $C(t,0)$.
The quantity $ \left| C(t,0) - q_0 \right|$, with $q_0 = \lim_{t \to +\infty} C(t,0) $, is plotted against $t$
in the case of quenches with $T' = 0.5 \ J_0 $ (first line) and $T' = 1.25 \ J_0 $ (second line).
The function $f(x) = a \ x^{-b}$ has been fitted to the envelope of $ \left| C(t,0) - q_0 \right|$ for $t \gg 1$
(dashed black line).
The numerical value of $b$ is indicated close to the fit. On the critical line the fit yields a rather small value, $\alpha_C=0.226$. Right beyond criticality,
there is a jump in the exponent $\alpha_C $ for $T'/J_0 < 1$,
see Fig.~\ref{fig:corr_t0_T0p50_env} and Fig.~\ref{fig:corr_t0_crit}.
Near the critical line, on its right ($J \gtrsim T'$), it is a bit difficult to extract the algebraic decay
of the envelope of $C(t,0) - q_0 $. However, we believe that given a sufficient long time $t$,
we should be able to recover $\alpha_C \simeq 1.5$ from the numerical data, as soon as $J > T'$.
}
\label{fig:corr_t0_T0p50}
\end{figure}

\begin{figure}[h!]
\vspace{0.5cm}
\begin{center}
\vspace{0.5cm}
\hspace{-0.75cm} (a) \hspace{6cm} (b) \hspace{3cm} $\;$
\\
\vspace{0.25cm}
 \includegraphics[scale=0.45]{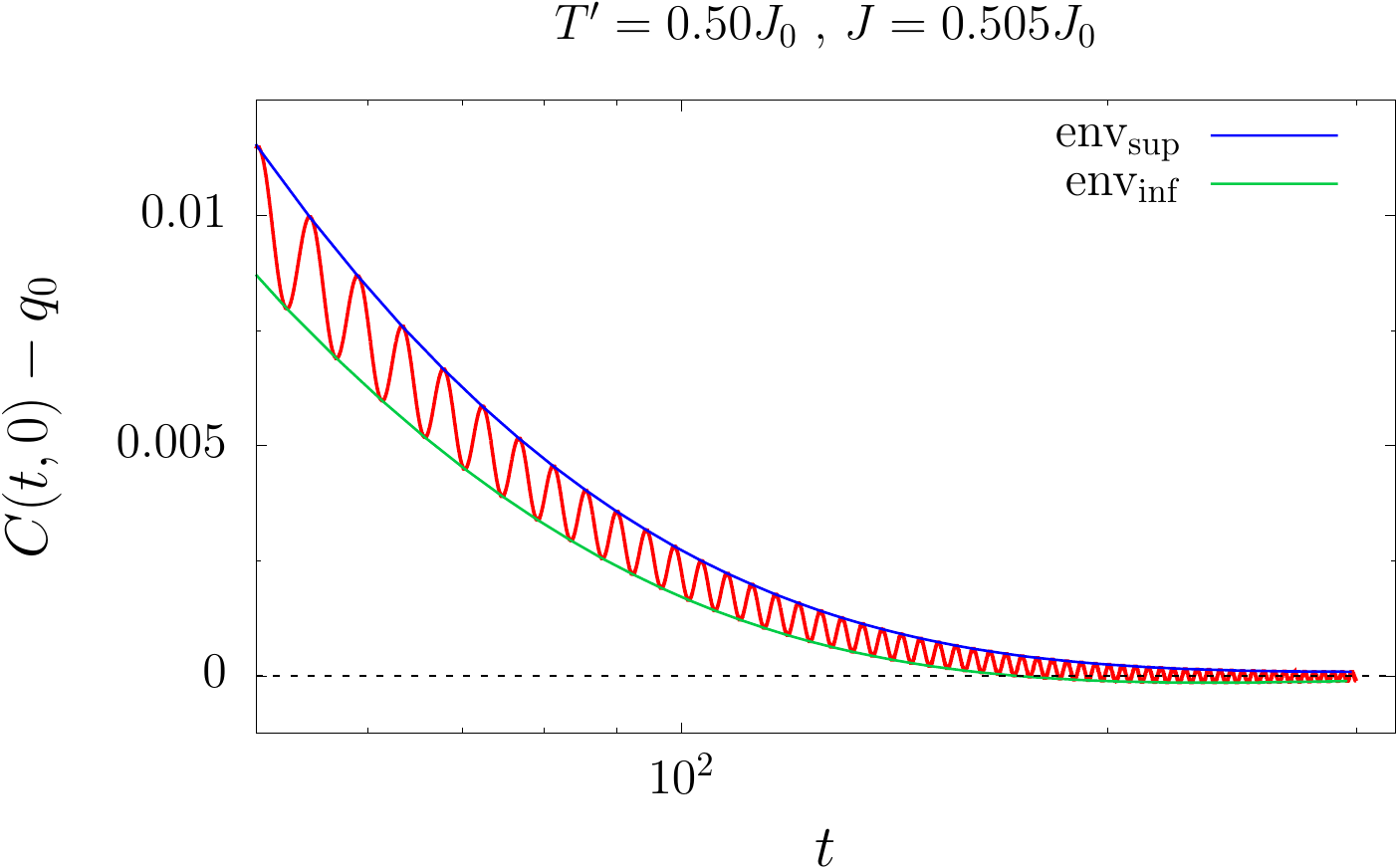}\quad%
 \includegraphics[scale=0.45]{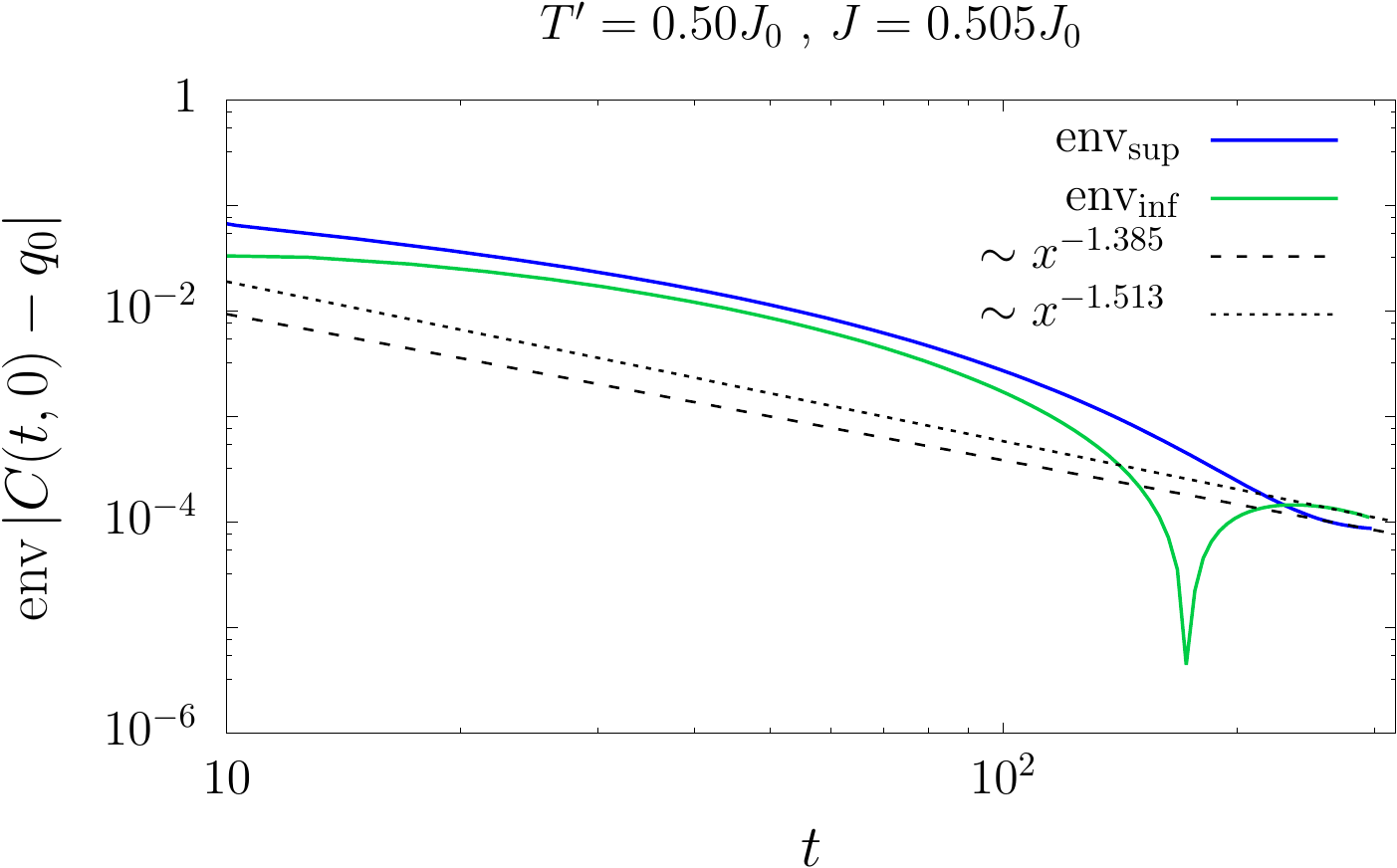}%

\vspace{0.5cm}
\hspace{-0.75cm} (c) \hspace{6cm} (d) \hspace{3cm} $\;$
\\
\vspace{0.25cm}
 \includegraphics[scale=0.45]{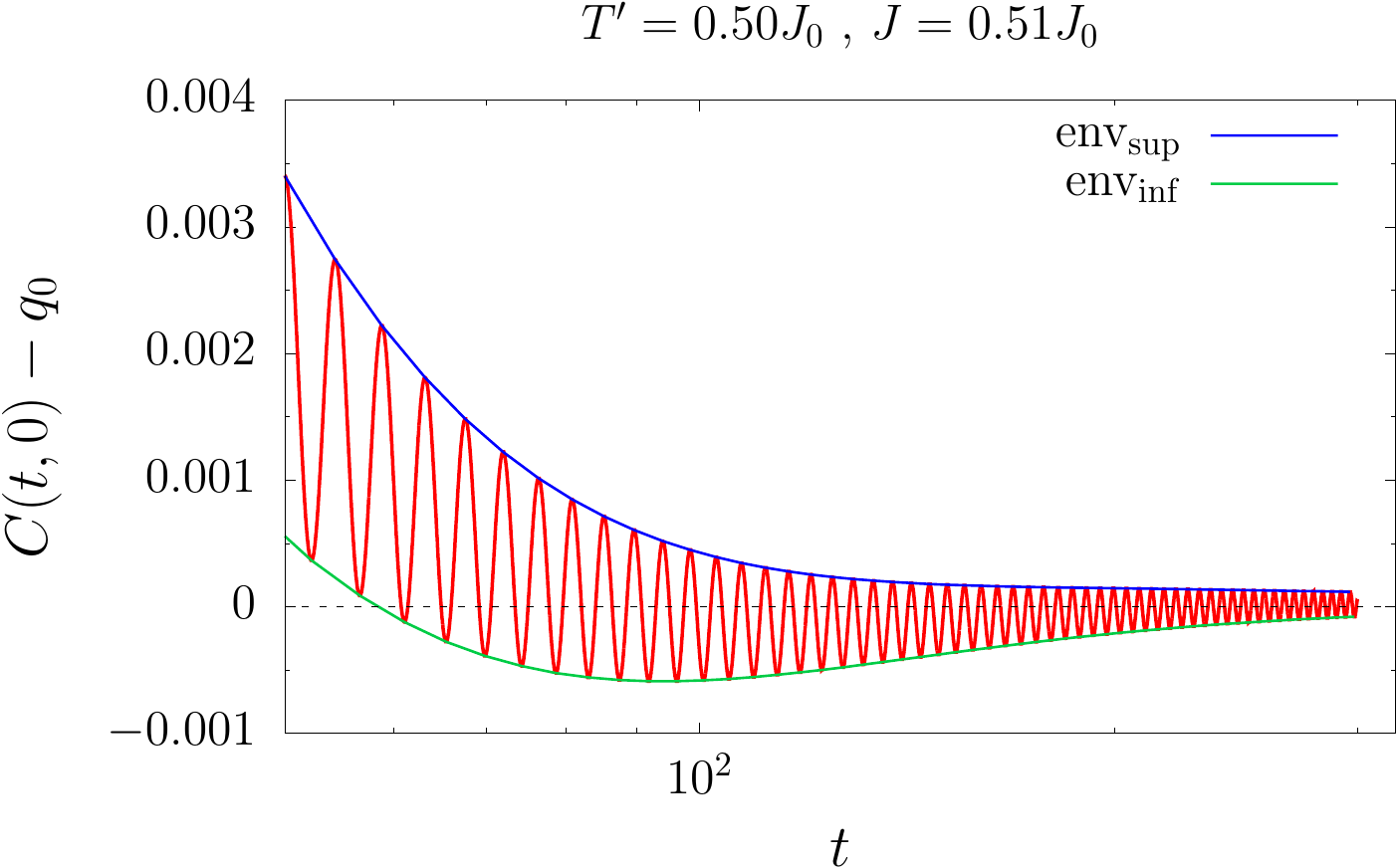}\quad%
 \includegraphics[scale=0.45]{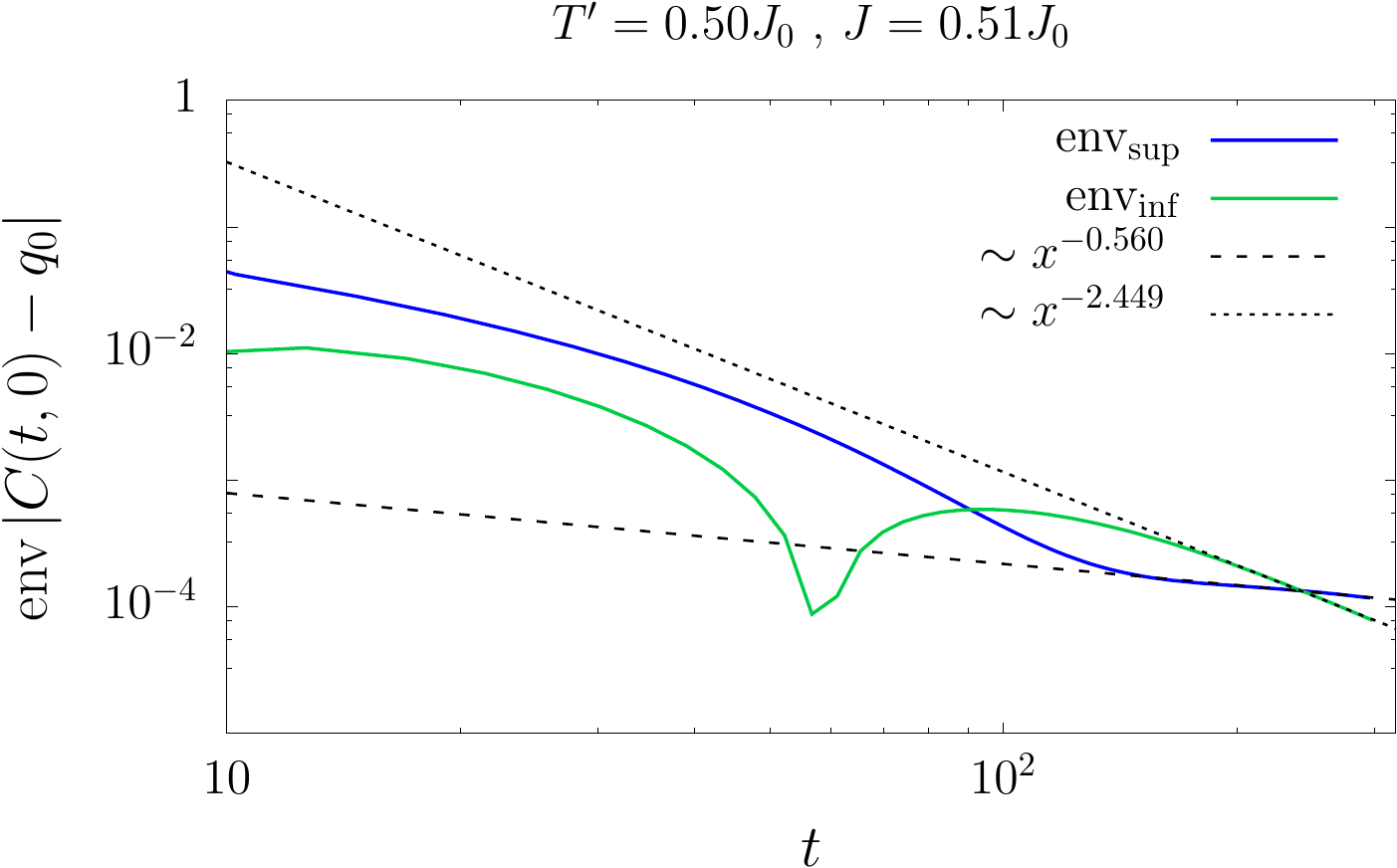}%

\vspace{0.5cm}
\hspace{-0.75cm} (e) \hspace{6cm} (f) \hspace{3cm} $\;$
\\
\vspace{0.25cm}
 \includegraphics[scale=0.45]{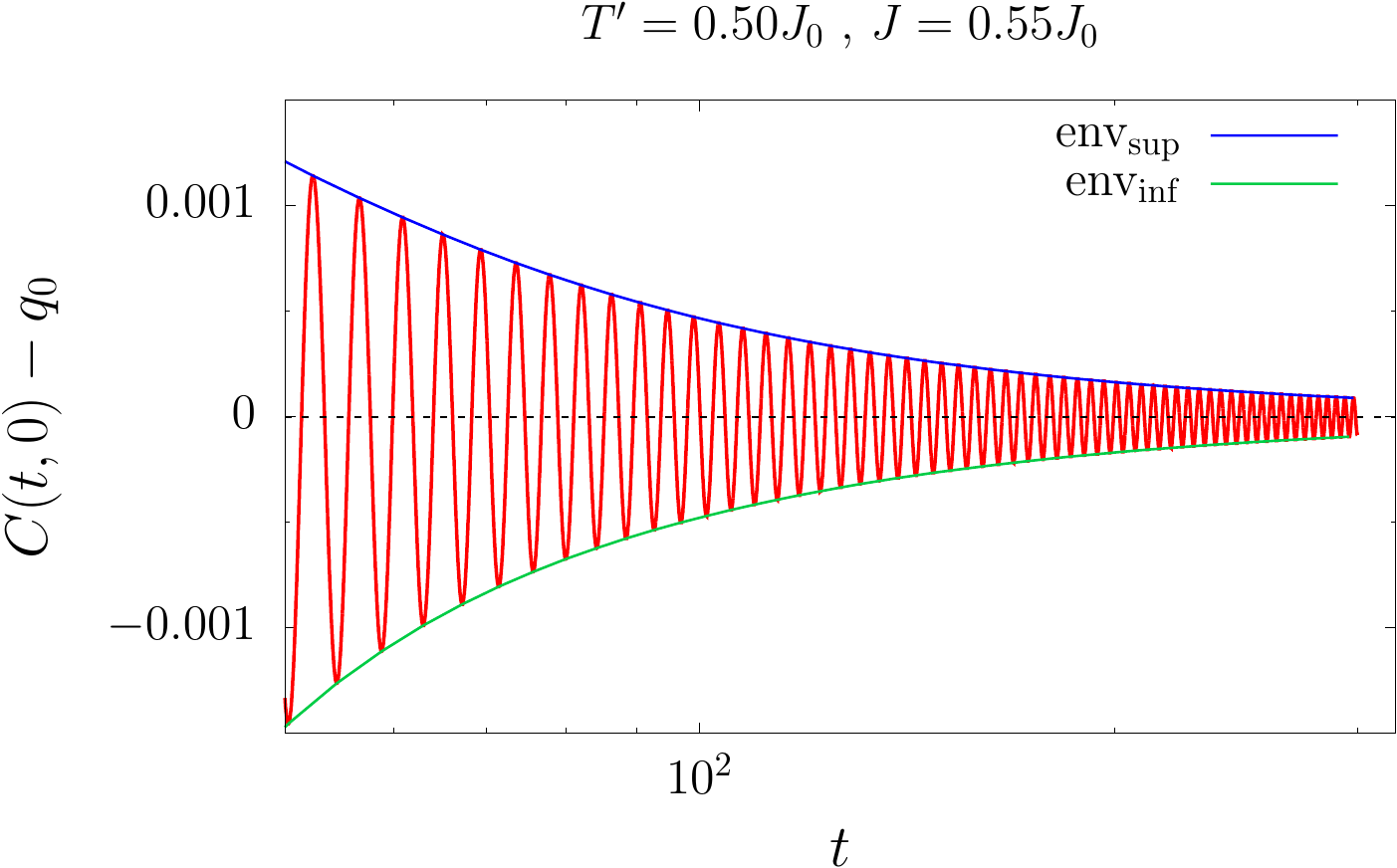}\quad%
 \includegraphics[scale=0.45]{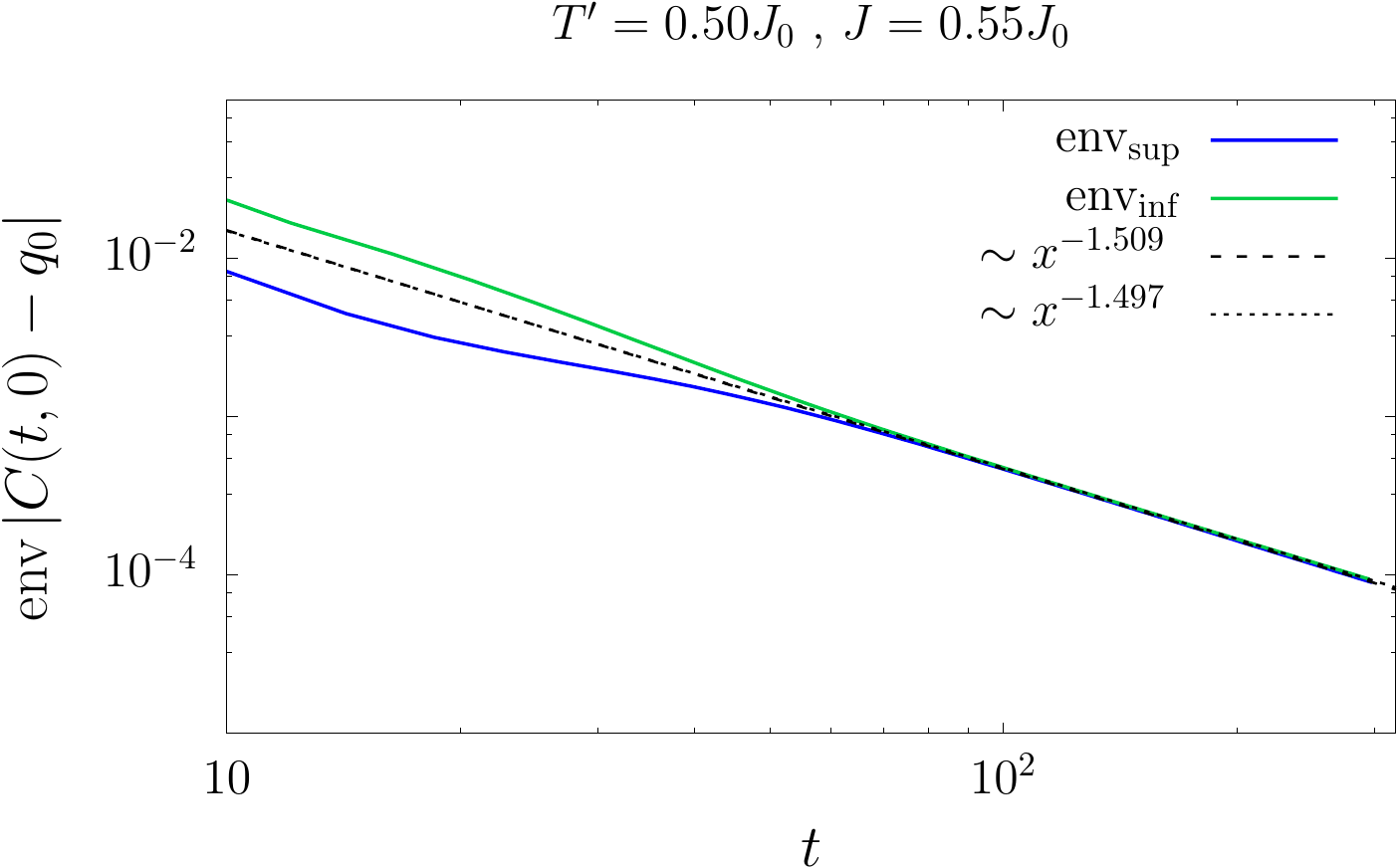}%
\end{center}
\caption{\small
Pre-asymptotic behaviour of $C(t,0)$ near the critical line.
On the left, $ C(t,0) - q_0$,
with $q_0 = \lim_{t \to +\infty} C(t,0) $, is plotted against $t$
for quenches from equilibrium at  $T' = 0.5 \ J_0 $.
On each row we present data for a value of $J/J_0$: from top to bottom,
$J/J_0 = 0.505, 0.51, 0.55$.
In the same plots, in the panels in the left column, we included the upper and lower envelopes,
indicated by a blue and a green line, respectively.
On the right column, we just plot these envelopes, $ \mathrm{env} | C(t,0) - q_0 | $, with the same colour code
against $t$ (using a double logarithmic scale)
and the corresponding fits (dashed line for the upper envelope, dotted line for the lower one).
}
\label{fig:corr_t0_T0p50_env}
\end{figure}

For reference we show in Fig.~\ref{fig:corr_t0_T0p50} the behaviour of this
correlation for $T'/J_0 =0.5$ and various choices of $J/J_0$. To notice here is the
case of the critical quench that has been approached from the disordered
side, and the asymptotic value $q_0=0$ was used. The exponent  $\alpha_C$
takes a very small value $\alpha_C\simeq 0.226$. Right beyond criticality
we see, in Fig.~\ref{fig:corr_t0_T0p50_env}, that the exponent jumps to $\alpha_C\simeq 3/2$,
the value it takes in the full trapped dynamics, with $q_0\neq 0$. There are also  differences
between the behaviour of the upper and lower envelopes in close to critical
quenches  due to the short times available.
Very close to criticality the data approach zero from above, and only at the
latest times reached numerically we see the oscillations go beyond 0 and the upper
and lower envelopes approach similar algebraic behaviours. When we get away from
criticality the amplitude of the oscillations around zero is larger at the same time scales
and the upper and lower envelope power laws get closer to each other.
We think that by exploring longer times we should be able to observe the two envelopes
with the same algebraic decay and recover $\alpha_C \simeq 1.5$ in all these cases.

\begin{figure}[h!]
\begin{center}
\hspace{-2cm} (a) \hspace{5cm} (b) \hspace{5cm}  (c) $\;$
\\
\vspace{0.2cm}
 \includegraphics[scale=0.36]{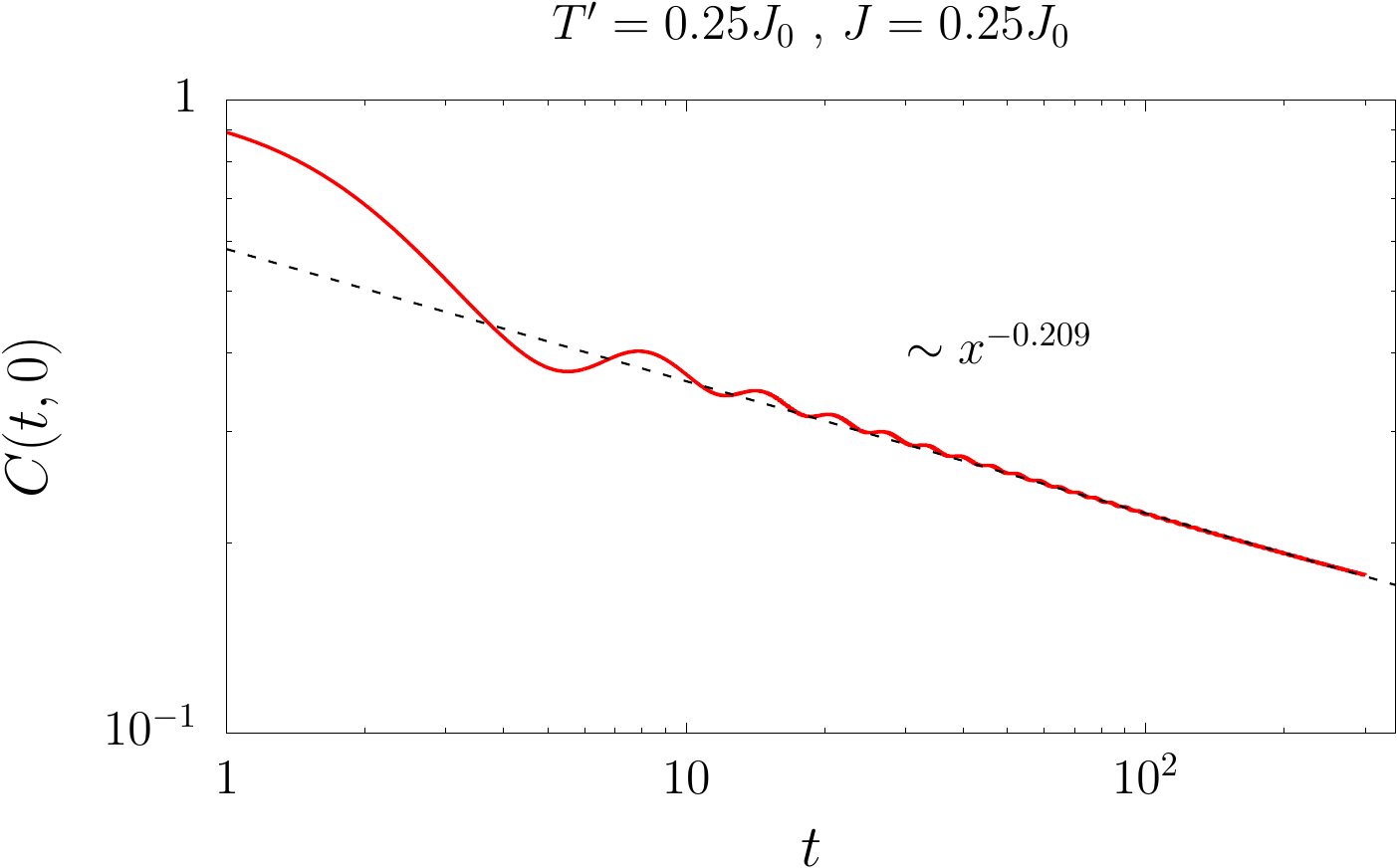}\quad%
 \includegraphics[scale=0.36]{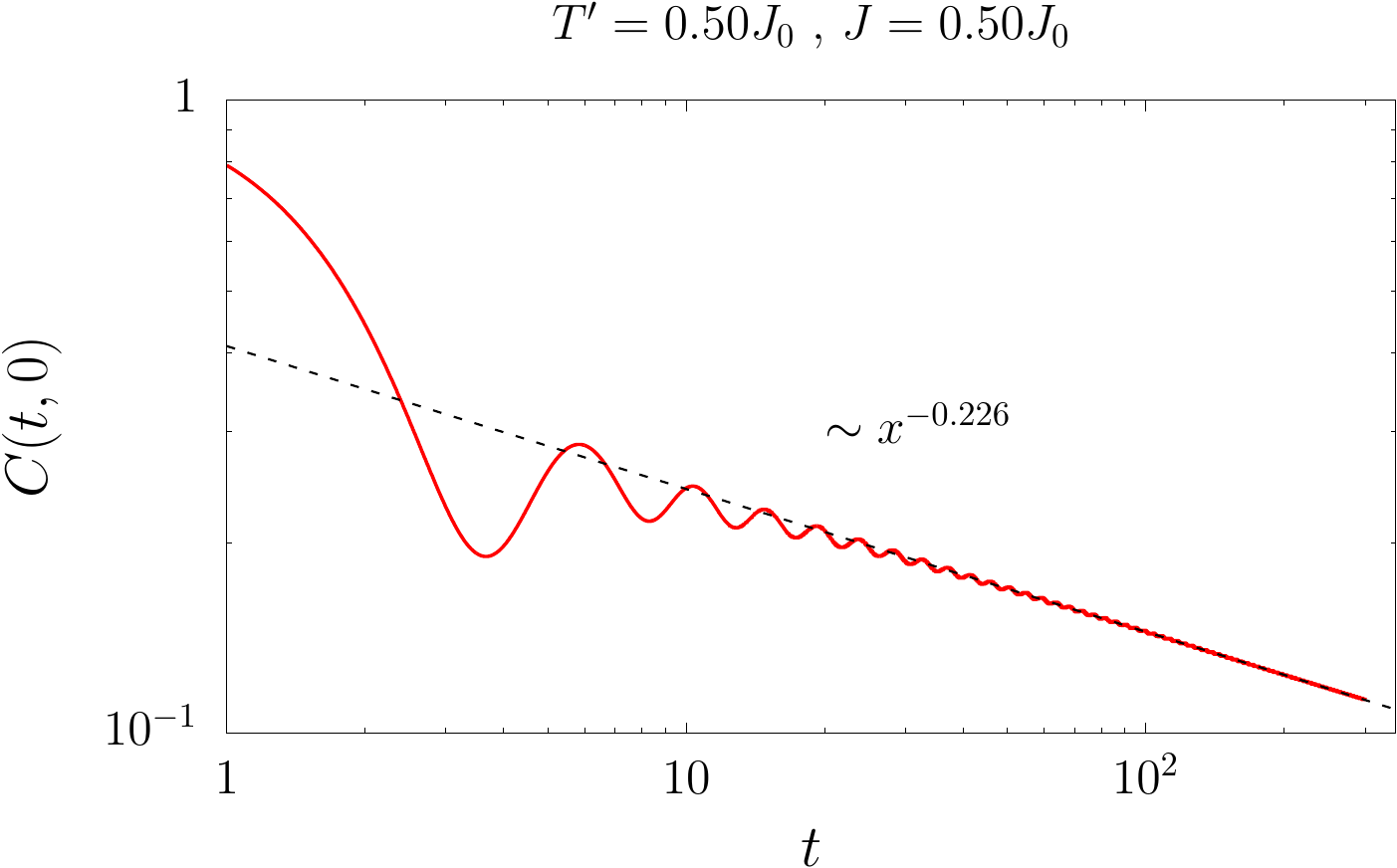}\quad%
 \includegraphics[scale=0.36]{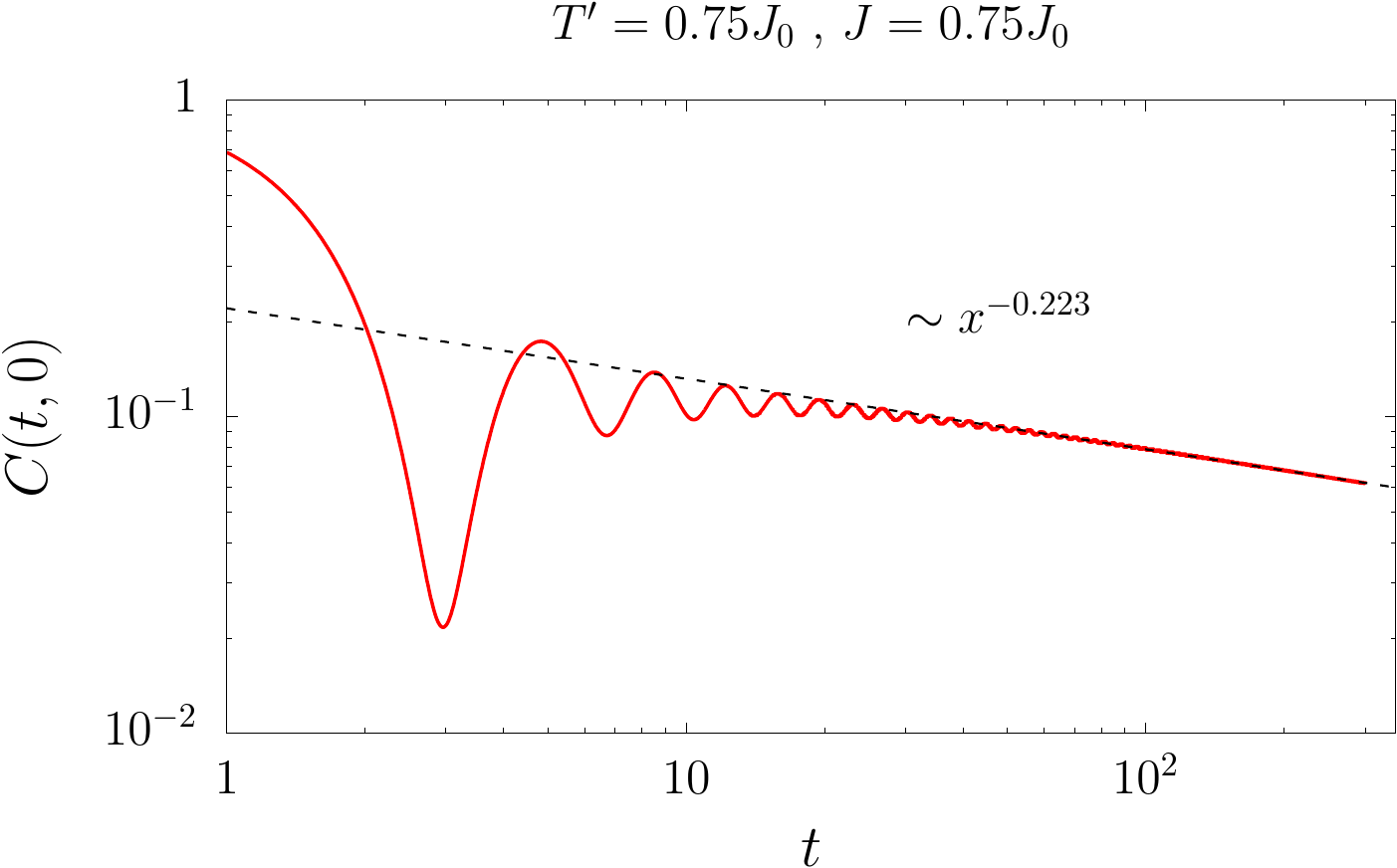}%
\end{center}
\caption{\small
Pre-asymptotic behaviour of $C(t,0)$, on the critical line $T' = J$, for $T'/J_0 = 0.25, 0.5, 0.75$.
The function $f(x) = a \ x^{-b}$ has been fitted to
$C(t,0)$ for $t \gg 1$ (dashed black line).
The numerical value of $b$ is indicated close to the fit.
}
\label{fig:corr_t0_crit}
\end{figure}

Interestingly enough, on the critical line $T'=J$, both for $T'<J_0$ and $T'>J_0$, the
oscillations are suppressed at sufficiently long times, and the correlation approaches $q_0$
from above, taking only positive values,
see Fig.~\ref{fig:corr_t0_T0p50} (b) and the three panels in Fig.~\ref{fig:corr_t0_crit}.  Moreover,
in all the fits, we took $q_0=0$ meaning that we were approaching the critical line
from the side $T'>J$.

The collection of exponent values $\alpha_C$ thus derived are shown in Fig.~\ref{fig:exp_decay_z} (b)
and Fig.~\ref{fig:exp_decay_z_ht} (b). A comparison with the exponent $\alpha_z$ shown in
the panels (a) of the same figures proves that $\alpha_C$ has a similar behaviour,
though not identical, to the one
of $\alpha_z$.  Both for $T'/J_0 <1$ and $T'/J_0>1$ the two exponents are constant and coincide,
$\alpha_z = \alpha_C=3/2$, beyond the transition at $T'=J$.
Instead, for both kinds of initial states $\alpha_C$ weakly depends on the parameters for
$J/J_0<1$. It is hard to establish beyond any doubt whether $\alpha_C=\alpha_z$ for these parameters
but the numerical data are compatible with this possibility. In the region in between $J/J_0=1$ and
$T'/J=1$ the determination of
$\alpha_C$ is more difficult. The data do not exclude the possibility of $\alpha_C$ being constant and
equal to $3/2$ in this region too but a very weak variation could also be hidden within the errorbars.

\subsubsection{The time-delayed self-correlation}
\label{sec:corr-small-tw}

The time-delayed correlation $C(t_1,t_2)$ depends, in principle, on the two times $t_1$ and $t_2$.
In the study in~\cite{CuLoNePiTa18} we did not find ageing phenomena in this model.
We want to investigate more deeply here the possible two-time dependencies and,
in particular, evaluate whether an interrupted ageing scenario, with the
two-time correlation decaying as
\begin{equation}
  C(t_1,t_2) - q_{\infty}(t_2) \sim  f(t_1-t_2) +
  \  \mathcal{A}(t_2)  \; g\left( \frac{t_1}{t_2} \right) \,
  \qquad\qquad
  \mbox{with}
  \qquad\qquad
  \mathcal{A}(t_2) \simeq a t_2^{-\alpha_C'}
   \mbox{for} \, t_2 \ll 1, \quad t_1 \gg t_2
 \label{eq:corr-st-asymptotics}
\end{equation}
and $ q_{\infty}(t_2) = \lim_{ \tau \to +\infty} C(t_2 + \tau,t_2) $, is possible.

For initial conditions in the ordered phase $T'/J_0<1$ or quenches from the
disordered phase $T'/J_0$ to parameters on the left of the critical line, that is to say
 the white zone in the phase diagram in Fig.~\ref{fig:phase-diagram} that we
called sector I,
we do not expect ageing nor interrupted ageing, that is to say, $a=0$. This is confirmed by
our numerical studies (not shown). We can therefore immediately study the stationary dynamics
in these cases.

For quenches into the orange sector of the phase diagram, that we called II,
one needs to be more careful and check whether there is, or not, interrupted ageing.
We can mention at least three reasons for this possibility. Firstly, it was  for this kind of
quench that we found ageing in the $p=3$ model~\cite{CuLoNe17}. Secondly, these
kind of energy extraction quenches from disordered initial conditions are the ones that
yield interrupted ageing in the isolated $O(N)$ model in the large $N$ limit~\cite{ScBi10,ScBi11,ScBi13,ChNaGuSo13,MaChMiGa15,ChTaGaMi16,ChGaDiMa17,Berges15, BoDeHoSa99, BoDeDe04}.
Thirdly, a current study that is being carried out on the dynamics of the isolated
one dimensional scalar field theory after quenches of different kind displays non trivial
two-time dependencies in various correlations after this kind of procedures~\cite{BaChCuGa19}.

\begin{figure}[h!]
\vspace{0.5cm}
\begin{center}
\hspace{-2cm} (a) \hspace{5cm} (b) \hspace{5cm}  (c) $\;$
\\
\vspace{0.2cm}
 \includegraphics[scale=0.375]{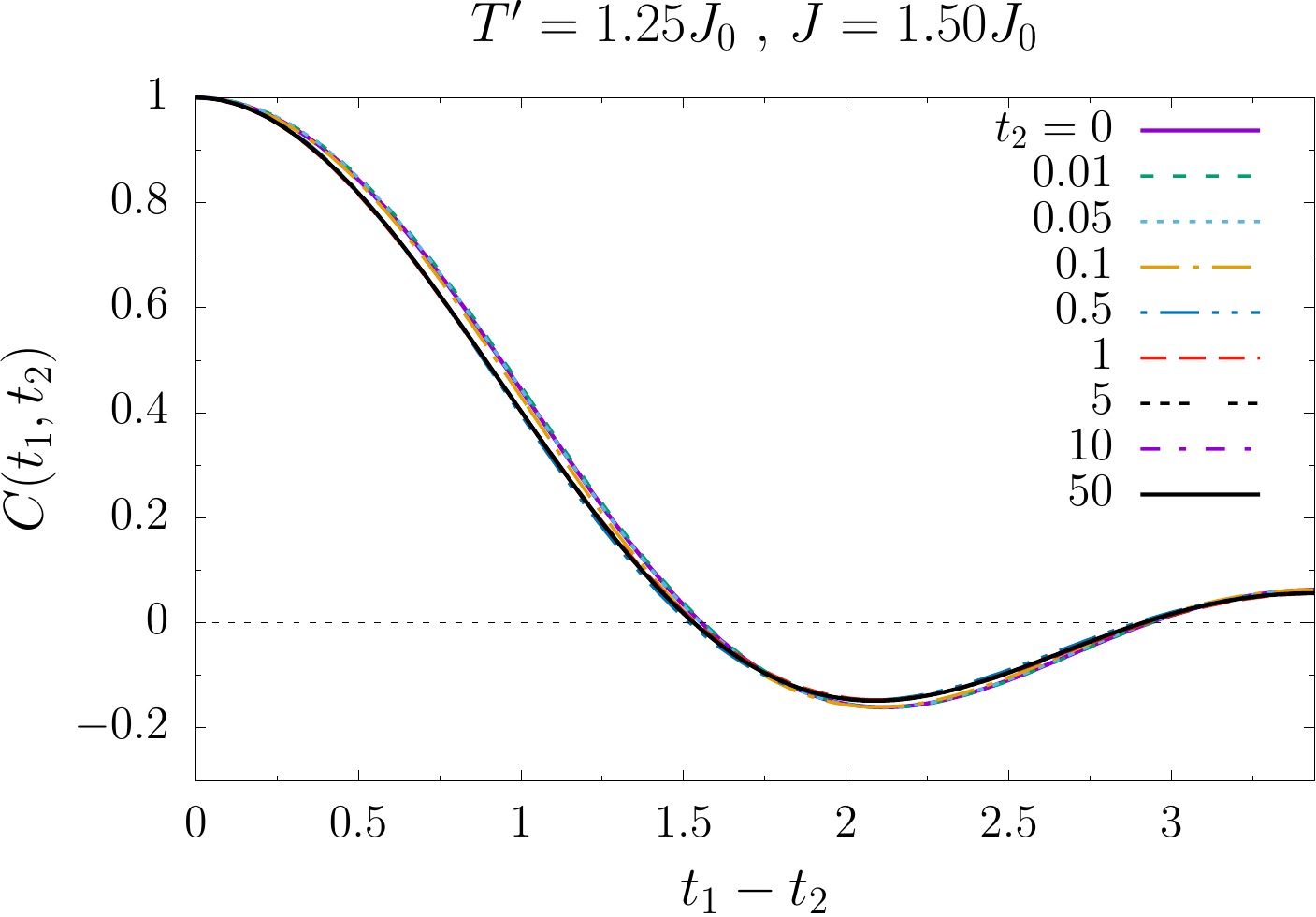}
 \includegraphics[scale=0.375]{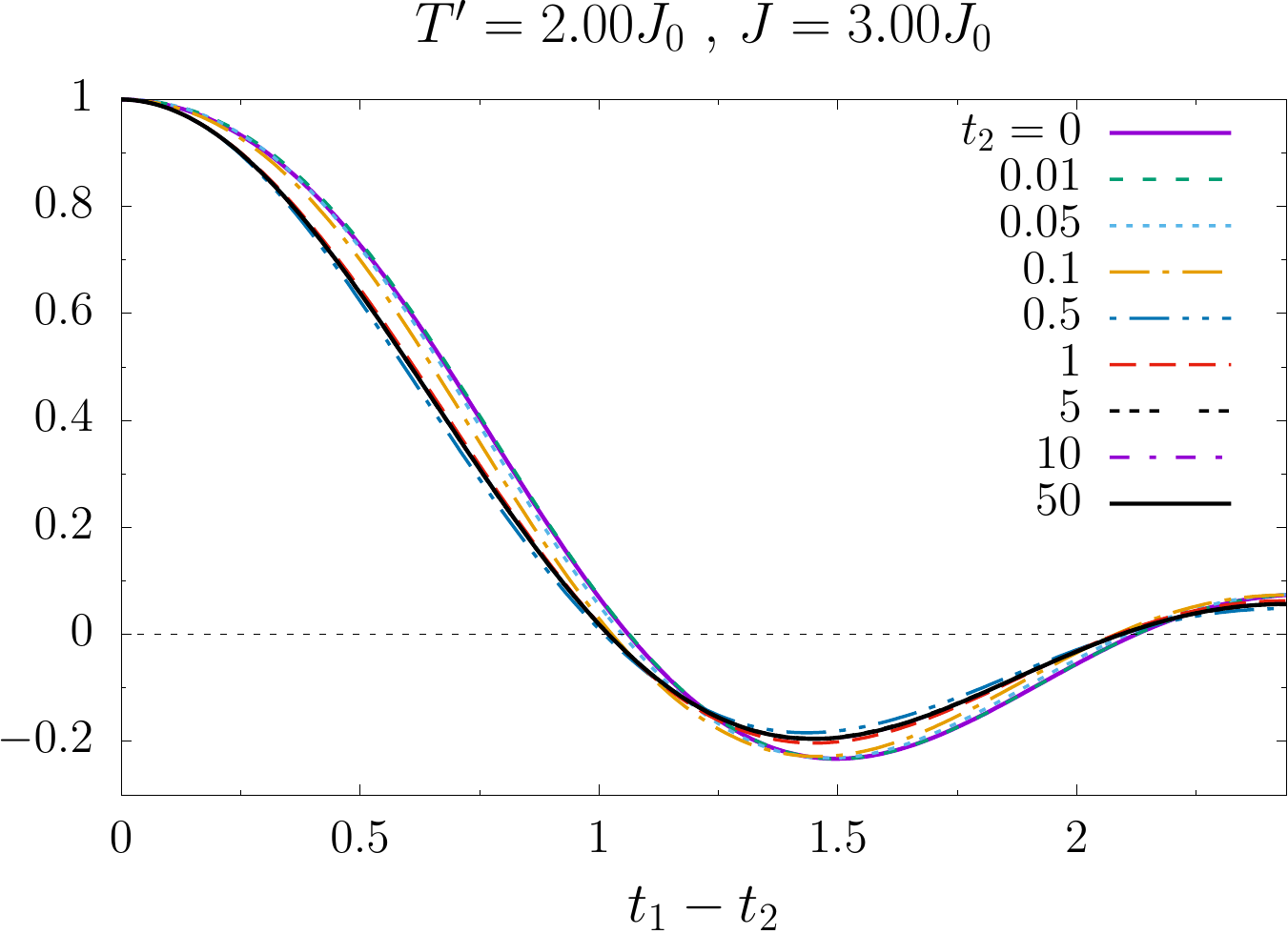}
 \includegraphics[scale=0.375]{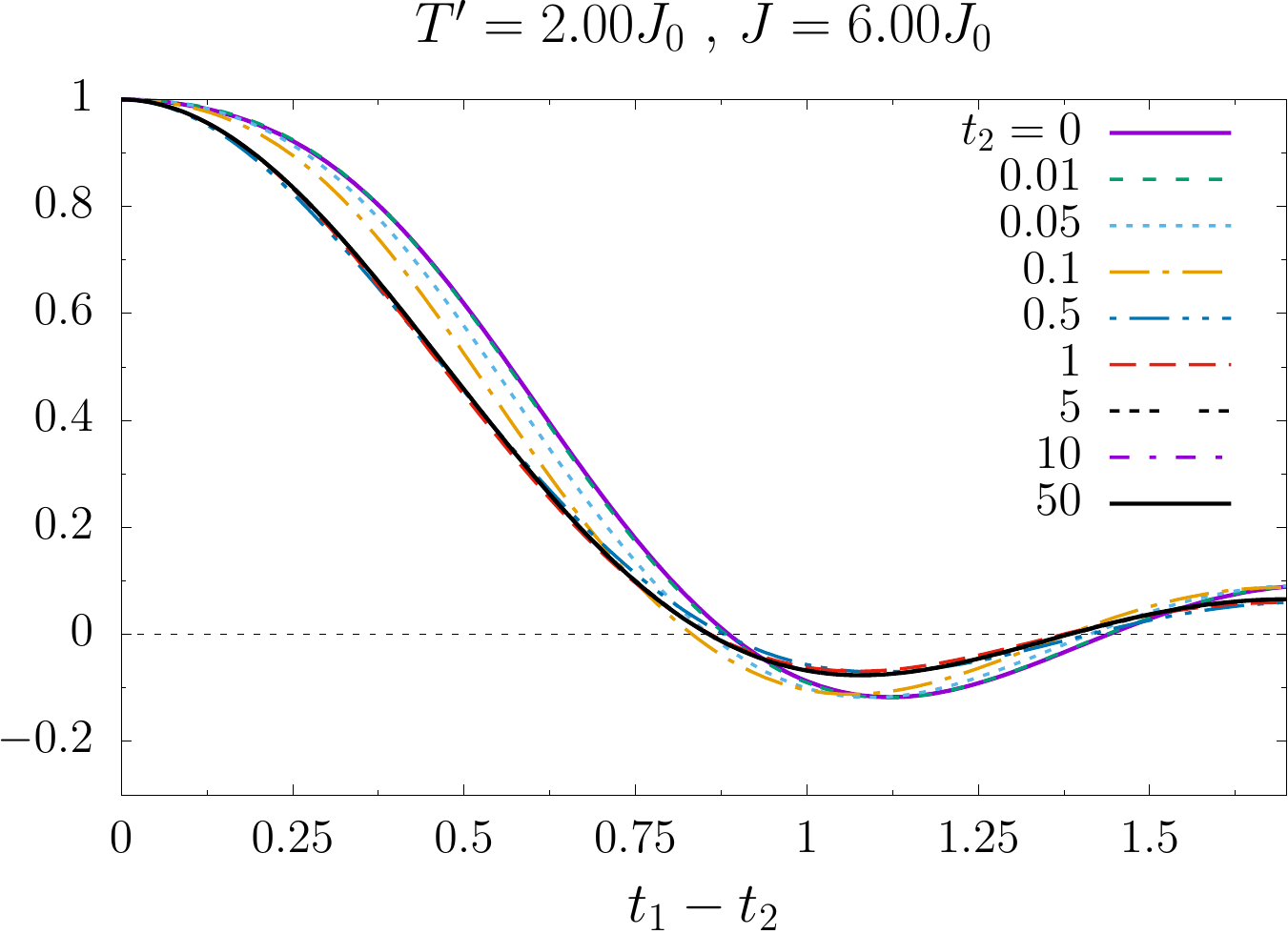}
\end{center}
\caption{\small
Time-delayed correlation $C(t_1,t_2)$, for different values of $t_2$ (indicated in the key),
plotted against $t_1 - t_2$, for three quenches in Sector II of the phase diagram
($T' / J_0 > 1$, \ $ T' < J $). In all cases the data rapidly approach, say after $t_2\approx 5$,
a stationary behaviour. 
}
\label{fig:corr_zoom}
\end{figure}

In Fig.~\ref{fig:corr_zoom} we plot $C(t_1,t_2)$ as a function of $t_1-t_2$ for various
$t_2$, say $t_2=0, 0.01, 0.05, 0.1, 0.5, 1, 5, 10, 50$
after quenches into the orange sector of the phase diagram, that is to say, for
parameters such that $T' / J_0 > 1$, $T' < J $. We show the data in linear scale and
for  short $t_1-t_2$ only, so as to avoid the many oscillations hiding the behaviour we
wish to highlight. In all cases the dynamics soon become stationary, with the
$t_2$ dependence being lost after, say, $t_2=5$. We conclude that there is no
relevant interrupted ageing in this problem.

For $t_2 \gg 1$, that is to say, for times $t_2$ such that $C$ has become stationary,
we can perform a similar analysis of the algebraic approach to the asymptotic
limit $q$
\begin{equation}
   \mathrm{env} \left[ C(t_1,t_2) - q \right] \ \sim
    \left( t_1- t_2 \right)^{-\alpha_{ \small C }''} \,
 \label{eq:corr-lt-asymptotics}
\end{equation}
with $ q = \lim_{t_2\to\infty} q_{\infty}(t_2) = \lim_{t_2\to\infty} \lim_{ \tau \to +\infty} C(t_2 + \tau,t_2) $.

\begin{figure}[h!]
\vspace{0.5cm}
\begin{center}
 \includegraphics[scale=0.7]{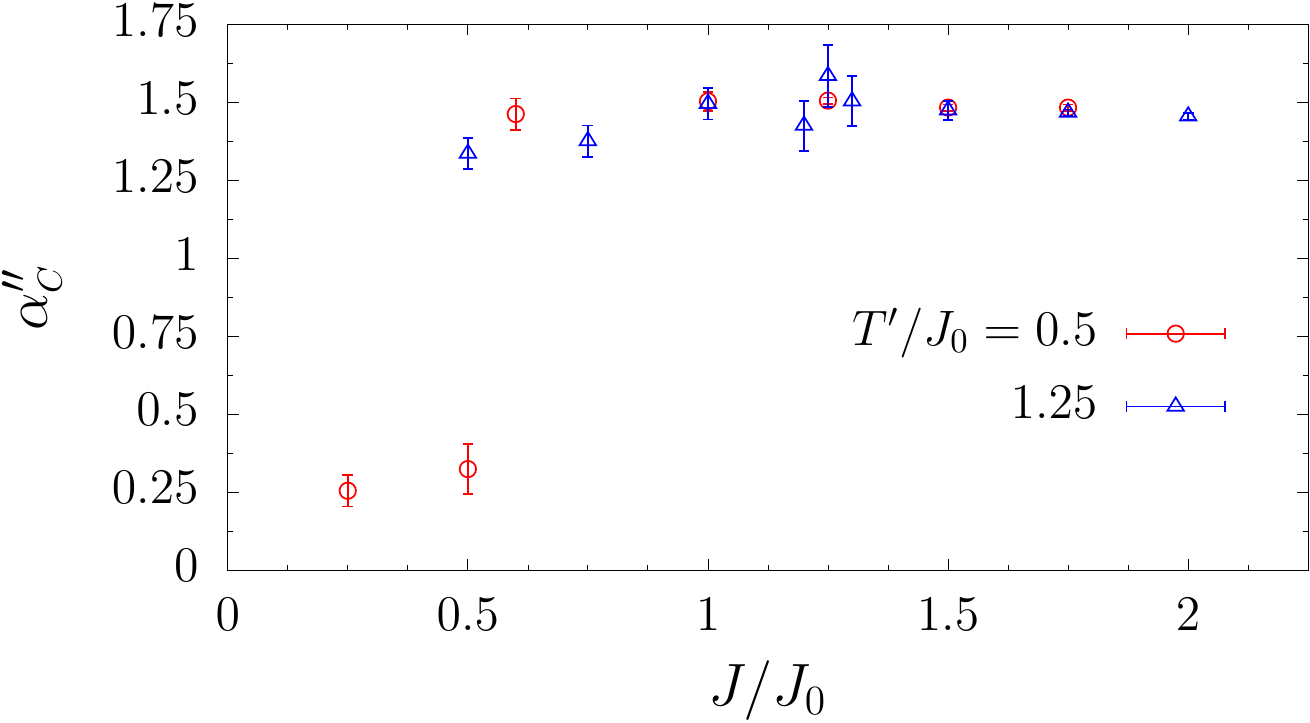}\quad%
\end{center}
\caption{\small
Exponent $\alpha^{\prime\prime}_C$ of Eq.~(\ref{eq:corr-lt-asymptotics}) obtained from fitting the envelope of
$ \left| C(t_1,t_2) - q \right|$ as function of $x = t_1 - t_2$ (for $t_2 \gg 1$),
plotted against the parameter $J/J_0$, for two different values of $T'/J_0$.
}
\label{fig:exp_decay_corr_lt}
\end{figure}

Figure~\ref{fig:exp_decay_corr_lt} summarises the results of the study of this envelope
and its fits with a power law.
Once again we find that for $T'/J_0 <1$ and $J > T'$ (on the right of the critical line)
the exponent fixes to $3/2$ while for $T'/J_0>1$ the data suggest that the behaviour
of $\alpha_{C}''$ is similar to the one of $\alpha_{C}$. That is,
for $ T' / J_0 > 1$, $\alpha_{C}''$ is equal to $3/2$ for $J > J_0$.
For $J < J_0$, $\alpha_{C}''$ seems to decrease but it still remains very close to $3/2$.

\section{Mode analysis}
\label{sec:mode-analysis}

The dynamics can also be addressed by writing the Langevin equations in the rotated basis
constructed with the  eigenvectors $\vec{v}_{\mu}$, $\mu = 1, . . . , N$, 
of the post-quench matrix $J_{ij}$. Indeed,  defining
\begin{equation}
s_{\mu}(t)=\vec{s}(t) \cdot \vec{v}_{\mu}
\; , 
\end{equation}
the $N$ rotated equations of motion read
\begin{equation}\label{eq:newton_eq0}
m \ddot{s}_{\mu}(t)+(z(t)-\lambda_{\mu})s_{\mu}(t)=0
\; .
\end{equation}
where $\lambda_\mu$ are the 
eigenvalues associated to the eigenvectors $\vec{v}_{\mu}$.

This set of equations has to be complemented with the initial conditions $s_\mu(0)$ and $\dot s_\mu(0)$. They are
close to the equations for a parametric oscillator, with the difference that here
the time-dependent frequency depends on the variables {\it via} the Lagrange multiplier. 
In this formulation the relation with Neumann's integrable classical system~\cite{Neumann} is explicit.

The uniform interaction quench corresponds to
\begin{equation}
\lambda^{(0)}_{\mu}=\frac{J_0}{J}\lambda_{\mu}
\; .
\end{equation}

\subsection{Exact solution}
\label{subsec:parametric-osc}

Equation~(\ref{eq:newton_eq0}) can be implicitly solved with the amplitude-phase {\it Ansatz}~\cite{Ermakov,Milne,Pinney,SoCa10}  in which the projection of the spin configuration is
written as
\begin{eqnarray}
s_\mu(t) &=& 
s_\mu(0) \sqrt{\frac{\Omega_{\mu}(0)}{\Omega_{\mu}(t)}}  \; \cos \int_0^t dt' \; \Omega_{\mu}(t')
+  \frac{\dot s_\mu(0)}{\sqrt{\Omega_{\mu}(t)\Omega_{\mu}(0)}} \; \sin \int_0^t dt' \; \Omega_{\mu}(t') 
\label{eq:spin-proj}
\end{eqnarray}
and the one of the spin's projection velocity as
\begin{eqnarray}
\dot s_\mu(t) &=& 
\left[
-\frac{1}{2} s_\mu(0) \sqrt{\frac{\Omega_{\mu}(0)}{\Omega_{\mu}(t)}} 
\frac{\dot\Omega_{\mu}(t)}{\Omega_{\mu}(t)} 
+\dot s_\mu(0) \sqrt{\frac{\Omega_\mu(t)}{\Omega_\mu(0)}}   \; 
\right]
\cos \int_0^t dt' \; \Omega_{\mu}(t')
\nonumber\\
&&
+ 
\left[ 
-\frac{1}{2} \frac{\dot s_\mu(0)}{\sqrt{\Omega_{\mu}(t)\Omega_{\mu}(0)}} 
\frac{\dot \Omega_\mu(t)}{\Omega_\mu(t)} 
- s_\mu(0) \sqrt{\Omega_\mu(0)\Omega_\mu(t)} 
\right]\; \sin \int_0^t dt' \; \Omega_{\mu}(t') 
\; ,
\label{eq:veloc-proj}
\end{eqnarray}
complemented with 
the equation ruling the evolution of $\Omega_{\mu}(t)$,
\begin{equation}\label{eq:omega0}
\frac{1}{2}\frac{\ddot\Omega_{\mu}(t)}{\Omega_{\mu}(t)}-\frac{3}{4}\left( \frac{\dot\Omega_{\mu}(t)}{\Omega_{\mu}(t)} \right)^2+\Omega^2_{\mu}(t)
= \omega^2_\mu(t) \equiv (z(t)-\lambda_{\mu})/m
\; , 
\end{equation}
with  the initial condition
\begin{equation}
\dot\Omega_{\mu}(0)=0
\; .
\end{equation}
Equations~(\ref{eq:spin-proj}) and (\ref{eq:veloc-proj}) 
 are reminiscent of the general solution of the harmonic oscillator problem,  here with a
time-dependent ``frequency'' $\Omega_\mu(t)$.
In the numerical calculations we choose the initial condition
\begin{equation}
\Omega_{\mu}^2(0) = \lambda_N-\lambda_{\mu}
\; ,
\end{equation}
that ensures that $\Omega_\mu(t)$ is real at all times.

The spherical constraint 
fixes the time-dependence of the Lagrange multiplier that can be expressed in different ways. For example,  
\begin{eqnarray}
&&
z(t) = 
     2\left[ e_{\rm kin}(t)-e_{\rm pot}(t) \right]
     = \frac{1}{N}\sum_{\mu}\left[  \frac{\langle p^2_{\mu}(t)\rangle}{m}+\lambda_{\mu}\langle s^2_{\mu}(t)\rangle  \right]
     =
     e_f + \frac{2}{N} \sum_\mu \lambda_\mu \langle s_\mu^2(t) \rangle
     \label{eq:z}
\; . 
\end{eqnarray}
In Ref.~\cite{CuLoNePiTa18} we gave more details on the numerical solution of the problem set in these terms. For our purposes here, that is finding the preasymptotic approach towards the limit value of 
$z(t)$, the finite $N$ solution is not well adapted, since we observe that even for relatively large number of modes, say $N=1024$, the power law is blurred at times that are quite short, say, $t=50$. In the next Subsection we propose an approximation that allows us to take the large $N$ limit within this formalism.

\subsection{Decay of the Lagrange multiplier}

The formalism in the previous subsection is exact but still quite complicated and, as we explained, not so 
well adapted to the numerical analysis of preasymptotic effects. We can make a series of assumptions, that we 
will check are verified, to go farther into a quasi-analytic description of the problem.
 
We therefore propose that if $z(t)$ is sufficiently close to its stationary limit, i.e., if $z(t)\simeq z_{f}$, then
\begin{eqnarray}
 &&  s_{\mu}(t) = \sqrt{\frac{2T_{\mu}}{z_f-\lambda_{\mu}}} \;
  \cos\left( \omega^{(f)}_{\mu}t+\phi_{\mu} \right) 
  \; , \\
 &&  p_{\mu}(t) = -\sqrt{2T_{\mu}m} \; \sin\left( \omega^{(f)}_{\mu}t+\phi_{\mu} \right) 
 \; ,
\end{eqnarray}
where $\omega^{(f)}_{\mu}=\sqrt{(z_f-\lambda_{\mu})/m}$ and $\phi_{\mu}$
is an unspecified phase. This form for $p_{\mu}(t)$ is consistent with the one obtained differentiating
$s_{\mu}(t)$ with respect to time, and this implies that the two phases $\phi_\mu$ should be the
same. This {\it Ansatz} assumes that when $z(t)$ is close enough to its stationary value, all the modes decouple and behave like independent harmonic oscillators, something that is verified in our calculations and that we discussed in~\cite{CuLoNePiTa18}.
The amplitude of the $s_\mu$ oscillation is given by $\sqrt{2T_{\mu}/(z_f-\lambda_{\mu})}$, where $T_{\mu}$ is the $\mu$th mode temperature. Inserting this {\it Ansatz} in the equation for
$z(t)$, Eq.~(\ref{eq:z}), we obtain
\begin{equation}
  z(t)=\frac{2}{N}\sum_{\mu} \, T_{\mu}\left[  \sin^2 \left(\omega^{(f)}_{\mu}t+\phi_{\mu} \right)+\frac{\lambda_{\mu}}{z_f-\lambda_{\mu}}\cos^2 \left(\omega^{(f)}_{\mu}t+\phi_{\mu} \right)  \right]
  \; .
  \label{eq:z-modes}
\end{equation}
If we are only interested in the asymptotic behaviour and we assume that $\phi_{\mu}$ is approximately the same for all modes we can take $\phi_{\mu}=0$. In the long time limit the factors $\sin^2(\omega^{(f)}_{\mu}t)$ and $\cos^2(\omega^{(f)}_{\mu}t)$ average to 
$1/2$ and a constant  limit is reached due to dephasing among the different mode contributions,
\begin{equation}
1 =  \frac{1}{N} \sum_\mu \frac{T_\mu}{z_f-\lambda_\mu} \; ,
\label{eq:eq_zf}
\end{equation}
which is just another way of expressing the spherical constraint, now very close to the one in canonical equilibrium in which $T_\mu=T$ for all $\mu$. This equation fixes the 
constant asymptotic value of the Lagrange multiplier  $z_f$. Moreover, we showed in Sec. 5.7 in Ref.~\cite{CuLoNePiTa18}
that in the limit in which the problem can be thought of as one of independent harmonic oscillators, 
the mode temperatures are given by
\begin{equation}
\label{eq_t_lambda}
  T_\mu=\frac{T'}{2}\left(  \frac{z_f-\lambda_\mu}{z_{\rm in}-\frac{J_0}{J}\lambda_\mu} +1 \right)
  \; , 
\end{equation}
with $z_{\rm in}$ the value of the Lagrange multiplier in the initial condition.
After some careful handling of the sum in Eq.~(\ref{eq:eq_zf}) for finite $N$, 
that is described in Sec.~5.7 of this reference,
one rewrites the constraint equation as 
\begin{equation}
1 =  \frac{1}{N} \sum_\mu \frac{T'}{z_f-\lambda_\mu} \; ,
\label{eq:eq_zf2}
\end{equation}
that is to say, the canonical equilibrium equation at temperature 
$T'$ for a model with variance of the disorder distribution equal to $J^2$. This result is verified numerically, see Table 1.

In search of the finite time corrections to this asymptotic limit, we then split Eq.~(\ref{eq:z-modes}) in two 
terms, one equal to the asymptotic limit $z_f$ and the other one equal to the finite time correction:
\begin{equation}
z(t) =  z_f + \frac{1}{N} \sum_\mu
\; T_\mu \; \frac{2\lambda_\mu-z_f}{z_f-\lambda_\mu} \;
\cos\left(2\omega^{(f)}_\mu t \right)
\equiv z_f+f(t)
\; . 
\label{eq:ft-discrete}
\end{equation}
The issue now is to compute the time dependence of the sum in the second term.

One can evaluate the sums in (\ref{eq:eq_zf2}) and (\ref{eq:ft-discrete}) in the $N\rightarrow\infty$ thermodynamic
limit, introducing the density of eigenvalues,
\begin{equation}
\rho(\lambda)=\frac{1}{2\pi J^2} \, \sqrt{(2J)^2-\lambda^2} \; \theta(2J-\vert{\lambda}\vert)
\; .
\end{equation}

Concerning the sum in (\ref{eq:eq_zf2}),
\begin{equation}
1= T' \int d\lambda \, {\displaystyle{\frac{\rho(\lambda)}{z_f-\lambda}}} 
= 
\frac{T'}{2J^2} \left(z_f+\sqrt{z_f^2 - (2J)^2} \right) 
\qquad\mbox{for} \;\; z_f>2J
\qquad\quad
\Rightarrow
\qquad\quad
z_f = T' + \frac{J^2}{T'}  \; ,
\end{equation}
the correct asymptotic limit in Sectors I and IV. We already know that this expression cannot be continued on the other side of the 
transition at $T'=J$ and that the Lagrange multiplier freezes at $z_f=2J$ for $T'<J$. In order to capture this behaviour one has to go beyond the 
approximation (\ref{eq_t_lambda}) for the mode temperatures and consider that the edge-mode one scales with $N$,
similarly to what happens in the canonical equilibrium calculation with the projection of the spin on the edge mode eigenvector. 

As regards the sum in the finite time correction $f(t)$ it becomes
\begin{equation}
f(t)\equiv \int d\lambda \, \rho(\lambda) 
\; T(\lambda) \; \frac{2\lambda-z_f}{z_f-\lambda} \;
\cos\left(2\omega^{(f)}(\lambda)t \right)
\label{eq:ft-eq}
\end{equation}
where $\omega^{(f)}(\lambda)=\sqrt{(z_f-\lambda)/m}$ and
\begin{equation}
T(\lambda) = \frac{T'}{2} \left( \frac{z_f-\lambda}{z_{\rm in} - J_0/J \, \lambda} +1 \right)
\; . 
\end{equation}
We have not found  an analytical solution to this integral but we computed it numerically. We follow this route below.

\begin{figure}[h!]
\vspace{0.5cm}
\begin{center}
\vspace{-0.2cm}
\hspace{-5cm} (a) \hspace{7cm} (b) \hspace{3cm}
\\
\vspace{0.2cm}
  \includegraphics[scale=0.57]{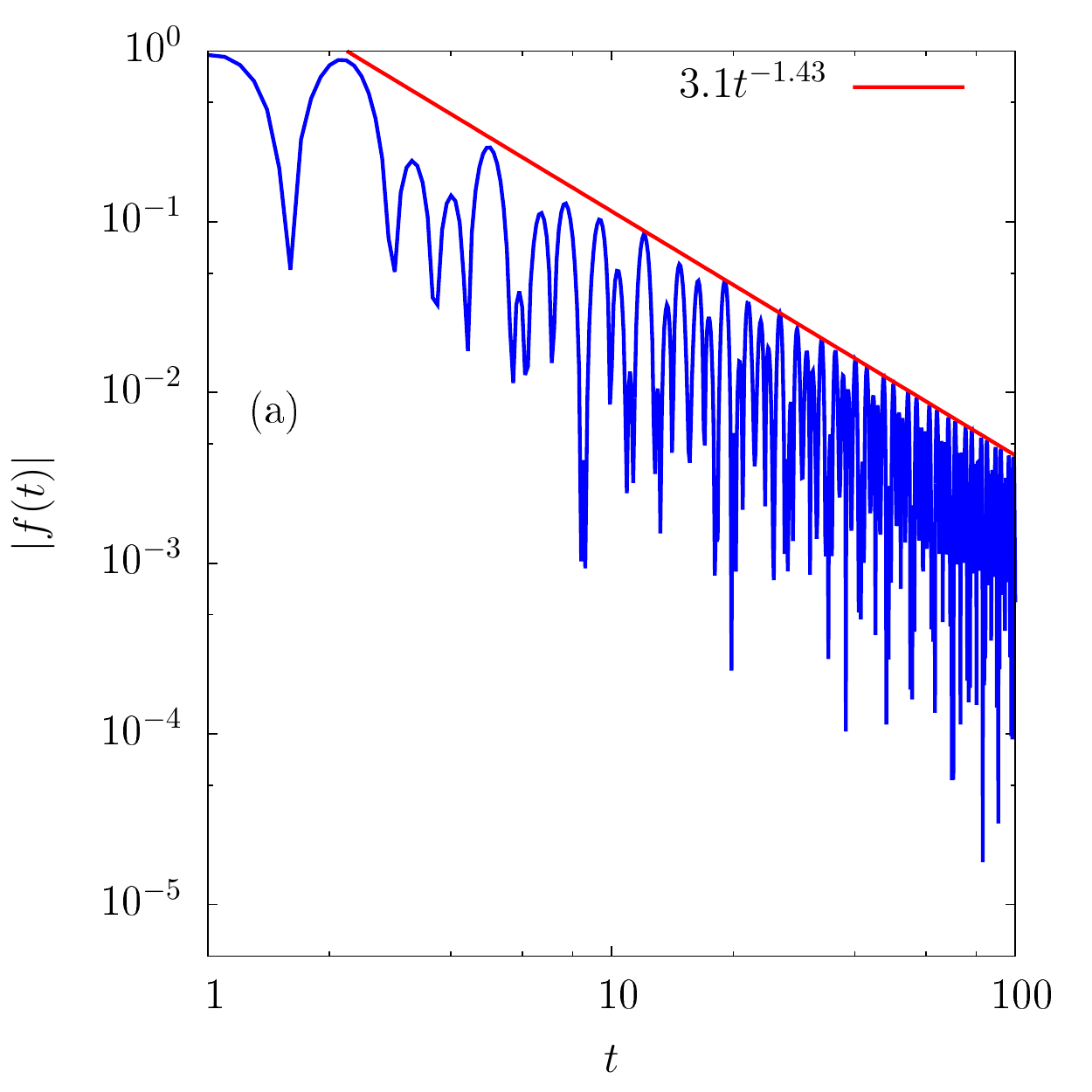}\quad%
    \includegraphics[scale=0.57]{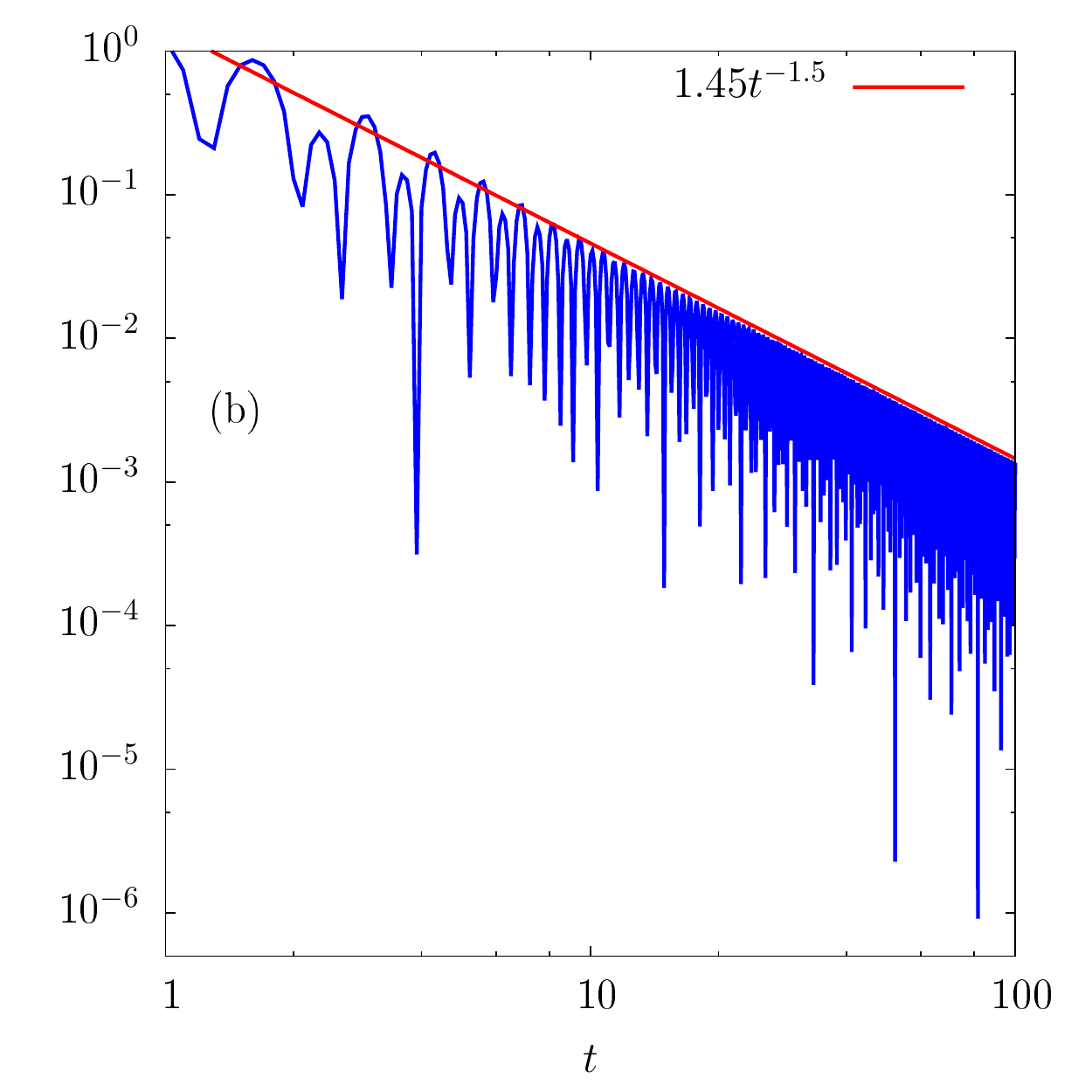}
\end{center}
\caption{\small
Decay of the function $f(t)$. (a) Sector I of the phase diagram. The red line is a decaying power law with exponent $1.43$.  (b) Sector II of the phase diagram. The red line is a decaying power law with exponent $3/2$. The results are in agreement with the solution of the mean-field equations presented in previous sections, in particular, see Fig.~\ref{fig:z_T1p25} and compare it to the data in panel (a).
} 
\label{fig:seci}
\end{figure}

In Fig.~\ref{fig:seci} (a) we show the decay of the function $f(t)$ evaluated from Eq.~(\ref{eq:ft-eq}) 
for $T'=1.25\, J_0$ and $J=0.75\, J_0$, a point which belongs to sector I of the phase diagram. 
The function $f(t)$ decays as a power law with an exponent $\alpha_z=1.43$, which is in agreement with the 
value $\alpha_z=1.45$ found using the mean field equations (see Fig.~\ref{fig:z_T1p25}). 
We note that in this case there is no problem in taking the $N\to\infty$ limit in the 
sum in (\ref{eq:ft-discrete}) since $z_{\rm in} > 2J_0$ and $z_f>2J$.
In panel (b) in the same figure 
we show the decay of $f(t)$ for $T'=1.25 \, J_0$ and $J=1.5 \, J_0$, in
Sector II (orange) in the phase diagram. $f(t)$ now decays as a power law with exponent $3/2$,  
that coincides  with  the result from the $N\to\infty$ equations that predicts $\alpha_z=3/2$ in this sector of the phase diagram.
In this case, the continuum approximation of the sum is not fully justified as $z_f=2J$. Still the agreement between the
algebraic decay of the numerical evaluation of Eq.~(\ref{eq:ft-eq}) and the slope $-3/2$ found with the 
Schwinger-Dyson approach is excellent.

\begin{figure}[h!]
\vspace{0.5cm}
\begin{center}
\vspace{-0.2cm}
\hspace{-5cm} (a) \hspace{7cm} (b) \hspace{3cm}
\\
\vspace{0.2cm}
  \includegraphics[scale=0.57]{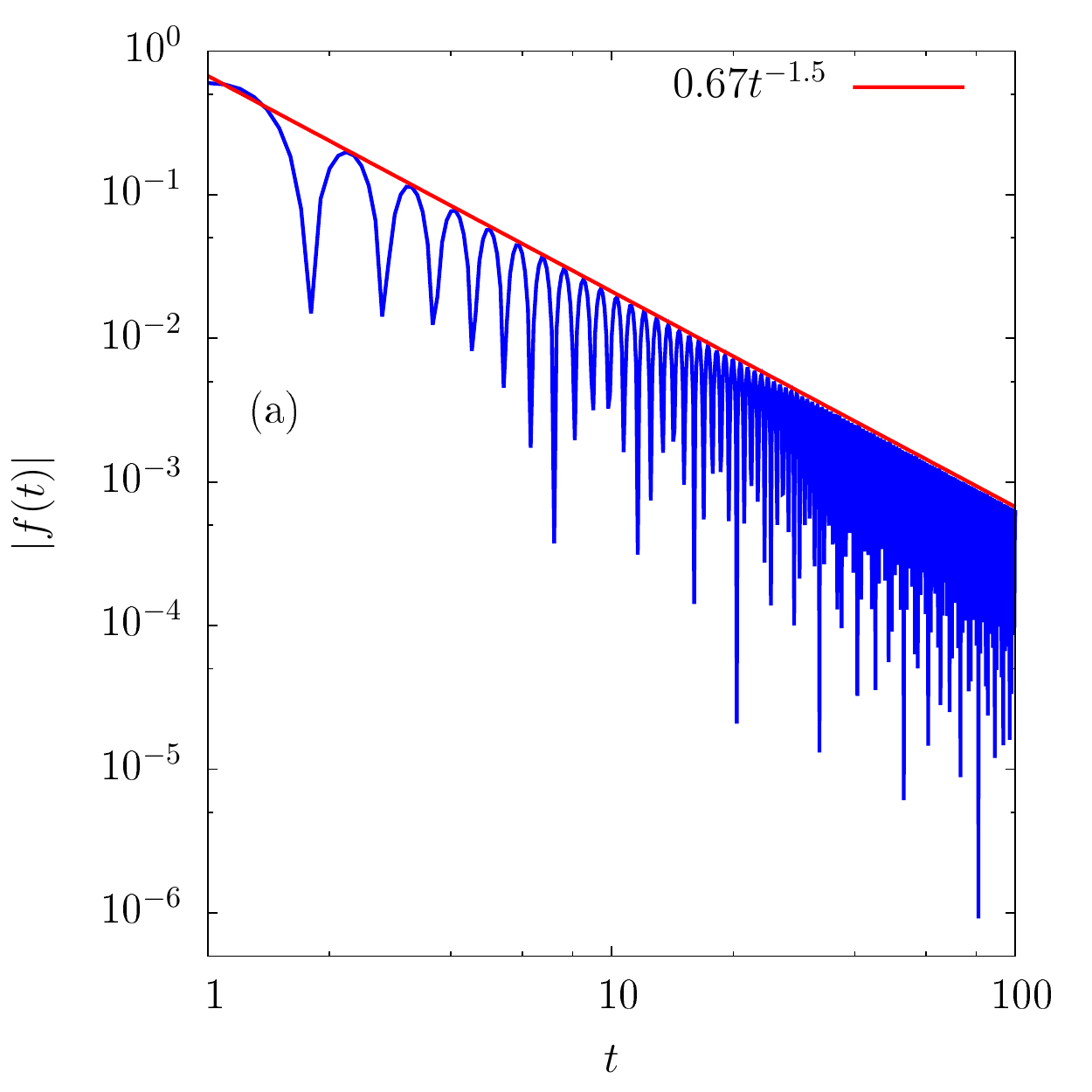}\quad%
    \includegraphics[scale=0.57]{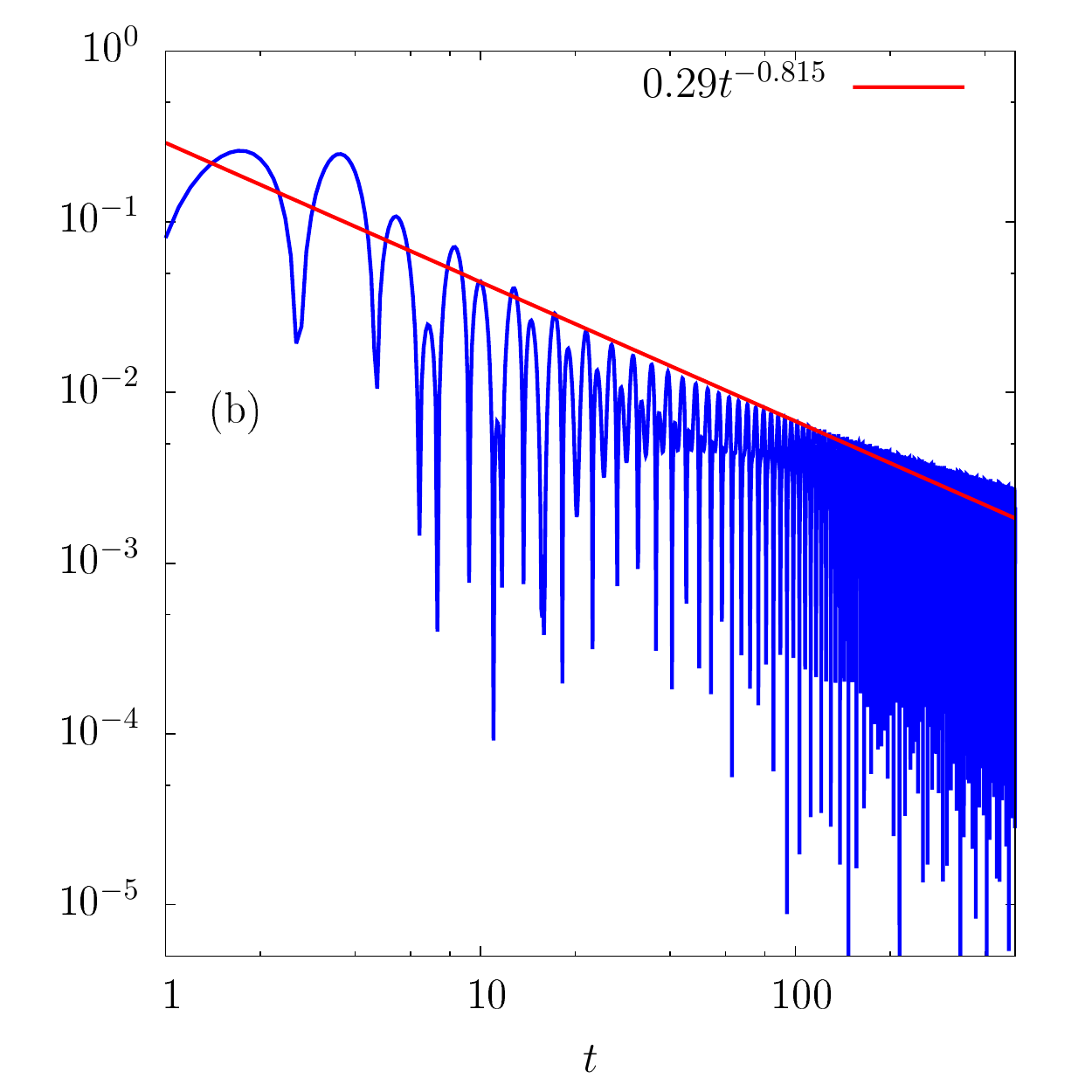}
\end{center}
\caption{\small
Decay of the function $f(t)$. (a) In Sector III of the phase diagram. The red line is a decaying power law with exponent $3/2$. The results are in agreement with the solution of the mean-field equations presented in previous sections. (b) In sector IV of the phase diagram. The red line is a decaying power law with exponent $0.815$.
}
\label{fig:seciii}
\end{figure}

In Fig.~\ref{fig:seciii} (a) we show the decay of the function $f(t)$ 
for $T'=0.5 \, J_0$ and $J=0.8 \, J_0$, in sector III of the phase diagram. 
We can clearly observe that $f(t)$ decays as a power law with the exponent $3/2$, which is in agreement with the results from the mean-field equations that predict $\alpha_z=3/2$ in this sector of the phase diagram. 
This is another case in which the continuum limit should not hold since both $z_0=2J_0$ and $z_f=2J$.

In panel (b) we show $f(t)$ for $T'=0.5 \, J_0$ and $J=0.25 \, J_0$, in sector IV of the phase diagram. The behaviour is not as clear in this case. One can 
try a power law with an exponent $\alpha_z=0.815$, 
the same value that was found for $N\rightarrow\infty$ (see Fig.~\ref{fig:z_T0p50}). The power law fits well the quasi-analytic results up to time $t=100$, approximately. For times larger than $t\simeq 100$ there are deviations from the power law behaviour. Since in this case $z_{\mathrm{in}}=2 J_0$, there is a singularity in $T(\lambda)$ at $\lambda=2J$. The deviation from the power law behaviour could be due to large numerical errors caused by this singularity.

Let us end this section with a series of comments on the behaviour of $f(t)$ in the various sectors of the phase diagram.

The case in which there is no good agreement between the power law behaviour predicted by the results from the Schwinger-Dyson $N\to\infty$ equations and the quasi-analytic results for finite $N$ is sector IV. In this sector $z_f>2J$ but $z_{in}=2J_0$ in the $N\to\infty$ limit.

We can end with a short discussion of the behaviour of  $f(t)$, more precisely, of 
each factor in the integrand that contribute in different 
ways:

--
The factor $\sqrt{(2J)^2-\lambda^2}$, originating in the density of eigenvalues, tends to cure the singularities independently of the value of $z_f$ and $z_{in}$.

--
The factor $T(\lambda)$ diverges if $z_{in}=2J_0$ and $z_f>2J$ which is the case in the problematic sector IV. If $z_{in}=2J_0$ and $z_f=2J$ it seems from the results in sector III that this factor does not cause trouble, the zero in the numerator being "cancelled" with the zero in the denominator.

--
The factor $(2\lambda-z_f)/(z_f-\lambda)$ diverges in sectors II and III, but the behaviour of the integral is smooth. Seemingly, the divergence is cured by the factor 
$\sqrt{(2J)^2-\lambda^2}$ coming from the eigenvalue density.

\section{Conclusions}
\label{sec:conclusions}

This paper complements the study of the conservative dynamics of the
$p=2$ disordered spherical model  or, in other words, the
Neumann integrable model, that we started in~\cite{CuLoNePiTa18}.
We focused here on the pre-asymptotic dynamics
of the model in the infinite system limit; more precisely, on the
way in which the approach to the constant asymptotic values of the Lagrange multiplier,
and correlation functions, is achieved.
We showed that in all phases the dynamics approach the constant values
algebraically with functions that decay as power laws but also oscillate in time.
The exponents take the same constant value $\alpha=3/2$ for $z(t)$, $\chi(t)$, $R(t)$, $C(t,0)$ and
$C(t-t')$ in all quenches such that $J>T'$, that is to say, to the right of the critical diagonal
$T'=J$ in the phase diagram. Instead, they depend on the parameters on the other side of
the diagonal $T'=J$. Apart from the algebraic decay, the functions are harmonic, with  the frequencies given by
$\omega_+$ and $\omega_-$.

The relaxational dynamics of the spherical $p=2$ model and the O(N) field theory share many points in common. In the former, the out of equilibrium dynamics after subcritical quenches correspond to the progressive alignment of the $N$ dimensional vector that collects all the spins, $\vec s = (s_1,\dots, s_N)$ on the direction of the eigenvector associated to the largest eigenvalue of the interaction matrix $J_{ij}$~\cite{CuDe95a,CuDe95b}. In the latter, the mechanism is very similar, and the progressive condensation takes place on the vanishing wave vector (tendency to local order)~\cite{CoLiZa02,Corberi-etal07}. Both models coarsen after subcritical quenches and the growing length scales have the same time-dependence $t^{1/2}$. Moreover,
the Lagrange multiplier imposing the spherical constraint, or the equivalent quantity imposing the 
self-consistent condition on the field modulus, approach their asymptotic values with the same $t^{-1}$
algebraic law.

One can then naturally ask whether the two models, endowed now with conservative dynamics, also 
share the same scaling properties. However, this is not the case. To start with, interrupted ageing properties after quenches to the critical dynamic phase transition of the quantum O(N) model~\cite{ScBi10,ScBi11,ScBi13,ChNaGuSo13,MaChMiGa15,ChTaGaMi16,ChGaDiMa17,Berges15, BoDeHoSa99, BoDeDe04} were derived in~\cite{MaChMiGa15}. In the model here considered, interrupted ageing in critical quenches, if present, 
does not last sufficiently long to be measurable and is therefore irrelevant.

\appendix
\section{Correlation with the initial configuration}
\label{app1}

We recall here a way to derive the relation $q_0 = \sqrt{q_{\rm in} q}$, obtained for models with a complex 
structure of metastable states, such as the $p$-spin spherical model with $p\geq 3$. This relation, as 
we showed numerically in the main text, also holds for the simpler $p=2$ model.

The analysis of $q_0$ enables us to draw a link with the TAP states.  
In fact, the long-term limit of the correlation function with the initial condition, Eq.~(\ref{eq:def-q0})
can also be written as
\begin{equation}
q_0 =\frac{1}{N}\sum_i \langle s_i(0)s_i(\infty)\rangle
=
\frac{1}{N}\sum_i \langle s_i(0)\rangle\langle s_i(\infty)\rangle
=\frac{1}{N}\sum_i m_i(0) m_i(\infty)
\; ,
\end{equation}
where the set $\{m_i(0)\}$ and  $\{m_i(\infty)\}$ are the local 
magnetisations of the TAP state in which the system is initialised
and the one  reached asymptotically, at $t\rightarrow+\infty$.
Recalling that
\begin{equation}
    m_i(0)=\sqrt{q_{\rm in}} \, \sigma_i(0) \hspace{0.2cm}\text{and}\hspace{0.2cm} 
    m_i(\infty)=\sqrt{q} \, \sigma_i(\infty)
\end{equation}
with
\begin{equation}
    \frac{1}{N}\sum_i \sigma_i(0)\sigma_i(0)=\frac{1}{N}\sum_i \sigma_i(\infty)\sigma_i(\infty)=1
    \; ,
\end{equation}
it follows that
\begin{equation}
q_0 = \frac{1}{N}\sqrt{q_{\rm in}q}\sum_i \sigma_i(0)\sigma_i(\infty)
\; .
\end{equation}
If we finally assume that the angular variables of the initial and final TAP states are 
identical, and that the unique effect of the quench has been to rescale their ``width'' 
the previous formula boils down to
\begin{equation}
q_0 = \sqrt{q_{\rm in}q}
\; . 
\end{equation}
This relation is checked, numerically, in Fig.~\ref{fig:asympt-qs} and the agreement is perfect, within our 
numerical accuracy.

As an example, the dissipative spherical $p$-spin model verifies this property for a 
quench with $T_{\rm in} \in [T_s,T_d]$. In fact, the dynamical equations yield, in this case, the two limits
\begin{equation}
\label{eq:1}
    q_0^2=q\Big[1-\mu (1-q)^2  q^{p-2}\Big]=q^* q 
    \;,
    \qquad\qquad
    \frac{T_{\rm in}}{T(1-q)}=\mu q_0^{{p}-2}
\end{equation}
where in the first equation we defined $q^*$ as the factor between square brackets and the 
\begin{equation}
    \mu\equiv \frac{pJ^2}{2{T}^2}
    \; . 
\end{equation}
Thus, injecting the second equation in~(\ref{eq:1}) into the first one it follows that
\begin{equation}
    \frac{T_{\rm in}}{T(1-q)}=\mu (q^* q)^{\frac{p}{2}-1}
    \; , 
\end{equation}
and then
\begin{equation}
    1
    =
    \sqrt{\mu \, \frac{pJ^2}{2{T^2_{\rm in}}}} \; (1-q) (q^* q)^{\frac{p}{2}-1}
    =\sqrt{\frac{pJ^2}{2{T^2_{\rm in}}}(1-q^*)}  \; {q^*}^{\frac{p}{2}-1}
    \; .
\end{equation}
This last equation is exactly the one that determines 
the order parameter when the system is at equilibrium with $T=T_{\rm in}$. Therefore, 
$q^*=q_{\rm in}$ and $q_0=\sqrt{q_{\rm in}q}$. 
If after the quench the system stays in the same TAP state, only the order parameter 
$q$ changes.

\vspace{0.5cm}
\noindent
{\bf Acknowledgements}

\vspace{0.5cm}

We acknowledge financial support from ECOS-Sud A14E01, PICS 506691 (CNRS-CONICET Argentina) and NSF under Grant No. PHY11-25915. LFC thanks the KITP Santa Barbara for hospitality during part of the preparation of this work. She is a member of Institut Universitaire de France.

\vspace{0.5cm}

\bibliographystyle{phaip}

\end{document}